\documentclass[11pt,a4paper]{article}
\usepackage{iftex}
\ifPDFTeX
  \usepackage[T1]{fontenc}
  \usepackage[utf8]{inputenc}
\fi
\usepackage[a4paper,margin=1in]{geometry}
\usepackage{amsmath,amssymb,mathtools}
\usepackage{XCharter}
\usepackage[charter,vvarbb]{newtxmath}
\usepackage{microtype}
\usepackage{graphicx}
\usepackage{booktabs,longtable,array,tabularx,multirow}
\usepackage{pdflscape}
\usepackage[dvipsnames]{xcolor}
\usepackage[font=small,labelfont=bf]{caption}
\usepackage{subcaption}
\usepackage{enumitem}
\usepackage{tikz}
\usetikzlibrary{arrows.meta,positioning,calc}
\usepackage[numbers,sort&compress]{natbib}
\usepackage{listings}
\usepackage[colorlinks=true,linkcolor=velblue,citecolor=velblue,urlcolor=velblue]{hyperref}

\definecolor{velclay}{HTML}{D97757}
\definecolor{velclaydeep}{HTML}{A34529}
\definecolor{velclaypale}{HTML}{F6E3D9}
\definecolor{velblue}{HTML}{216EAE}
\definecolor{velbluepale}{HTML}{D6E4F0}
\definecolor{velslate}{HTML}{141413}
\definecolor{velmuted}{HTML}{3D3D3A}
\definecolor{velgrid}{HTML}{E3E1DA}

\newcommand{\VEL}{Vitrify Exact Learning}
\newcommand{\code}[1]{\texttt{#1}}
\newcommand{\qq}[1]{\textquotedblleft #1\textquotedblright}
\newcommand{\dd}{\mathrm{d}}
\newcommand{\ee}{e}
\newcommand{\pFq}[2]{{}_{#1}F_{#2}}
\newcommand{\MeijerG}[5]{G^{\,#1,#2}_{\,#3,#4}\!\left(#5\right)}
\newcommand{\tento}[1]{10^{#1}}
\newcommand{\timestento}[1]{\times 10^{#1}}
\DeclareMathOperator{\Ai}{Ai}
\DeclareMathOperator{\Bi}{Bi}
\DeclareMathOperator{\erf}{erf}
\DeclareMathOperator{\erfc}{erfc}
\DeclareMathOperator{\Ei}{Ei}
\DeclareMathOperator{\sech}{sech}
\DeclareMathOperator{\arsinh}{arsinh}
\DeclareMathOperator{\artanh}{artanh}
\DeclareMathOperator{\logit}{logit}
\DeclareMathOperator{\probit}{probit}
\DeclareMathOperator{\rms}{rms}
\newcommand{\velneed}[1]{\par\dimen0=\pagegoal \advance\dimen0 by -\pagetotal \ifdim\dimen0<#1\relax\clearpage\fi}
\newcolumntype{L}[1]{>{\raggedright\arraybackslash}p{#1}}
\newcolumntype{C}[1]{>{\centering\arraybackslash}p{#1}}
\newcolumntype{R}[1]{>{\raggedleft\arraybackslash}p{#1}}

\title{Vitrify Exact Learning: exact differential equations,\\ changes of variable and closed forms from numerical data}
\author{Benedict W. J. Irwin,\thanks{Corresponding author: \href{mailto:ben@vitrify.ai}{ben@vitrify.ai}. ORCID: \href{https://orcid.org/0000-0001-5102-7439}{0000-0001-5102-7439}.}\hspace{0.7em}Alex Mathiasen and Tom Etminan\\[3pt] \normalsize Vitrify Labs}
\hypersetup{pdftitle={Vitrify Exact Learning: exact differential equations, changes of variable and closed forms from numerical data},
  pdfauthor={Benedict W. J. Irwin, Alex Mathiasen, Tom Etminan}}
\date{November 2026}

\begin{document}
\maketitle

\begin{abstract}
\noindent Symbolic regression searches a space of expression trees for a formula that fits data. Many functions of the sciences
have no finite expression over the usual operations and still satisfy a linear differential equation with polynomial
coefficients. \VEL{} takes numerical values of a function and proceeds in three steps. It assumes that the function, in its own
variable or after a change of variable or a pushforward of a density, is annihilated by an operator with polynomial
coefficients. It finds that operator as the null vector of a matrix of weak-form measurements, in which the data are integrated
against test functions and never differentiated, makes the coefficients exact, and tests the exact operator on data that the
fit did not read against a noise level propagated from the data. It then solves the exact operator symbolically and keeps the
closed form that matches the data. Each step has precedent; the combination takes values on an interval, from a quadrature, a
float64 table or a sample, to an exact equation and a closed form. On noise-free benchmarks it recovers 45 of 53 classic
targets, 108 of the 120 AI Feynman equations and 74.1\% of the Feynman problems of SRBench, against 33, 69 and 55.8\% for the
best other programs, and 50 of 132 special-function targets generated after its code was frozen, against 23. From 40-digit
values it returns Meijer $G$ densities, hypergeometric forms for the deflection of light by a black hole, and Appell and
Lauricella functions of two variables. With noise in several variables, with sparse data and for equations above fourth order
it mostly returns no model.
\end{abstract}

\section{Introduction}\label{sec:intro}

A calculation often ends in numbers whose law is not known: an integral evaluated by quadrature for each value of a
parameter, a series summed term by term, a density estimated from a simulation. What is wanted is the exact expression behind
the numbers, one that still holds outside the range that was computed. Symbolic regression is the usual approach. It searches a
space of expression trees, with variables and constants at the leaves and operations such as addition, sine and exponential
at the nodes, for a tree that fits the data \cite{koza1992gp,schmidt2009distilling,cranmer2023pysr,lacava2021srbench}. The
search is combinatorial and NP-hard \cite{virgolin2022nphard}, and its answer has to be a finite tree over the chosen
operations, which most of the functions of a handbook of special functions \cite{olver2010handbook,dlmf} are not.

\begin{figure}[t]
\centering
\includegraphics[width=\textwidth]{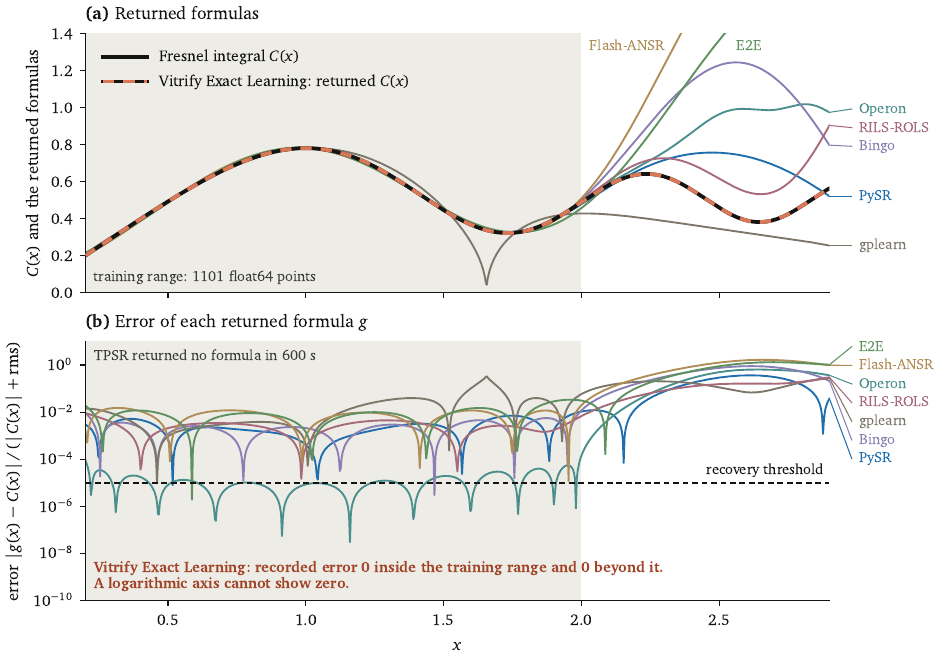}
\caption{The Fresnel integral $C(x)$ and the formulas that seven symbolic-regression programs returned for 1101 float64 points
of it on $[0.2,\,2]$ (grey band). (a)~The formulas continued to $x = 2.9$; \VEL{} returned $C(x)$ itself (dashed). (b)~The
error $|g - C| / (|C| + \rms)$ of each returned formula $g$, where rms is the root mean square of the training values. The
dashed line is the recovery threshold $\tento{-5}$ used throughout. The recorded error of \VEL{} is 0 inside the training
range and beyond it, which a logarithmic axis cannot show. TPSR returned no formula within 600\,s.
Table~\ref{tab:fresnel} gives the numbers and Section~\ref{app:fresnel} the seven formulas.}
\label{fig:fresnel_fits}
\end{figure}

Figure~\ref{fig:fresnel_fits} shows what happens then. The Fresnel integral $C(x) = \int_0^x \cos(\pi t^2/2)\,\dd t$ has no
finite expression in sines, exponentials and logarithms. Seven symbolic-regression programs given 1101 points of it on
$[0.2,\,2]$ each returned a formula; six fit the points to within 3.4\%, and all seven are wrong by 27\% to 167\% within half
a range-width beyond the data. Each returned the nearest thing that it could build.

We ask a different question of the same data: which linear differential equation does the function obey? Most special
functions of physics and statistics satisfy one with polynomial coefficients; they are called holonomic, or D-finite
\cite{stanley1980dfinite,zeilberger1990holonomic,kauers2023dfinite}. The Fresnel integral has no formula tree, and its
equation has three terms,
\begin{equation}\label{eq:fresnel}
  x\,C'''(x) - C''(x) + \pi^2 x^3\,C'(x) = 0 .
\end{equation}
Once an order and a degree are fixed, finding such an equation is linear algebra, and once the equation is exact, its solutions
can be found by computer algebra. The method of this paper, \VEL{}, is three steps:
\begin{center}
\setlength{\fboxsep}{0pt}\fcolorbox{velclay}{white}{\begin{minipage}{\dimexpr\linewidth-2\fboxrule\relax}\vspace{-4pt}
\begin{alignat}{2}
  &\text{1. Ansatz:} \quad && f(x) = |\varphi'(x)|^{k}\,g\bigl(\varphi(x)\bigr), \qquad
     L g = \sum_{i=0}^{r} \sum_{j=0}^{d} c_{ij}\,y^{j}\,\partial_y^{i}\,g = 0 , \label{eq:step1}\\
  &\text{2. Annihilator:} \quad && \hat c = \operatorname{exact}(\ker M), \qquad
     M_{t,(i,j)} = \int g(y)\,(-\partial_y)^{i}\bigl[y^{j} B_t(y)\bigr]\,\dd y , \label{eq:step2}\\
  &\text{3. Closed form:} \quad && f = |\varphi'|^{k} \sum_{l} C_l\; g_l \circ \varphi, \qquad
     \{g_l\}\ \text{a basis of}\ \ker L_{\hat c},\ \text{found symbolically} . \label{eq:step3}
\end{alignat}
\vspace{-2pt}\end{minipage}}
\end{center}
The function is assumed holonomic in some variable $y = \varphi(x)$, either as a plain substitution ($k = 0$) or as the
pushforward of a density ($k = 1$). The coefficients $c$ of its annihilating operator $L$ are a null vector of a matrix $M$ of weak-form measurements: the
data are integrated against test functions $B_t$ that vanish at the ends of a window, so no derivative of the data is taken.
The null vector at the least order $r$ and degree $d$ is made exact, as fractions and named constants, and the exact operator
must pass a test on data that the fit did not read. The solutions of the exact operator are then constructed symbolically, and
the data choose the combination. From the 1101 float64 points of Figure~\ref{fig:fresnel_fits} this returns
equation~\eqref{eq:fresnel} and $C(x)$ in 57\,s.

Section~\ref{sec:overview} describes the three steps, Section~\ref{sec:related} relates them to earlier work and states what
is new, Section~\ref{sec:results} gives examples and benchmarks, and Section~\ref{sec:mainlimits} the limits. The
Supplementary Information that follows the references holds the full method, every example and every benchmark table. On
noise-free benchmarks the method recovers more targets than eight open symbolic-regression programs, including 50 of 132
special-function targets that were generated after its code was frozen, against 23 for the best other program. With noise,
with sparse data and for equations above fourth order it mostly returns no model.

The name exact learning comes from earlier work of the first author \cite{irwin2020natural,irwin2021natural,irwin2020exactlearning},
which fitted the Mellin transform of a function with a product of gamma functions and then looked for exact values of the
fitted parameters. Its notes proposed to use the holonomic equation of the function instead; this paper takes up that proposal.

\section{The three steps}\label{sec:overview}

Figure~\ref{fig:pipeline} shows the steps for a function of one variable. Section~\ref{sec:method} of the Supplementary
Information gives every detail; the implementation is about 4{,}200 lines of Python on mpmath \cite{mpmath} and SymPy
\cite{meurer2017sympy}.

\begin{figure}[t]
\centering
\begin{tikzpicture}[
  font=\footnotesize,
  box/.style={draw=velslate, rounded corners=2pt, align=center, inner sep=4pt, minimum height=13mm, text width=30mm, fill=white},
  data/.style={box, fill=velbluepale},
  result/.style={box, fill=velclaypale},
  arr/.style={-{Stealth[length=2mm]}, semithick, draw=velmuted},
  node distance=6mm and 6mm]
\node[data] (data) {values of $f$: callable at $N_p$ digits, float64 table or draws};
\node[box, right=of data] (mom) {window moments $\nu_m$ with realisations of their error};
\node[box, right=of mom] (weak) {weak matrix $M$ for order $r$ and degree $d$};
\node[box, right=of weak] (svd) {smallest singular value at its noise level?};
\node[box, below=11mm of svd] (exact) {exact coefficients: fractions, constants, evidence in digits};
\node[box, left=of exact] (hold) {exact operator on the held-out window: residual $\le 10\,\times$ noise?};
\node[result, left=of hold] (cert) {certificate: map, operator and grade};
\node[result, left=of cert] (solve) {closed form from the solver, which reads the exact operator};
\draw[arr] (data) -- (mom);
\draw[arr] (mom) -- (weak);
\draw[arr] (weak) -- (svd);
\draw[arr] (svd) -- node[right]{yes} (exact);
\draw[arr] (exact) -- (hold);
\draw[arr] (hold) -- node[above]{yes} (cert);
\draw[arr] (cert) -- node[above]{best} (solve);
\draw[arr] (svd.north) -- ++(0,4mm) -| (weak.north);
\node[anchor=south, text=velmuted] at ($(weak.north)!0.5!(svd.north)+(0,4mm)$) {no: next $(r, d)$};
\draw[dashed, draw=velmuted, rounded corners=3pt] ($(mom.north west)+(-2.5mm,9.5mm)$) rectangle ($(exact.south east)+(2.5mm,-7.5mm)$);
\node[anchor=south west, text=velmuted, font=\footnotesize\itshape] at ($(cert.south west)+(-1.5mm,-7.3mm)$)
  {repeated for each change of variable $y = \varphi(x)$ and weight $k$};
\end{tikzpicture}
\caption{The steps of \VEL{} for a function of one variable. The data enter only through moments on windows. At each order
the search over the degree stops at the first support whose smallest singular value is at its noise level, and it moves to
the next order when that null vector is not unique or cannot be made exact. The coefficients are made exact before the
operator is tested on a window that the fit did not read.}
\label{fig:pipeline}
\end{figure}
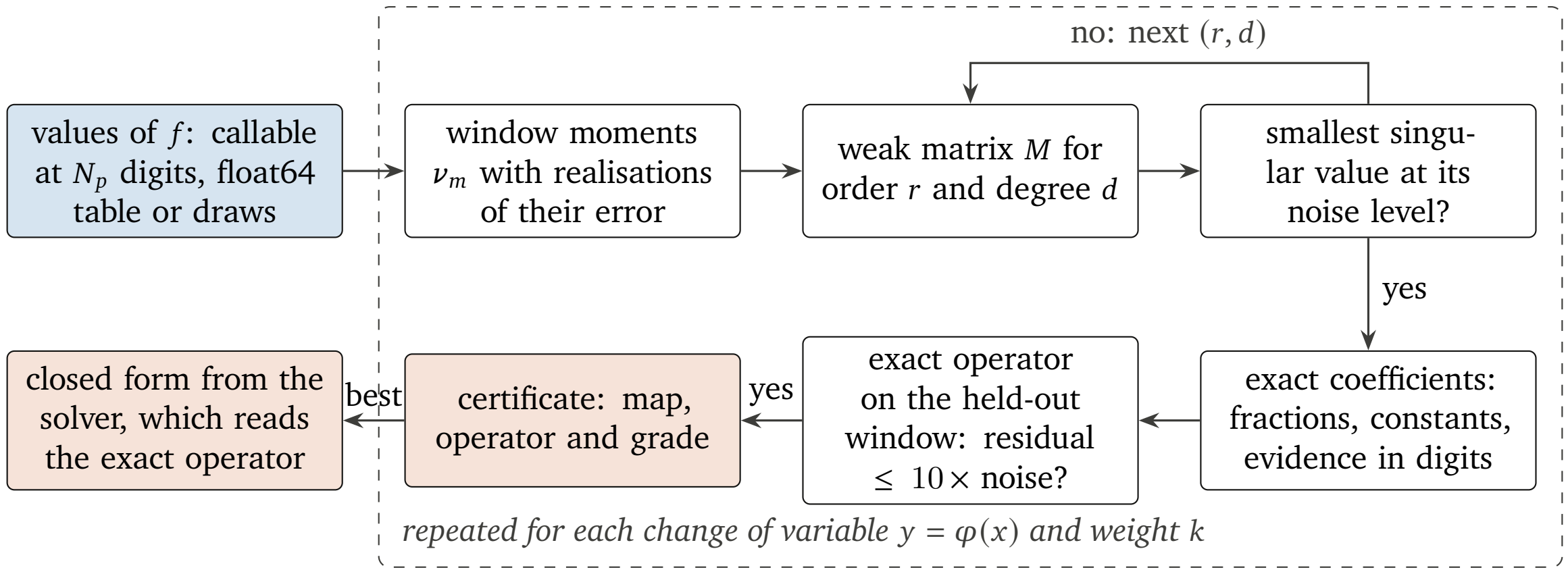

\paragraph{1. Ansatz.} A function is holonomic when $L f = 0$ for an operator $L = \sum_{i \le r} p_i(x)\,\partial_x^i$ with
polynomial coefficients \cite{stanley1980dfinite,zeilberger1990holonomic}. The class contains the exponential and the
trigonometric functions, the error function, Bessel, Airy and Struve functions, elliptic integrals, orthogonal polynomials,
the generalised hypergeometric functions $\pFq{p}{q}$ and Meijer's $G$-function, and it is closed under sums, products,
derivatives, antiderivatives and algebraic substitution. Many densities are outside it in $x$ and inside it in another
variable: the lognormal density has no such operator in $x$ and a first-order one in $y = \log x$. The ansatz~\eqref{eq:step1}
therefore runs the search on the transported data
\begin{equation}\label{eq:transportmain}
  q(y) = f\bigl(\varphi^{-1}(y)\bigr)\,\bigl|(\varphi^{-1})'(y)\bigr|^{k} ,
\end{equation}
for each map $\varphi$ of a bank of twenty (the identity, $\log x$, $\ee^{x}$, $1/x$, powers, logit, probit, $\arsinh$,
$\tan$ and others), for a power $x^{\rho}$ whose exponent is fitted and identified, and for maps that the caller supplies.
With $k = 1$, $q$ is the pushforward of the density $f$ under $\varphi$; with $k = 0$ it is a substitution. An operator found
in $y$ is pulled back to an operator in $x$ when its coefficients become polynomials there. When they do not, the
representation $f = |\varphi'|^{k} g \circ \varphi$ with $L_y g = 0$ is the result, which extends the class beyond holonomic
functions: $\tan x$ is returned as the solution of $y\,g' - g = 0$ in $y = \tan x$.

\paragraph{2. Annihilator.} On a window $[c - h,\,c + h]$ with local coordinate $s$ the only numbers read from the data are
the moments $\nu_m = \int_{-1}^{1} s^m\,q(c + h s)\,\dd s$. The test functions are $B_t(s) = (1 - s^2)^r P_t(s)$, with $P_t$
the Legendre polynomial of degree $t$. They and their first $r - 1$ derivatives vanish at both ends of the window, so
integration by parts moves every derivative onto $B_t$ without boundary terms, and each entry of the weak matrix $M$
of~\eqref{eq:step2} is a combination of moments with rational coefficients. If $L$ annihilates $q$ then $M c = 0$. Three
windows covering 4--70\% of the interval are fitted, and the fourth, 70--96\%, is held out.

Each kind of data also gives realisations of the error of its moments: the difference of two quadrature levels for a
callable, the difference of two interpolants for a table, batch means for draws. The same linear maps applied to these
realisations give the noise level $\eta$ of every later quantity, so no tolerance is set by hand for a particular function.
The search visits $(r, d) = (1, 0), (1, 1), \ldots$ up to order 4 and degree 8, and stops at the first support at which the
smallest singular value $\sigma_1$ of the column-scaled matrix is within $10\,\eta(v_1)$, the next one is not, and the null
vector $v_1$ can be made exact. The result is the operator of least order and, at that order, of least degree. When this
search finds nothing, operators with two or three terms, $\sum_k x^{k m} P_k(x\partial_x)$, are searched separately; with two
terms they are the equations of $\pFq{p}{q}$ and Meijer $G$-functions of $x^m$, and since they have far fewer unknowns they are
searched up to order 6.

$\operatorname{exact}(\cdot)$ in~\eqref{eq:step2} divides $v_1$ by a pivot entry and replaces each ratio $\rho$, known to
within $\tau$, by the fraction of least height $H$ in $[\rho - \tau,\,\rho + \tau]$ when
$e = -\log_{10}\bigl(\tau / \max(|\rho|, 1)\bigr) - 2 \log_{10} H$, the number of digits by which it beats chance, is large
enough; otherwise PSLQ \cite{ferguson1999pslq} looks for an algebraic number or a rational combination of $\pi$, $\ee$,
$\log 2$, $\sqrt{\pi}$ and other constants. At most two entries may stay numerical, as an energy eigenvalue does. The exact
operator is then applied to the held-out window and passes when its residual there is at most ten times its noise level; for
a callable the test is repeated with moments computed at $2N_p$ digits, where $N_p$ is the working precision. Every run in this
paper uses $N_p = 40$, so the second test is at 80 digits, and a callable without a result is tried once more at $N_p = 90$. A
result that passes with noise below $\tento{-8}$ of the size of its terms and at least 10 digits of evidence is graded
\code{verified}; one that passes otherwise is a \code{conjecture}.

\paragraph{3. Closed form.} The solver reads the exact operator and nothing else, and builds candidate solutions by rules:
exponentials of integrals for order 1; exponential polynomials for constant coefficients; Frobenius series about each
singular point, which for the two-term form $\prod_j(\vartheta - b_j) f = \mu z \prod_j (\vartheta - \alpha_j) f$, with
$z = t^m$ and $\vartheta = z\,\dd/\dd z$, are hypergeometric series $z^b\,\pFq{p}{q-1}(\ldots;\,\mu z)$; Meijer $G$-functions for
every split of the same parameters, which covers solutions with logarithms; Airy and Bessel forms; symmetric squares;
antiderivatives, including binomial integrands, whose primitives are Appell and Lauricella functions
\cite{appell1926,lauricella1893}; and gauge factors $\exp(a x^2 + b x) \prod_j q_j^{e_j}$. SymPy's \code{hyperexpand}
\cite{roach1996hypergeometric,meurer2017sympy} names the series where it can. A candidate is kept when it satisfies the
operator numerically; the data select the smallest set whose span contains them, and the constants of the combination go
through the same identification step. When no rule applies, the operator and $r$ initial values fitted to the data still
determine the function, which is then evaluated anywhere by analytic continuation of its Taylor series
\cite{vanderhoeven1999holonomic,mezzarobba2010numgfun}.

\paragraph{Other inputs.} Independent draws from a density enter through sample moments on wide windows, and the held-out
test becomes Hotelling's $T^2$ \cite{hotelling1931}. Exact terms of a sequence give a recurrence by exact linear algebra, and
exact moments of a density give a recurrence that becomes a differential operator for the density. For float64 tables in
several variables a separate entry point searches for algebraic relations over elementary atoms and for scaling symmetries,
and passes one remaining variable to the operator search. For a callable in two variables the same weak form on boxes gives a
system of partial differential operators, and two two-term operators with independent shifts fix a double hypergeometric
series (Section~\ref{sec:bivariate}).

\section{Related work and what is new}\label{sec:related}

Every part of the three steps has precedent. Table~\ref{tab:related} sets the closest methods side by side, and
Section~\ref{sec:background} of the Supplementary Information describes them and symbolic regression in more detail.

\begin{table}[t]
\centering\footnotesize
\caption{Methods that infer an equation or a formula from numbers. \emph{Input}: what the method reads. \emph{Equation}: what
it returns. \emph{Coefficients}: real numbers fitted to the data, or exact numbers. \emph{Map}: whether a change of variable
is searched. \emph{Closed form}: whether the method returns the function in named terms. \emph{Test}: how the result is
checked.}
\label{tab:related}
\renewcommand{\arraystretch}{1.12}\setlength{\tabcolsep}{3pt}
\begin{tabular}{@{}L{30mm}L{27mm}L{29mm}L{16mm}L{10mm}L{15mm}L{19mm}@{}}
\toprule
Method & Input & Equation & Coefficients & Map & Closed form & Test \\
\midrule
Symbolic regression \cite{koza1992gp,cranmer2023pysr,lacava2021srbench} & points & expression tree & real & no & the tree & held-out points \\
Pearson's system \cite{pearson1895} & moments of a density & $f'/f$ a ratio of polynomials of degree 1 and 2 & real & no & by type & none \\
Modulating functions \cite{shinbrot1957,pearson1985modulating,preisig1993modulating} & input--output records, weak form & linear ODE of a system & real & no & no & fit residual \\
SINDy, implicit SINDy, SINDy-PI \cite{brunton2016sindy,mangan2016implicit,kaheman2020sindypi} & trajectories, differentiated & sparse sum over a library & real, sparse & no & no & held-out data \\
Weak SINDy, WENDy \cite{messenger2021wsindy,messenger2021wsindypde,bortz2023wendy} & trajectories, weak form & sparse sum over a library & real, sparse & no & no & held-out data \\
Moment inversion \cite{batenkov2009moment} & moments on an interval & linear ODE, polynomial coefficients, and jumps & real & no & no & none \\
Guessing \cite{salvy1994gfun,kauers2015ore,kauers2022guessing} & exact series or sequence terms & linear ODE or recurrence, polynomial coefficients & exact & no & separate algorithms & further terms \\
Differential approximants \cite{guttmann1972approximants} & series coefficients & linear ODE, polynomial coefficients & real & no & no & agreement of approximants \\
Integer relations \cite{ferguson1999pslq,bailey2001integer,bailey2008bessel} & high-precision numbers & relation among constants & exact & no & case by case & more digits \\
Mellin-space fitting \cite{irwin2020natural,alaa2019metamodels} & Mellin transform or a model & gamma products, Meijer $G$ & real, then rounded & no & by construction & none \\
\addlinespace[2pt]
\textbf{\VEL{}} & values on an interval (callable, table, draws), sequences, moments & linear ODE, polynomial coefficients, in $\varphi(x)$ & exact, with evidence & bank, fitted power & solver & held-out window against propagated noise, at $N_p$ and $2N_p$ \\
\bottomrule
\end{tabular}
\end{table}

\paragraph{The ansatz.} Fitting a linear differential equation with polynomial coefficients to a function is old. Pearson
fitted $f'/f = (x - a)/(b_0 + b_1 x + b_2 x^2)$ to the moments of a density in 1895 \cite{pearson1895}, which is our ansatz at
order 1 and degree 2, and the beta, gamma and Student densities are his types. Differential approximants fit such equations to
series coefficients to locate singular points \cite{guttmann1972approximants}. Batenkov reconstructed a function that satisfies
an unknown equation of this kind on each piece of an interval from its moments, through a recurrence that the moments obey
\cite{batenkov2009moment}. Our moment input (Section~\ref{sec:other}) is that recurrence for a single piece. The change of
variable, with the weight that makes it a pushforward of a density, has no counterpart in these methods.

\paragraph{The weak form.} Integrating data against test functions that vanish at the ends of an interval, so that the
derivatives of an equation fall on the test functions, is Shinbrot's method of modulating functions \cite{shinbrot1957}. Pearson
and Lee used Fourier modulating functions to identify linear systems from input--output records
\cite{pearson1985modulating}, Preisig and Rippin reviewed the method \cite{preisig1993modulating}, and weak SINDy and WENDy
use it to identify nonlinear dynamics from noisy trajectories \cite{messenger2021wsindy,messenger2021wsindypde,bortz2023wendy}.
The choice of test functions matters for noise: weak SINDy places many low-order bumps $(t - a)^p (b - t)^q$ across the domain,
and recent work constructs them from the data \cite{tran2026testfunctions}. We use three windows and Legendre polynomials of
high degree times $(1 - s^2)^r$, which suits values known to many digits and may be one reason for the poor results under
noise (Section~\ref{sec:mainlimits}).

\paragraph{The null vector and sparsity.} Writing an equation as a null vector of a matrix of candidate terms is the implicit
form of SINDy \cite{mangan2016implicit}; SINDy-PI was introduced because that form is sensitive to noise
\cite{kaheman2020sindypi}. Computer-algebra guessing solves the same kind of system exactly
\cite{salvy1994gfun,kauers2015ore,kauers2022guessing}, and for sequences a transformer has been trained to propose recurrences
directly \cite{dascoli2022recurrent}. Our method does not use sparse regression. Its sparsity is the
smallest support: the first order and, at that order, the first degree at which the matrix has a null vector that is unique,
plus the families of two- and three-term operators, and each decision is a comparison of a singular value with its propagated
noise.

\paragraph{Exact numbers.} Recognising an exact number from its digits by PSLQ is standard experimental mathematics
\cite{ferguson1999pslq,bailey2001integer}. Bailey, Borwein, Broadhurst and Glasser computed moments of Bessel functions to
high precision, recognised closed forms for several of them, and stated the identities that they could not prove as
conjectures checked to high precision \cite{bailey2008bessel}. Our identification step uses the same tools, with a count of the digits by which a fraction beats
chance and a calibration of the false-identification rate (Table~\ref{tab:calibration}).

\paragraph{The closed form.} Solving an exact operator is computer algebra: Kovacic's algorithm \cite{kovacic1986},
hypergeometric solutions after a change of variable \cite{imamoglu2017hypergeometric}, Roach's expansion of hypergeometric
and Meijer $G$-functions \cite{roach1996hypergeometric,roach1997meijer}, factorisation of differential operators
\cite{chyzak2022factorization} and rigorous continuation \cite{vanderhoeven1999holonomic,mezzarobba2010numgfun,mezzarobba2016rigorous}.
Our solver is a set of rules of this kind and is not complete. We ran the factorisation, the hyperexponential solutions, the
irreducibility test and the continuation of ore\_algebra \cite{kauers2015ore} on the 21 certified operators for which our
solver found no closed form (Section~\ref{sec:orecheck}). It found no closed form either. It proved two of the operators
irreducible and gave certified error bounds for the continued solutions, which agree with ours to $\tento{-33}$ or better. Its
guesser returned the same recurrence as our learner for all 22 sequence and moment inputs that have one.

\paragraph{What is new.} The guessing programs need exact series or sequence terms at one point, and the weak-form methods
return equations with real coefficients for a dynamical system. What we have not found in earlier work is the combination that
takes numerical values of one function on an interval to an exact equation and a closed form:
\begin{enumerate}[leftmargin=*,itemsep=1pt,topsep=3pt]
\item weak-form moments on windows of the interval in the place of Taylor coefficients, so that guessing runs on values from a
quadrature, a float64 table or a sample;
\item realisations of the error of the data, carried through every linear map, which decide when a singular value is zero,
when a coefficient is a fraction and whether the exact operator holds on the held-out window;
\item a search over substitutions and pushforwards, with the pullback of the operator to $x$, which extends the ansatz to
functions that are not holonomic in $x$;
\item a symbolic solution of the exact operator, with the data choosing the combination of solutions and its constants made
exact in turn.
\end{enumerate}
The search of the literature behind this statement was limited, and we would welcome references to earlier work of this kind.

\section{Results}\label{sec:results}

Every number below is from a recorded run of one frozen version of the learner, with $N_p = 40$ for callables. A target is
\emph{recovered} when the returned function agrees with it to $\tento{-5}$, relative to $|f|$ plus the rms of the training
values, both inside the training range and beyond it, up to half a range-width past its upper end. The rule does not compare
expression trees, so a correct form with fitted constants passes and an approximant that fits the training range fails
(Section~\ref{sec:protocol}).

\begin{table}[tp]
\centering\small
\caption{Some identities returned by \VEL{}. \emph{Input}: a callable at $N_p = 40$ digits or a float64 table of 1101 points,
on the interval or box given. The operator is the one returned; $a_1$ is a coefficient that stayed numerical. Every result is
graded \code{verified}. Sections~\ref{sec:examples} and~\ref{sec:benchmarks} and Tables~\ref{tab:examples},~\ref{tab:physical_operators}
and~\ref{tab:bivariate} give the tests, the times and the errors against the target.}
\label{tab:highlights}
\renewcommand{\arraystretch}{1.35}
\begin{tabular}{@{}L{34mm}L{57mm}L{57mm}@{}}
\toprule
Quantity, input & Operator returned & Closed form returned \\
\midrule
Morse oscillator, $n = 2$; callable on $[-0.5,\,3]$ &
  in $y = \ee^{x}$: $2 y^2 (y - 2)(y - 1)\,g' + (7 y^3 - 39 y^2 + 58 y - 24)\,g = 0$ &
  $124416 \sqrt{3}\,(\ee^{x} - 2)(\ee^{x} - 1)\,\ee^{-11 x/2 - 6 \ee^{-x}}$ \\
Quantum bouncer, ground state; callable on $[0.1,\,6]$ &
  $\psi'' + (a_1 - x)\,\psi = 0$, $a_1 = 2.33810741045976703848919725245$ &
  $\Ai(x - a_1)$; $a_1$ is minus the first zero of $\Ai$ \\
Fluorescence recovery after photobleaching \cite{soumpasis1983frap}; callable on $[0.2,\,6]$ &
  $t^3 f'' - 4 t f' + 2 f = 0$ &
  $\ee^{-2/t}\,\bigl[I_0(2/t) + I_1(2/t)\bigr]$ \\
Density of a product of four standard normals; float64 table on $[0.2,\,3]$ &
  $x^3 f'''' + 6 x^2 f''' + 7 x f'' + f' - x f = 0$ &
  $\dfrac{1}{4 \pi^2}\,\MeijerG{4}{0}{0}{4}{\dfrac{x^2}{16}\;\middle|\;\begin{smallmatrix}\text{--}\\ 0, 0, 0, 0\end{smallmatrix}}$ \\
Deflection of light against $u = M/b$; callable on $[0.01,\,0.12]$ &
  $u (27 u^2 - 1)\,\alpha'''' + 3 (81 u^2 - 1)\,\alpha''' + 501 u\,\alpha'' + 192\,\alpha' = 0$ &
  $4 u\,\pFq{3}{2}\bigl(\tfrac23, 1, \tfrac43;\,\tfrac32, \tfrac32;\,27 u^2\bigr) + \pi \bigl[\pFq{2}{1}\bigl(\tfrac16, \tfrac56;\,1;\,27 u^2\bigr) - 1\bigr]$ \\
Period of the oscillator $q^2/2 - q^5/5$ at energy $x$; callable on $[0.01,\,0.18]$ &
  order 4, degree 4 &
  $2\pi\,\pFq{4}{3}\bigl(\tfrac{1}{10}, \tfrac{3}{10}, \tfrac{7}{10}, \tfrac{9}{10};\,\tfrac13, \tfrac23, 1;\,\tfrac{1000 x^3}{27}\bigr)$ \\
Thermonuclear rate integral $\int_0^\infty \ee^{-y - x/\sqrt{y}}\,\dd y$; callable on $[0.2,\,4]$ &
  $x f''' + 2 f = 0$ &
  $\pi^{-1/2}\,\MeijerG{3}{0}{0}{3}{\dfrac{x^2}{4}\;\middle|\;\begin{smallmatrix}\text{--}\\ 0, \frac12, 1\end{smallmatrix}}$ \\
$\Pi(n \mid m)$, two variables; callable on $[0.05,\,0.55]^2$ &
  three operators, including $2 (m - n)\,\partial_n \partial_m f + \partial_n f - 2\,\partial_m f = 0$ &
  $\dfrac{\pi}{2}\,F_1\bigl(\tfrac12;\,\tfrac12, 1;\,1;\,m, n\bigr)$ \\
Carlson's $R_F(x, y, 1)$, two variables; callable on $[0.3,\,0.8] \times [0.35,\,0.9]$ &
  in $u = 1 - x$, $v = 1 - y$: includes $2 (u - v)\,\partial_u \partial_v f - \partial_u f + \partial_v f = 0$ &
  $F_1\bigl(\tfrac12;\,\tfrac12, \tfrac12;\,\tfrac32;\,1 - y,\,1 - x\bigr)$, Lauricella's $F_D$ \\
\bottomrule
\end{tabular}
\end{table}

\subsection{The Fresnel integral}\label{sec:mainfresnel}

\begin{figure}[t]
\centering
\includegraphics[width=\textwidth]{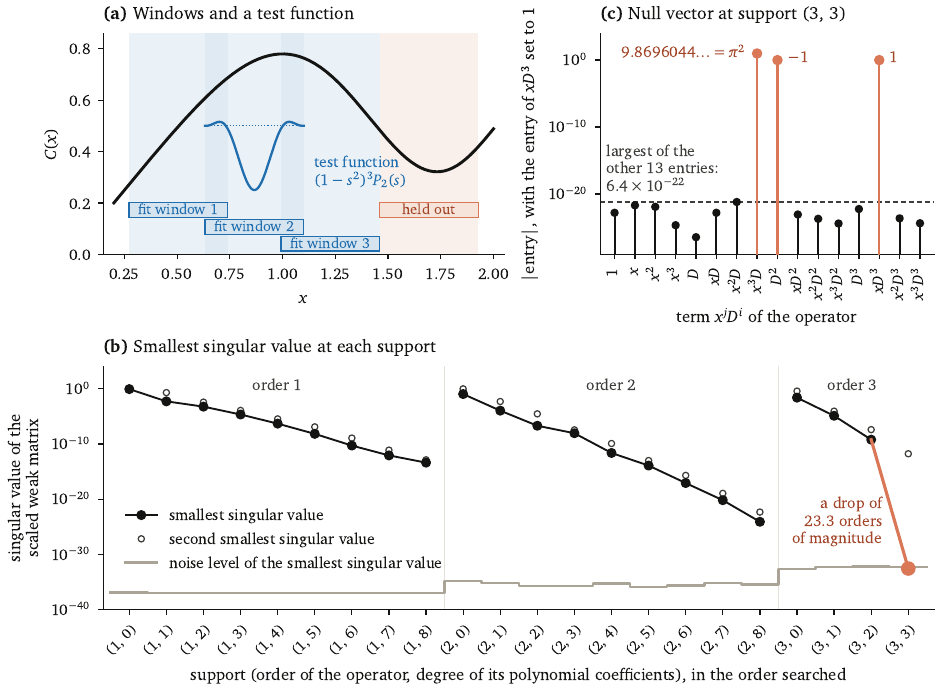}
\caption{The search for the operator of the Fresnel integral $C(x)$ from values at 40 digits. (a)~$C(x)$ on $[0.2,\,2]$
with the three fit windows, the held-out window and one of the test functions. (b)~The smallest singular value of the scaled
weak matrix (filled), the second smallest (hollow) and the noise level of the smallest (grey line) at each of the 22 supports
visited. (c)~The entries of the null vector at order 3 and degree 3, divided by the entry of $x D^3$.}
\label{fig:fresnel_trace}
\end{figure}

From values at 40 digits the search visits 22 supports (Figure~\ref{fig:fresnel_trace}). At orders 1 and 2 the smallest
singular value falls steadily as columns are added and stays far above its noise level. At order 3 and degree 3 it drops by
23 orders of magnitude in one step, to $\tento{-32.5}$ against a noise level of $\tento{-32.2}$, while the next one stays at
$\tento{-11.8}$. Divided by the entry of $x D^3$, the null vector has 13 entries below $6.4 \timestento{-22}$, which become 0,
an entry $-1.0000000000000000000000374$, which becomes $-1$, and an entry $9.8696044010893586\ldots$, which the identification
step names $\pi^2$. The exact operator~\eqref{eq:fresnel} leaves a residual of $9.7 \timestento{-33}$ against a noise level of
$9.2 \timestento{-33}$ on the held-out window at 40 digits, and $8.3 \timestento{-73}$ against $9.9 \timestento{-73}$ at 80
digits: the residual follows the noise down by forty orders of magnitude, which an approximate operator would not do. In
$z = x^4$ the operator has the two-term form, whose series the solver names as $C(x)$ and $S(x)$, and the data select $C(x)$
with the constant 1. From the float64 table the same support is found with a gap of a factor of 80, the coefficient
$\pi^2$ is first known only as $9.8693 \pm 0.0021$, and it is identified after it is re-estimated through the candidate
solutions. Section~\ref{sec:fresnel} follows both runs step by step.

\subsection{Identities found from numbers}\label{sec:highlights}

Table~\ref{tab:highlights} lists some of the results. Each was returned from a routine that evaluates the function and
nothing else: no potential, no energy, no parameter family. In the quantum examples the energy appears as a coefficient of
the Schr\"odinger equation that stays numerical, and its 30 recorded digits agree with the known value. The deflection of
light by a Schwarzschild mass, an elliptic integral computed by quadrature, came back as a fourth-order equation whose
singular point $27 u^2 = 1$ is the critical impact parameter, and as a sum of two hypergeometric functions whose series begins
with Einstein's $4 M/b$. The density of a product of four normal variables has four coincident local exponents, so no
hypergeometric series reaches it; it came back as a Meijer $G$-function from 1101 float64 values. For functions of two
variables the weak form on boxes gives a system of partial differential operators; for the complete elliptic integral of
the third kind it kept three, among them $2 (m - n)\,\partial_n \partial_m f + \partial_n f - 2\,\partial_m f = 0$, and two
further two-term operators fixed the ratios of a double series that is Appell's $F_1$. Section~\ref{sec:examples} gives the
operators, the tests and the times of these and other examples.

\subsection{Benchmarks}\label{sec:mainbenchmarks}

\begin{figure}[t]
\centering
\includegraphics[width=\textwidth]{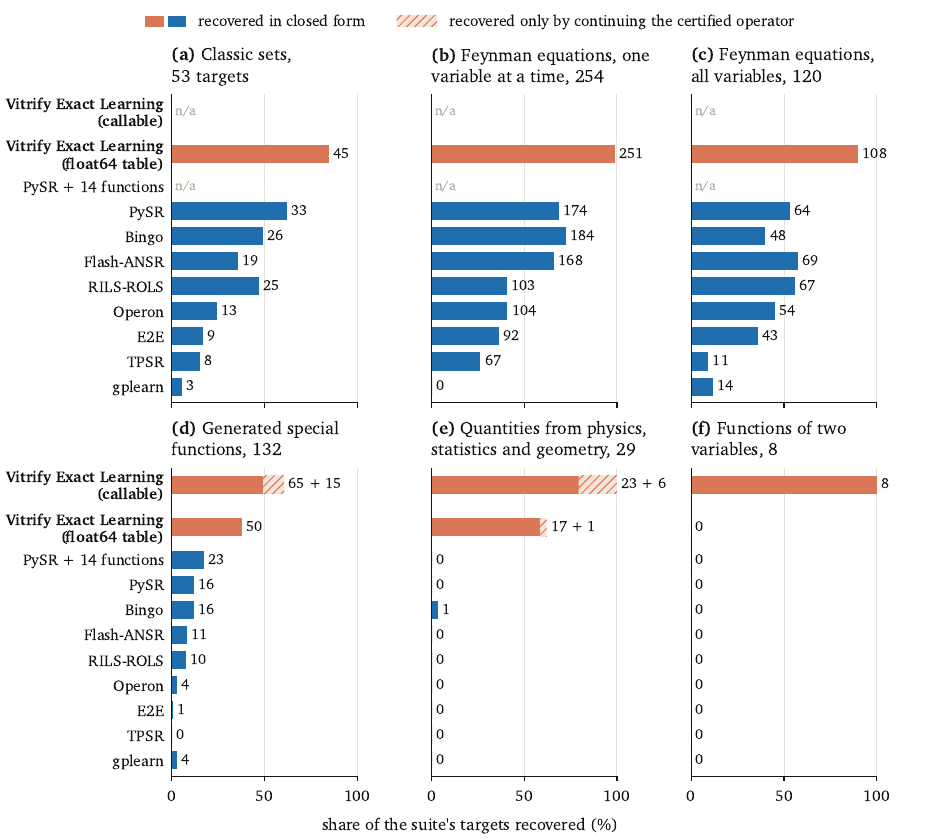}
\caption{Share of the targets of each suite that each program recovered, with the counts. The hatched part counts targets for
which \VEL{} returned a verified operator without a closed form and the continued solution of that operator passes the same
test. The rows marked \qq{callable} read the target at 40 digits, which the other programs cannot do; all other rows received
the same float64 points. \qq{n/a} marks a row that was not run on a suite.}
\label{fig:benchmarks}
\end{figure}

We ran eight open symbolic-regression programs (PySR, Operon, Bingo, gplearn, RILS-ROLS, Flash-ANSR, E2E and TPSR) on the same
float64 points as \VEL{}, and PySR also with fourteen special functions added to its operations (Table~\ref{tab:methods}).
Figure~\ref{fig:benchmarks} gives the overview and Table~\ref{tab:mainsummary} the headline counts.

\begin{table}[t]
\centering\small
\caption{Headline counts from noise-free float64 tables, with the best of the other programs. SRBench rows give the symbolic
solution rate under SRBench's protocol; the other rows are targets recovered. The published suites were in view during
development; the generated set was drawn after the learner was frozen. Section~\ref{sec:benchmarks} gives every suite.}
\label{tab:mainsummary}
\begin{tabular}{@{}L{72mm}R{18mm}R{14mm}L{38mm}@{}}
\toprule
Suite & \VEL{} & Best other & Program \\
\midrule
Classic sets at their published points, of 53 & 45 & 33 & PySR \\
Feynman equations, one variable at a time, of 254 & 251 & 184 & Bingo \\
Feynman equations in all variables, of 120 & 108 & 69 & Flash-ANSR \\
Special functions, pilot set, of 31 & 28 & 0 & -- \\
SRBench Feynman, symbolic solution rate & 74.1\% & 55.8\% & AI Feynman (published) \\
SRBench Strogatz, symbolic solution rate & 100\% & 29.5\% & GP-GOMEA (published) \\
Generated special functions, of 132 & 50 & 23 & PySR + 14 functions \\
Physical quantities, of 29 & 18 & 1 & Bingo \\
Functions of two variables, of 8 (callable for \VEL{}) & 8 & 0 & -- \\
\bottomrule
\end{tabular}
\end{table}

\paragraph{Published suites.} On the classic one-variable sets, the AI Feynman equations taken one variable at a time and in
all their variables, and a pilot set of special functions, \VEL{} recovers more targets than each other program
(Table~\ref{tab:mainsummary}). These suites were in view during development, and the entry point for tables in several
variables was written with them at hand. Of the 389 noise-free one-variable tables, PySR returned an expression for every
table, 239 of them recovered and 42 that match inside the training range and fail beyond it. \VEL{} returned 371 expressions,
all recovered, and no model for the other 18.

\begin{figure}[t]
\centering
\includegraphics[width=\textwidth]{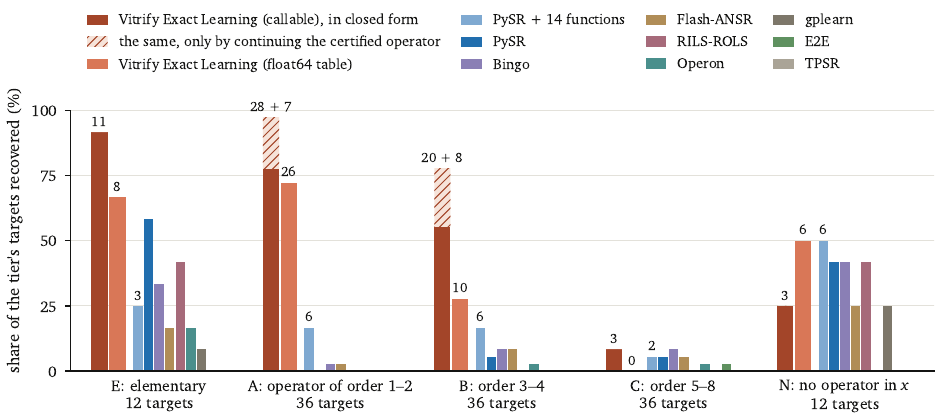}
\caption{The generated set: share of each tier's targets recovered, with the counts for \VEL{} and for PySR + 14 functions.
The hatched part counts targets for which only the continued solution of the verified operator passes the test.}
\label{fig:generated_tiers}
\end{figure}

\begin{figure}[t]
\centering
\includegraphics[width=\textwidth]{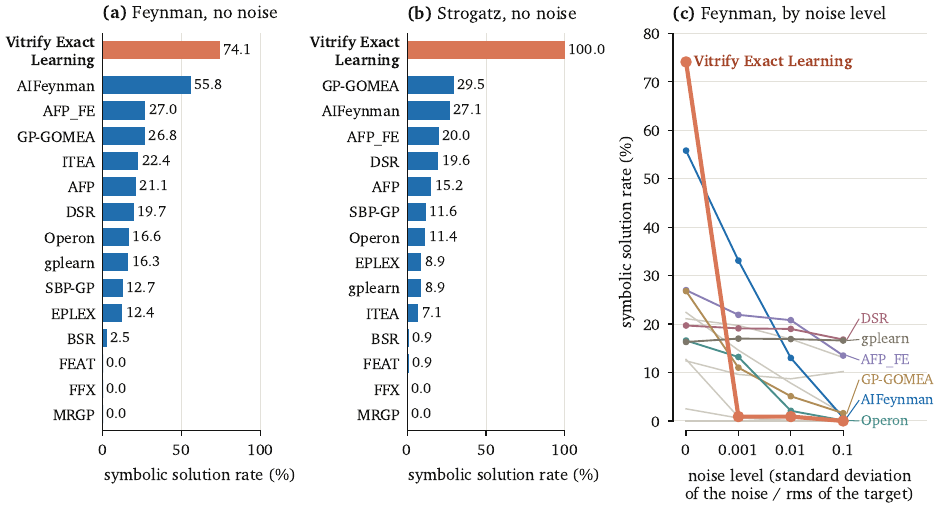}
\caption{SRBench, problems with known answers. (a)~Symbolic solution rate on the Feynman problems without noise for \VEL{}
and the 14 published methods. (b)~The same on the Strogatz problems. (c)~The rate on the Feynman problems against the noise
level.}
\label{fig:srbench}
\end{figure}

\paragraph{SRBench.} Under SRBench's protocol and scoring code \cite{lacava2021srbench}, \VEL{} solves 74.1\% of the
Feynman problems and all 14 Strogatz problems without noise, against published best rates of 55.8\% and 29.5\%
(Figure~\ref{fig:srbench}). By agreement with the test rows to $\tento{-9}$ the Feynman rate is 90.8\%; the symbolic test does
not reduce some square roots and exact constants to the form of the true model. With noise the rates fall to 0.9\%, 0.9\% and
0.0\% at the three levels, where AI Feynman keeps 33.1\%, 13.0\% and 0.7\%.

\paragraph{A generated set of special functions.} To test on functions that the method was not developed on, we froze the
learner and then drew 132 targets from a generator of 59 families of special functions, combined with rational parameters in
nine shapes, in five tiers by the order of the operator that annihilates them (Figure~\ref{fig:generated_tiers}). From float64
tables \VEL{} recovers 50 of the 132, and PySR + 14 functions recovers 23. From callables \VEL{} returns a closed form that
passes for 65 targets and a verified operator whose continued solution passes for 15 more. In tiers A to C every expression
that \VEL{} recovered from a table satisfies the target's own operator, while none of the recoveries of the programs without
special functions does: theirs are approximants that hold over the tested range. The tier of order 5 to 8 is beyond the search
bounds; three of its 36 targets are recovered from callables, all through two-term operators.

\paragraph{Physical quantities and two variables.} The 29 quantities of the physical set are periods of anharmonic
oscillators, lattice Green functions, densities of products of random variables, stable laws, the Voigt profile and elliptic
integrals of an ellipsoid, chosen by the authors. All 29 receive a verified operator from callables, in 1.5 to 62.9\,s, and
23 a closed form; from float64 tables 18 are recovered, and of the other programs only Bingo recovers one. The eight functions
of two variables, all double hypergeometric series, come back verified from callables at 40 digits as Appell, Lauricella and
Humbert functions; from 1000 float64 points neither \VEL{} nor any other program recovers any of them.

\section{Limits}\label{sec:mainlimits}

\paragraph{Noise and sparse data.} The method is built for values known well. Under SRBench's noise levels it solves almost
none of the Feynman problems, most of which have several variables, for which the method has no noise model. On the 25
observational data sets of SRBench's 2025 edition it returned no expression. In one variable with noise of $\tento{-3}$ on
the generated set, PySR + 14 functions recovers 31 targets and \VEL{} 27, and 60 of the 77 results that the learner graded
\code{fit} do not extrapolate. Our test functions, three windows with Legendre polynomials of high degree, suit values known to
many digits. Many low-order test functions translated across the interval, as in weak SINDy \cite{messenger2021wsindypde}, or
test functions constructed from the data \cite{tran2026testfunctions} may behave better under noise; we have not tested this.

\paragraph{Precision and search bounds.} A float64 table determines operators of order 1 and 2 and some of order 3 and 4;
the two-variable results need values to 40 digits. The search covers operators up to order 4 and degree 8, two-term operators
up to order 6, and order 2 and degree 4 in a transported variable. The change of variable is selected from a bank, a fitted
power and the caller's maps, and the costs that rank the maps were set by hand.

\paragraph{The solver.} The solver is a set of rules. It has none for operators of Heun type, for second-order operators that
need a change of variable to reach a hypergeometric form, or for solutions with complex exponents. For 15 of the 80 verified
operators of the generated set and 6 of the 29 of the physical set it found no closed form, and ore\_algebra's factorisation
did not supply one (Section~\ref{sec:orecheck}).

\paragraph{What {\normalfont\code{verified}} means.} The grade is the outcome of a numerical test on held-out data on the same
interval, which is weaker than a proof. In the runs of this paper one float64 table, the body-centred cubic lattice, received
a false \code{verified} certificate with two fitted parameters; two Feynman slices were returned as the positive square root of
an exact rational function, which is exact on the data and loses the sign beyond a point that the data do not reach; and five
conjectures with fitted constants are wrong, one inside the range of its data and four beyond it. Section~\ref{sec:limits} lists these and the limits of the comparison.

\section{Conclusion}\label{sec:conclusion}

\VEL{} assumes that a function satisfies a linear differential equation with polynomial coefficients in some variable, finds
that equation as the null vector of a matrix of weak-form measurements, makes it exact, tests it on held-out data, and solves
it symbolically. On accurate values it returned results that the tree searches compared here did not: special functions
generated after its code was frozen, Meijer $G$ densities, hypergeometric forms for periods and for the deflection of light, and
Appell and Lauricella functions of two variables. With noise in several variables, with few points and with observed data,
symbolic regression remains the better tool. Exact learning is for numbers that are known well: integrals evaluated by
quadrature, sums of series, solutions computed to many digits, exact sequences and moments.

Several extensions follow from the limits. The solver needs rules for operators of Heun type, which would give closed forms
for lattice Green functions and for wavefunctions of anharmonic wells; a computer-algebra system such as ore\_algebra could
serve as an optional back end for factorisation and for continuation with certified error bounds. Test functions designed for
noise, and an error model for tables in several variables, may extend the method to measured data. The change of variable is
selected from a bank, and a scored search over maps would enlarge the class. The ansatz is linear in the function. An ansatz
that is polynomial in the function and its derivatives keeps step 2 linear in the unknown coefficients and would reach
nonlinear equations such as those of the Painlev\'e transcendents. Products of derivatives cannot all be moved onto the test
functions, so the weak form needs more care there, and the unknowns and the solver grow accordingly.

\paragraph{Code and data.} Every figure and table is generated from recorded runs by scripts kept with the source, which assert
the counts that the text quotes. The learner will be available as a tool on the Vitrify Labs server for the Model Context
Protocol (MCP), \url{https://mcp.vitrify.ai/mcp}, and free credits are given on registration. A basic version of the code will
be released on the Vitrify Labs GitHub, as the repository Exact Learning, upon publication in a peer-reviewed journal; the
advanced features will remain on the MCP server. The code of the earlier Mellin-transform method is public
\cite{irwin2020exactlearning}.

\clearpage
\bibliographystyle{unsrtnat}
\bibliography{refs}

\clearpage
\setcounter{section}{0}
\setcounter{figure}{0}
\setcounter{table}{0}
\setcounter{equation}{0}
\renewcommand{\thesection}{S\arabic{section}}
\renewcommand{\thefigure}{S\arabic{figure}}
\renewcommand{\thetable}{S\arabic{table}}
\renewcommand{\theequation}{S\arabic{equation}}
\renewcommand{\theHsection}{S\arabic{section}}
\renewcommand{\theHfigure}{S\arabic{figure}}
\renewcommand{\theHtable}{S\arabic{table}}
\renewcommand{\theHequation}{S\arabic{equation}}
\begin{center}
{\Large\bfseries Supplementary Information}\\[6pt]
{\large Vitrify Exact Learning: exact differential equations, changes of variable and closed forms from numerical data}\\[6pt]
Benedict W. J. Irwin, Alex Mathiasen and Tom Etminan, Vitrify Labs
\end{center}
\suppressfloats[t]
\noindent This Supplementary Information gives the method of the main text in full (Sections~\ref{sec:background}
to~\ref{sec:bivariate}), the worked example on the Fresnel integral (Section~\ref{sec:fresnel}), the examples in detail
(Section~\ref{sec:examples}), every benchmark with its protocol (Section~\ref{sec:benchmarks}), the limits in detail
(Section~\ref{sec:limits}), a check of the operators without a closed form with the computer-algebra package ore\_algebra
(Section~\ref{sec:orecheck}), and the tables target by target (Sections~\ref{app:bank} to~\ref{app:cases}). Figures,
tables and equations without the prefix S are those of the main text; the references are those of the main text.

\section{Background and related work}\label{sec:background}

\subsection{Symbolic regression}\label{sec:sr}

\begin{figure}[t]
\centering
\includegraphics[width=\textwidth]{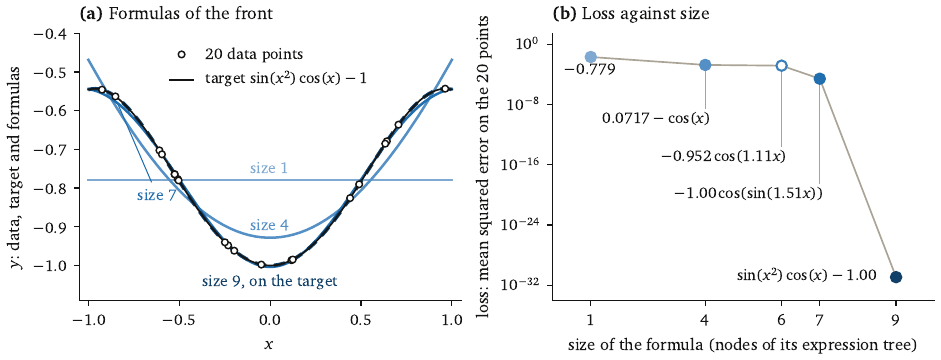}
\caption{A symbolic-regression search that succeeds. PySR on the 20 published points of the benchmark function Nguyen-5,
$\sin(x^2)\cos x - 1$ on $[-1,\,1]$. (a)~The points, the target and the best formulas of sizes 1, 4, 7 and 9; the formula of
size 9 lies on the target. (b)~The recorded loss of the best formula at each size, with the formulas. The point of size 6 is
hollow because panel~(a) omits it. The search took 28.0\,s.}
\label{fig:sr_front}
\end{figure}

Symbolic regression looks for an expression $g$ with $g(x_k) \approx f_k$ on given points, among the expressions that can be
built from a set of operations. In genetic programming \cite{koza1992gp} a population of expression trees is mutated and
recombined and the fittest trees survive; Schmidt and Lipson used it to recover conservation laws from motion-tracking data
\cite{schmidt2009distilling}. Current programs of this kind, among them PySR \cite{cranmer2023pysr}, Operon
\cite{burlacu2020operon}, Bingo \cite{randall2022bingo} and gplearn \cite{stephens2026gplearn}, differ in how they keep the
population diverse and in how they fit the constants of a tree, for which nonlinear least squares is now common
\cite{kommenda2020nls}. They report a front of formulas, the best fit found at each size. RILS-ROLS
\cite{kartelj2023rilsrols} replaces the population by an iterated local search with ordinary least squares for the constants.
Figure~\ref{fig:sr_front} shows one such search on a standard test function.

A second family proposes formulas with a neural network. Deep symbolic regression trains a recurrent network by a
risk-seeking policy gradient to emit good trees for one data set \cite{petersen2021dsr}. Pretrained transformers map a set of
points to a formula in one pass after training on millions of synthetic formulas \cite{biggio2021nesymres,kamienny2022e2e};
TPSR adds a Monte-Carlo tree search over the decoder of such a model \cite{shojaee2023tpsr}, and Flash-ANSR samples and
refines many candidate formulas from one \cite{saegert2026flashansr}. AI~Feynman combines a neural-network fit with tests for
units, symmetries and separability that split a problem in several variables into smaller ones
\cite{udrescu2020aifeynman,udrescu2020aifeynman2}. Recent work guides the search with a large language model
\cite{grayeli2024lasr,shojaee2025llmsr}. The authors of the second paper argue that results of language models on the
textbook equations are partly recitation of known formulas, and a later benchmark was built to prevent it
\cite{shojaee2025llmsrbench}. A recent review covers the field \cite{makke2024review}.

The standard test sets are small formulas over elementary operations: the Nguyen and Keijzer functions
\cite{uy2011nguyen,keijzer2003}, the 100 equations of the Feynman Lectures with 20 harder ones \cite{udrescu2020aifeynman},
and the Strogatz systems of two-state dynamics \cite{lacava2016strogatz}. SRBench runs many methods on the last two through
the Penn Machine Learning Benchmarks \cite{olson2017pmlb,romano2022pmlb} with one protocol and one scoring code
\cite{lacava2021srbench}, and its 2025 edition adds observational data sets \cite{aldeia2025srbench}.

Two properties of this formulation matter here. The search space grows combinatorially with the size of the tree, and the
problem is NP-hard \cite{virgolin2022nphard}. And the answer must be a finite tree over the chosen operations. A special
function can be added to the operations, as we do for one baseline in Section~\ref{sec:benchmarks}, but then every function
that might occur has to be supplied in advance, and with rational parameters, as in $K_{1/4}(x)$ or
$\pFq{2}{1}(2/5, 4/3;\,3/4;\,x)$, there are infinitely many.

A separate line of work identifies the equations of motion of a dynamical system from a trajectory. SINDy selects a sparse
set of terms from a library of candidate right-hand sides \cite{brunton2016sindy}; its weak form integrates the trajectory
against test functions in place of differentiating it \cite{messenger2021wsindy}, an idea that goes back to the modulating
functions of Shinbrot \cite{shinbrot1957}; ODEFormer does the task with a pretrained transformer \cite{dascoli2024odeformer}.
We use the same weak form for a different purpose. The equation sought here is linear in the function, its coefficients are
polynomials in the independent variable with exact numbers in them, and it serves as a finite description of one function,
from which a closed form is then derived.

\subsection{Holonomic functions}\label{sec:holonomic}

A function $f$ of one variable is holonomic, or D-finite, when it satisfies
\begin{equation}\label{eq:operator}
  L f = 0, \qquad L = \sum_{i=0}^{r} p_i(x)\,D^i, \qquad D = \frac{\dd}{\dd x},
\end{equation}
with polynomials $p_i$ that are not all zero \cite{stanley1980dfinite,zeilberger1990holonomic}. We call $r$ the order of the
operator $L$ and the largest degree of the $p_i$ its degree, and we say that $L$ annihilates $f$. The class contains the
algebraic functions, the exponential, logarithm, sine and cosine, and most of the special functions of a handbook
\cite{olver2010handbook,dlmf,erdelyi1953htf}: the error function, Bessel and Airy functions, the complete elliptic
integrals, the incomplete ones as functions of the modulus or of the sine of the amplitude, the orthogonal polynomials, the
generalised hypergeometric functions $\pFq{p}{q}$ and Meijer's
$G$-function \cite{meijer1936,beals2013meijer}. Fox's $H$-function \cite{fox1961,mathai2010hfunction} belongs to it when its
parameters are rational, since it is then a $G$-function of a power of its argument.

The class is closed under addition, multiplication, differentiation, integration and composition with algebraic functions,
and an operator for the result can be computed from the operators of the parts
\cite{stanley1980dfinite,lipshitz1989dfinite,kauers2011tetrahedron,kauers2023dfinite}. A function is holonomic exactly when
its Taylor coefficients satisfy a linear recurrence with polynomial coefficients, which carries the definition over to
sequences \cite{petkovsek1996ab}. An operator together with finitely many initial values is therefore an exact, finite data
structure for a function. Zeilberger's holonomic systems approach proves identities between special functions by computing
with these data structures \cite{zeilberger1990holonomic,chyzak2000zeilberger,koutschan2014holonomic}, and the solution of an
operator can be evaluated to any precision anywhere on its Riemann surface by analytic continuation of its Taylor series
\cite{vanderhoeven1999holonomic,mezzarobba2010numgfun}. For several variables the corresponding object is a system of partial
differential operators whose solution space has finite dimension. The holonomic gradient method uses such a system to
evaluate and to maximise a likelihood numerically \cite{nakayama2011hgm}. That is the forward direction, from a known system
to numbers; this paper is about the inverse.

Some common functions are outside the class. The gamma function satisfies no algebraic differential equation at all
(H\"older's theorem), and the same holds for the Riemann zeta function; $\tan x$ has infinitely many poles, which a solution
of~\eqref{eq:operator} cannot have. For such inputs the right answer of a method that searches for~\eqref{eq:operator} is
that there is none.

\paragraph{Guessing.} In computer algebra the step from data to an operator is called guessing. Given the first terms of a
sequence or of a power series as exact rational numbers, the programs gfun \cite{salvy1994gfun}, the guesser of ore\_algebra
\cite{kauers2015ore} and their successors \cite{kauers2022guessing} solve a linear system for the coefficients of a
recurrence or a differential equation and check it on further terms. The differential approximants of lattice statistics fit
a differential equation to the known coefficients of a series in the same way and use it to locate singular points
\cite{guttmann1972approximants}. These methods need exact coefficients at one expansion point. The input considered here is
weaker: values on an interval, to 40 digits from a quadrature or to 16 from a table, or only draws from a density. No series
is available, the linear system is numerical, its solution has to be made exact afterwards, and the result needs an error
model and a test before it can be believed. Section~\ref{sec:related} of the main text compares the closest methods and
states which part of the combination we have not found in earlier work.

\paragraph{Constants.} For a single real number the corresponding task is integer-relation detection: the lattice reduction
of Lenstra, Lenstra and Lov\'asz \cite{lenstra1982lll} and the PSLQ algorithm \cite{ferguson1999pslq} find integers $n_k$
with $\sum_k n_k c_k = 0$ for given real $c_k$, which identifies an algebraic number from its powers and a combination of
known constants from their values \cite{bailey2001integer}. The Ramanujan Machine searches numerically for continued-fraction
formulas of constants and states its matches as conjectures \cite{raayoni2021ramanujan}. We use PSLQ inside the method, with
a rule for how much numerical evidence an identification needs (Section~\ref{sec:exact}).

\paragraph{From an operator to a closed form.} Kovacic's algorithm decides whether a second-order operator with rational
coefficients has a solution built from exponentials, integrals and algebraic functions \cite{kovacic1986}, and there are
algorithms for solutions in terms of hypergeometric and Bessel functions after a change of variable
\cite{imamoglu2017hypergeometric,debeerst2008bessel}. Roach's method rewrites hypergeometric and Meijer $G$-functions in
terms of named functions by walking through contiguous parameter values \cite{roach1996hypergeometric,roach1997meijer}; it
is implemented in SymPy \cite{meurer2017sympy}, which our solver calls. The solver of Section~\ref{sec:solver} is a set of
rules of this kind and is far from complete.

\subsection{Earlier exact learning}\label{sec:earlier}

The term exact learning was introduced by the first author for the task of assigning an exact expression to a function that is
known only numerically \cite{irwin2020natural,irwin2021natural}. That work represents a function on $(0, \infty)$ by its
Mellin transform,
\begin{equation}\label{eq:mellin}
  \phi(s) = \int_0^\infty x^{s-1} f(x)\,\dd x ,
\end{equation}
which it called the fingerprint of $f$. For a large class of functions $\phi$ is a ratio of products of gamma functions of
linear forms in $s$, times $c^{s}$: this is the Mellin--Barnes representation that defines the $G$- and $H$-functions
\cite{mathai2010hfunction,beals2013meijer,flajolet1995mellin}, and Ramanujan's master theorem relates $\phi$ to the
coefficients of the power series of $f$ \cite{amdeberhan2012rmt}. The method computed $\phi(s)$ numerically at chosen points
$s$ and fitted the parameters of an assumed form by quasi-Newton optimisation. The public code
\cite{irwin2020exactlearning} fits forms such as $\phi(s) = (1 + p_0 s + p_1 s^2)\,p_2^2\,(p_3^2)^{s}\,\Gamma(p_4 + p_5 s)$ and
then proposes exact values for the fitted numbers, for example $4/3$, $1/2$ and $2^{1/4}$. Symbolic metamodelling fits the
parameters of a Meijer $G$-function to a black-box model by gradient descent \cite{alaa2019metamodels}, which is a fit within
a parametric family of the same kind. Mellin transforms and Meijer kernels have also been used for density estimation on the
half-line \cite{geenens2021mellin}.

The approach has three limits. The family must be chosen in advance and the fit is a non-convex optimisation in real
parameters. The transform~\eqref{eq:mellin} must exist, which excludes functions known only on an interval and functions
that grow or oscillate without decay. And nothing in the procedure tests the result.

The notes that accompany the code \cite{irwin2020exactlearning} point to another route. If $f$ satisfies~\eqref{eq:operator},
then $\phi$ satisfies a linear recurrence in $s$ with polynomial coefficients, because under the transform $x\,D$ becomes
multiplication by $-s$ and multiplication by $x$ becomes the shift $s \mapsto s + 1$. The coefficients of that recurrence can
be solved for by linear algebra from values of $\phi$. For $f = \ee^{-x}$, with $\phi(s) = \Gamma(s)$, the recurrence is
$\phi(s) - (s - 1)\,\phi(s - 1) = 0$. The notes also write Heun's equation as an annihilating operator and treat Mellin
transforms over a window $[a, b]$.

The method of this paper takes that route and leaves Mellin space. It measures the function on windows of the interval where
it is known, so no transform over the half-line has to exist. It finds the operator by linear algebra, makes its coefficients
exact with a count of the evidence, and tests it on data that the fit did not read. The two pictures meet in the solver. An
operator of the form $P(\theta) f = \mu\,x^{m} Q(\theta) f$ with $\theta = x D$ and polynomials $P$ and $Q$ gives
$P(-s)\,\phi(s) = \mu\,Q(-s-m)\,\phi(s+m)$, a first-order recurrence, whose solutions are the gamma-function products of the
earlier work. These operators are the ones for which Section~\ref{sec:solver} writes down hypergeometric and Meijer
$G$-functions directly.

\section{Method}\label{sec:method}

Figure~\ref{fig:pipeline} of the main text shows the steps. The implementation is about 4{,}200 lines of Python on mpmath \cite{mpmath}, SymPy
\cite{meurer2017sympy}, NumPy \cite{harris2020numpy} and SciPy \cite{virtanen2020scipy}. In the operator search no list of
named function families is fitted at any step: the closed form is derived from the operator, and the operator from the data.
The entry point for float64 tables (Section~\ref{sec:relations}) first tries relations over a fixed set of elementary
functions and then calls the operator search.

\subsection{Inputs and what is returned}\label{sec:inputs}

The learner for one variable accepts five kinds of data.
\begin{itemize}[leftmargin=*,itemsep=1pt,topsep=3pt]
\item A \emph{callable}: a routine that returns $f(x)$ to any requested precision, as a quadrature or a series does. It is read
at a working precision of $N_p$ significant digits on an interval that the caller gives, and the result is checked at
$2N_p$ digits. The working precision $N_p$ is a parameter of the learner; every run in this paper uses $N_p = 40$, so the
check is made at 80 digits.
\item A \emph{float64 table}: points $x_k$ with values $f_k$ in double precision, as another program would write them.
\item \emph{Draws}: independent samples from a density.
\item A \emph{sequence}: the leading terms $a_0, a_1, \ldots$ as exact integers or fractions.
\item \emph{Moments}: the exact moments $m_n = \int x^n f(x)\,\dd x$ of a density.
\end{itemize}
Two further entry points build on it: one for float64 tables in several variables (Section~\ref{sec:relations}) and one for
a callable in two variables (Section~\ref{sec:bivariate}).

The result is a list of \emph{certificates}, best first, and a list of rejected candidates with the reason for each. A
certificate holds the change of variable $y = \varphi(x)$ and its weight $k$ (Section~\ref{sec:transport}); the operator in
$y$ with exact coefficients; the same operator in $x$ when it has polynomial coefficients there; any coefficient that was
left numerical, with its value and error; the residual and the noise level of the test on held-out data at each precision;
the evidence for the exact coefficients in decimal digits; a grade; and, when the solver finds one, the closed form with its
largest error against the data.

\subsection{Weak-form measurements}\label{sec:measure}

Differentiating data amplifies its errors, and an operator of order four would need four derivatives. The method avoids
this by a weak form. Let $q$ be the data on a window $[c - h,\,c + h]$ with local coordinate $s \in [-1, 1]$, $y = c + h s$.
The only numbers that are taken from the data are the moments
\begin{equation}\label{eq:moments}
  \nu_m = \int_{-1}^{1} s^m\,q(c + h s)\,\dd s , \qquad m = 0, 1, 2, \ldots
\end{equation}
To test an operator of order $r$ we use the test functions
\begin{equation}\label{eq:test}
  B_t(s) = (1 - s^2)^r\,P_t(s) = \sum_k b_{tk}\,s^k , \qquad t = 0, 1, \ldots,
\end{equation}
where $P_t$ is the Legendre polynomial of degree $t$. The first $r - 1$ derivatives of $B_t$ vanish at both ends of the
window, so $i \le r$ integrations by parts leave no boundary terms:
\begin{equation}\label{eq:parts}
  \int_{-1}^{1} B_t(s)\,s^l\,\frac{\dd^i q}{\dd s^i}\,\dd s
  = (-1)^i \sum_k b_{tk}\,(l + k)^{\underline{i}}\;\nu_{l + k - i} ,
  \qquad n^{\underline{i}} = n (n - 1) \cdots (n - i + 1) .
\end{equation}
With $y^j D^i = \sum_{l=0}^{j} \binom{j}{l} c^{\,j-l} h^{\,l-i}\,s^l\,\dd^i/\dd s^i$ this gives, for every term $y^j D^i$ that
an operator of order $r$ and degree $d$ can contain, the number
\begin{equation}\label{eq:weak}
  M_{t,(i,j)} = \int_{-1}^{1} B_t(s)\,\bigl[y^j D^i q\bigr]\,\dd s
\end{equation}
as a combination of the moments with coefficients that depend only on the window. The $b_{tk}$ are rational numbers, so no
approximation enters beyond the moments themselves. The numbers~\eqref{eq:weak} form the weak matrix $M$, with one row for
each test function on each window and one column for each term. If $L = \sum_{i,j} c_{ij}\,y^j D^i$ annihilates $q$, then
\begin{equation}\label{eq:null}
  M c = 0 ,
\end{equation}
and the search for an operator becomes the search for a null vector. An operator of order $r$ and degree $d$ tested with $T$
functions per window reads the moments up to $m = d + T + 2 r - 1$.

\paragraph{Moments and their errors.} For a callable the moments are computed by Gauss--Legendre quadrature with
$3 \cdot 2^{\ell - 1}$ nodes, and the level $\ell$ is raised from 5 until two successive levels agree to a relative
$\tento{3-N_p}$ at the working precision of $N_p$ digits (the highest level is 9, and 7 for a table). For a float64 table the callable is a local
Lagrange interpolant through the nearest points. Its degree is the one of 4, 8, 12, 16, 24, 32 and 48 for which the
interpolant through all points and the interpolant through every second point agree best at 14 probe points. For draws the
moment of the density is a sample mean, since $\mathrm{E}\bigl[s^m\,\mathbf{1}(y \in \text{window})\bigr] = h\,\nu_m$.

Each kind of data also supplies \emph{realisations of the error} of its moments. For a callable these are the difference
between the last two quadrature levels and a random perturbation of the size of the rounding error. A table adds the moments
of the difference between its two interpolants. For draws the sample is split into 48 batches, and the deviations of the
batch means from the overall mean, divided by $\sqrt{47}$, are realisations whose mean outer product is the covariance of
the moments. Every later quantity is a linear function of the moments, or is linearised about them. Applying the same
linear map to the error realisations gives a matrix $E_k$ for each realisation, and from the $E_k$ the noise level of that
quantity. The thresholds below are multiples of noise levels obtained in this way. No tolerance is set for a particular
function, and the same code handles values good to 40 digits, to 16 digits and to 3.

\paragraph{Windows.} The operator is fitted on three overlapping windows that cover the fractions 0.04 to 0.30, 0.24 to 0.50
and 0.44 to 0.70 of the interval, and it is tested on a fourth window that covers 0.70 to 0.96
(Figure~\ref{fig:fresnel_trace}a). No quadrature node of the fit lies in the fourth window, and no node of any window lies
in the outer 4\% at each end. For a table, the interpolant near the end of a window is built on the nearest table points,
some of which lie beyond that end.

\subsection{Operator search}\label{sec:search}

The search visits supports in the order $(r, d) = (1, 0), (1, 1), \ldots, (1, 8), (2, 0), \ldots$ up to order 4 and degree
8. For each order the matrix is built once, for degree 8, with $\lceil 9 (r + 1) / 3 \rceil + 4$ test functions on each of
the three windows, and the supports of lower degree are subsets of its columns. The columns are scaled to unit norm, so
that a singular value is a residual relative to the size of the terms. A column that is zero within its error, such as the
derivative of constant data, is divided by ten times that error instead, so that it stays small.

Let $\sigma_1 \le \sigma_2$ be the two smallest singular values of the scaled matrix on a support and $v_1$, $v_2$ their
right singular vectors. If an exact null vector $v$ exists and the measured matrix differs from the exact one by $E$, the
smallest singular value of the measured matrix is at most $|E v|$. The noise level of $\sigma_1$ is therefore taken as
\begin{equation}\label{eq:noise}
  \eta(v_1) = \operatorname{rms}_k\,|E_k v_1| + \tento{3-N_p} ,
\end{equation}
with the second term for rounding at $N_p$ digits. A support has a null direction that the method accepts when
\begin{equation}\label{eq:accept}
  \sigma_1 \le \kappa\,\eta(v_1), \qquad
  \sigma_2 > \kappa\,\eta(v_2), \qquad
  \frac{\max\bigl(\eta(v_1),\,\sigma_1\bigr)}{\sigma_2} < \frac18 ,
\end{equation}
with $\kappa = 10$ for callables and tables and $\kappa = 3$ for draws, whose error realisations measure the sampling error
itself where the others are loose bounds. The three conditions say that the smallest singular value is zero within its
error, that the next one is not, so that the direction is unique, and that the direction is known to better than one part in
eight. At each order the search stops at the first degree at which the first condition holds. If the other two hold there
as well and the null vector can be made exact (Section~\ref{sec:exact}: at most two entries left numerical, at least 1.5
digits of evidence and a leading coefficient that is not zero), that support gives the operator. Because supports are
visited by order and then by degree, it is the operator of least order and, at that order, of least degree. Otherwise the
search moves to the next order, since every support of higher degree at the same order contains polynomial multiples of the
same direction.

The weak matrices of smooth data have singular values that decay steadily as columns are added
(Figure~\ref{fig:fresnel_trace}b), so a small singular value is not evidence on its own. What counts is the number of
decimal digits by which the smallest singular value drops below the trend of the others, and this number is added to the
evidence of Section~\ref{sec:exact}.

\paragraph{Operators with two and three terms.} A float64 table of a function whose operator has order 3 or 4 often leaves
several null directions on the full support, because the singular values fall below $\tento{-16}$ after about a dozen
columns. When no order gives a candidate, callables and tables are searched again for operators of the form
\begin{equation}\label{eq:banded}
  \sum_{k=0}^{K-1} x^{k m}\,P_k(\theta), \qquad \theta = x D ,
\end{equation}
with $K = 2$ and then $K = 3$ terms and steps $m \ge 1$. Such an operator of order $r$ has $K (r + 1)$ unknown
coefficients, 6 to 15 for orders 2 to 4, where the full support of order 4 and degree 8 has 45. Two terms are the equations
of the generalised hypergeometric functions of $x^m$; three terms add one more singular point.

\paragraph{The wide search.} When no change of variable gives any certificate, two-term operators alone are searched up to
order 6 with steps up to 6, in $x$, in the powers of $x$ of the bank and in $1/x$, with weight $k = 0$; the search ends at
the first of these variables that gives a certificate, and it is not run for draws. An operator of order 6 has 14 unknowns
there, where the full support of the same degree has 91. These are the equations of the Meijer $G$-functions of a power of
$x$, which include the Fox $H$-functions with rational parameters. The densities of the stable law of index $3/2$ and of
Holtsmark's distribution (Section~\ref{sec:physical}) are found in this way, with operators of order 5.

\paragraph{Precision.} All linear algebra runs at the working precision $N_p$, for tables and draws as well, so that rounding inside the
method never competes with the error of the data. For a callable the exact operator is tested again with moments computed
at $2N_p$ digits, where only numerical parameters are refitted. If a callable gives no certificate and $N_p$ is below 80,
the whole search is repeated once with $N_p = 90$, so with the second test at 180 digits: at 40 digits the weak matrix of a
smooth function on a short interval has several singular values at the rounding level once the support passes about 16
columns, and some operators of order 3 and 4 become unique only with more digits.

\subsection{Exact coefficients}\label{sec:exact}

The null vector is a list of real numbers, each with error realisations obtained by first-order perturbation of the singular
vector. The entries are divided by a pivot entry. Each ratio $\rho$ is known to within
$\tau = \kappa \times (\text{its rms error}) + \tento{3-N_p}\,(1 + |\rho|)$, where the second term is the rounding at $N_p$
digits, and is replaced by the fraction of least denominator in $[\rho - \tau,\,\rho + \tau]$, which a Stern--Brocot descent
finds. A ratio within $\tau < 1$ of zero is set to zero.

Any interval contains fractions, so a rule is needed for when to believe one. Fractions whose numerator and denominator are
at most $H$ are spaced about $3/H^2$ apart on average, so a random real number of size up to one lies within $\tau$ of one
with probability of about $0.6\,\tau H^2$, and a larger number with less. The method takes this probability as
$\tau H^2 / \max(|\rho|, 1)$, which is on the cautious side, and counts
\begin{equation}\label{eq:evidence}
  e = -\log_{10} \frac{\tau}{\max(|\rho|, 1)} - 2 \log_{10} H
\end{equation}
decimal digits of evidence for a fraction of height $H$. The fraction is accepted when
\begin{equation*}
  e \ge \max\bigl(0.5,\ \min(8,\ -\tfrac14 \log_{10} \tau)\bigr) ,
\end{equation*}
a margin that grows with the precision of the data. The pivot is
the entry that leaves the fewest ratios without an accepted fraction. The \emph{evidence} of an operator is the sum of $e$
over its entries plus the drop of the null singular value below the trend (Section~\ref{sec:search}). A candidate from a
callable or a table needs at least 1.5 digits.

A ratio with no acceptable fraction, when at least 16 of its digits are known, goes through four integer-relation searches
with PSLQ \cite{ferguson1999pslq}: for an algebraic number of degree 2 to 4; for a rational combination of 1, $\pi$, $\ee$,
$\log 2$, $\sqrt{\pi}$, $\sqrt{2}$, $\sqrt{3}$ and Euler's constant $\gamma$; for a ratio of two linear forms in one of
those constants; and for a product of rational powers of 2, 3, 5, 7, $\pi$ and $\ee$. Among $n$ unrelated numbers some
integer combination with coefficients up to $H$ falls below about $H^{1-n}$ by chance, so each search bounds its coefficients
by the largest $H$ for which the tolerance is still 8 digits below that level, and by 200, 200, 50 and 60 in the four
searches; with few digits the bound falls below 2 and the search is skipped. Numbers of the forms $q\,\pi^k$ and $\sqrt{q\,\pi^k}$ with a fraction $q$ and $|k| \le 3$ are tried
already from 8 digits, which names $4\pi/3$ or $1/\sqrt{2\pi}$ from a float64 value.

An entry that matches none of these stays a numerical parameter $a_1, a_2, \ldots$ with its value and error. At most two
are allowed. This is the intended result for an operator that contains a number without a closed form, such as an energy
eigenvalue (Section~\ref{sec:quantum}).

Table~\ref{tab:calibration} shows how often the identification step gives a false exact form. Applied to 300 random
numbers, which have none, it returned an exact form for none of them when the numbers were known to a relative $\tento{-12}$
or better, for 4 at $\tento{-6}$ and for 20 at $\tento{-3}$. A single constant known only to the accuracy of a table or of
a sample can therefore receive a false fraction a few percent of the time. This is one reason why results from draws are
graded as conjectures at best, and why the constant of a density is taken from its normalisation integral
(Section~\ref{sec:other}).

\begin{table}[t]
\centering\small
\caption{Calibration of two decisions of the method. Left: exact forms returned by the identification step for 300 random
numbers, which have none, at five relative errors. Right: the hold-out test for draws applied to operators fixed in
advance, on 500{,}000 draws in each of 40 runs; an operator is rejected at $p < 0.001$.}
\label{tab:calibration}
\begin{tabular}{@{}cc@{}}
\toprule
Relative error & Given an exact form, \\
of the number & of 300 \\
\midrule
$\tento{-35}$ & 0 \\
$\tento{-20}$ & 0 \\
$\tento{-12}$ & 0 \\
$\tento{-6}$ & 4 \\
$\tento{-3}$ & 20 \\
\bottomrule
\end{tabular}
\hspace{8mm}
\begin{tabular}{@{}llcc@{}}
\toprule
Draws & Operator tested & $p < 0.05$ & Rejected \\
\midrule
exponential(1) & $D + 1$, the true one & 1 of 40 & 0 of 40 \\
exponential(1) & $D + 51/50$ & 40 of 40 & 40 of 40 \\
gamma(3, 1) & $x D + x - 2$, the true one & 3 of 40 & 0 of 40 \\
gamma(3, 1) & $x D + x - 41/20$ & 40 of 40 & 40 of 40 \\
\bottomrule
\end{tabular}
\end{table}

\subsection{Hold-out test and grades}\label{sec:holdout}

The exact operator, with coefficient vector $c$, is applied to the fourth window. Its residual there is $|M c|$ divided by
the size of its terms, $\bigl(\sum_j |M_{\cdot j}|^2 c_j^2\bigr)^{1/2}$, and the noise level is the same combination of the
error realisations, $\operatorname{rms}_k |E_k c|$, divided by the same size, plus the rounding term $\tento{3-N_p}$. The
operator passes when the residual is at most $\kappa$ times the noise. For a callable the test is repeated at $2N_p$
digits and must pass at both precisions. Nothing in
the test was fitted to the fourth window, so it tests whether the equation that was found on 70\% of the interval extends
to the rest of it.

Two refinements came from failures, which we report in Section~\ref{sec:history}. First, when the terms of the operator
are themselves small on the window, the residual is divided by the size they would have for data that vary over one window:
the term $y^j D^i$ on a window of half-width $h$ that reaches $|y| = Y$ counts as $|q|\,Y^j / h^i$. Without this length
scale the operator $D$ passes for any data on a window that is wide enough in the transported variable. Second, an operator
with fitted parameters gets two tests, and both must pass. The parameters were fitted on other windows and have errors that
the noise of the fourth window does not contain. The first test keeps the fitted values and widens the noise by those
errors; it tests the values. The second estimates the parameters again on the fourth window, which is a linear
least-squares problem, and tests the exact part of the operator against the unwidened noise; it keeps its power when a
parameter is poorly determined.

A certificate that passes receives one of three grades, and data without one receive the fourth.
\begin{description}[leftmargin=1.5em,itemsep=1pt,topsep=3pt]
\item[\normalfont\code{verified}] The residual is within $\kappa$ times the noise at every precision, the noise, including the part
from fitted parameters, is below $\tento{-8}$ of the size of the terms, and the evidence is at least 10 digits. For a
sequence, the held-out equations hold exactly.
\item[\normalfont\code{conjecture}] The test passes with more noise or with less evidence. Every result from draws with exact
coefficients has this grade, because its test is a significance test.
\item[\normalfont\code{fit}] For draws, an operator that passes its test and still has numerical coefficients. For a table with noise
(Section~\ref{sec:relations}), an expression with fitted constants that matches the held-out rows to the noise level.
\item[\normalfont\code{no-model}] No candidate passed.
\end{description}
Certificates are ordered by grade, then operators with exact coefficients before operators with fitted parameters, then by
the cost
\begin{equation}\label{eq:cost}
  \text{cost of the map} + 4 \times \text{order} + \text{degree} + 2 \times \text{number of fitted parameters},
\end{equation}
and at equal cost the untransformed variable comes first. \code{verified} is a numerical statement about held-out data on
the same interval. It is weaker than a proof, and Section~\ref{sec:limits} returns to this.

\subsection{Changes of variable}\label{sec:transport}

\begin{figure}[t]
\centering
\includegraphics[width=\textwidth]{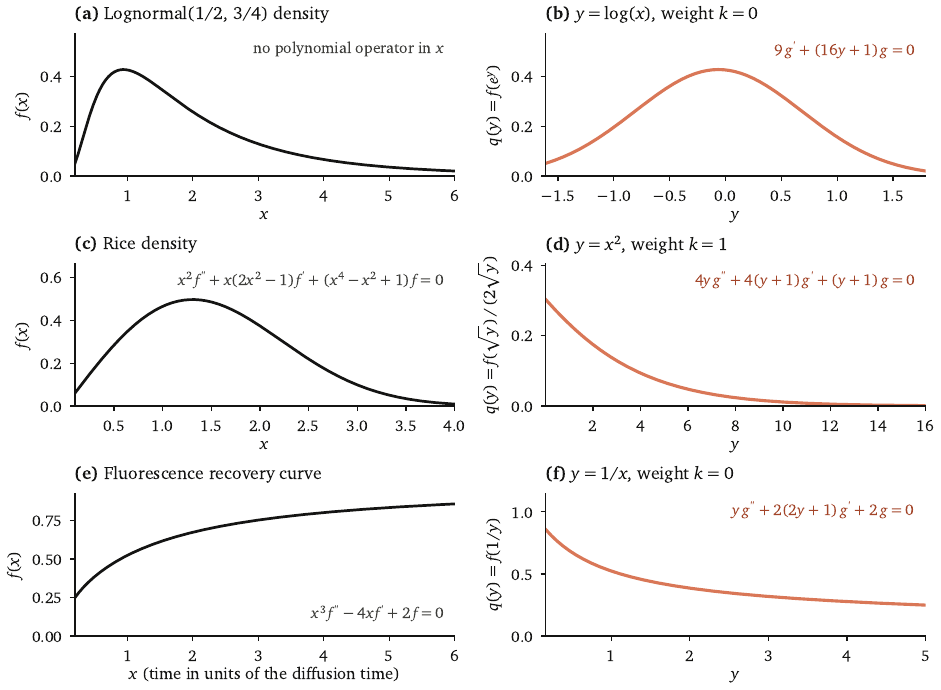}
\caption{Three changes of variable that the method selected. Left: the data $f(x)$. Right: the transported data
$q(y)$ of equation~\eqref{eq:transport}, which satisfies the equation written in the panel. (a,\,b)~The lognormal$(1/2,
3/4)$ density has no polynomial operator in $x$ and a first-order one in $y = \log x$. (c,\,d)~The Rice density has a
second-order operator of degree 4 in $x$ and one of degree 1 for its pushforward under $y = x^2$. (e,\,f)~The fluorescence
recovery curve of Section~\ref{sec:frap} has a simpler operator in $y = 1/x$.}
\label{fig:transport}
\end{figure}

The lognormal density has no operator of the form~\eqref{eq:operator} in $x$. Its logarithm is a quadratic in $\log x$, and
as a function of $y = \log x$ the density satisfies a first-order equation with coefficients of degree one
(Figure~\ref{fig:transport}a,\,b). For a monotone map $y = \varphi(x)$ and a weight $k$ the search of the previous sections
runs on the transported data
\begin{equation}\label{eq:transport}
  q(y) = f\bigl(\varphi^{-1}(y)\bigr)\,\left|\frac{\dd \varphi^{-1}}{\dd y}\right|^{k} ,
\end{equation}
and an operator that annihilates $q$ certifies the representation
\begin{equation}\label{eq:representation}
  f(x) = |\varphi'(x)|^{k}\,g\bigl(\varphi(x)\bigr), \qquad L_y\,g = 0 .
\end{equation}
With $k = 0$ this is a plain substitution. With $k = 1$ the function $q$ is the pushforward of $f$: if $f$ is the density of
a random variable $X$, then $q$ is the density of $\varphi(X)$. Both weights are tried for every map whose domain contains
the interval of the data. Before the interval is mapped, 2\% of it is cut from each end, where a map may tend to infinity;
the identity is not cut. Draws are pushed through $\varphi$ directly, which samples the pushforward.

The bank holds twenty maps, listed with their domains in Table~\ref{tab:bank}: the identity (cost 0); $\log x$, $\ee^x$,
$1/x$, $x/(1-x)$ and six powers of $x$ (cost 1); and the logit, probit, $\artanh$, $\arsinh$, $\tan$, $\arctan$, $\arcsin$,
$\sinh$ and $\tanh$ (cost 2). The search limits in a transported variable are order 2 and degree 4. The caller can supply
further maps, for example a bank map rescaled from $(0, 1)$ to an interval $(a, b)$.

\paragraph{Pulling back.} Let $W = |\varphi'|^{k}$, so that $f(x) = W(x)\,g(\varphi(x))$. The operator
\begin{equation}\label{eq:pullback}
  A = \frac{1}{\varphi'}\,D_x - \frac{W'}{W \varphi'}
\end{equation}
satisfies $A^i f = W\,g^{(i)}(\varphi(x))$ for every $i$. With $L_y = \sum_i p_i(y)\,D_y^i$, the equation $L_y\,g = 0$
becomes $\sum_i p_i(\varphi(x))\,A^i f = 0$. This is an operator in $x$ that acts on $f$, with coefficients that are
functions of $x$. When a common factor makes them polynomials, the result is reported as the operator in $x$, and certificates with the same operator in $x$ are merged under
the cheapest one: a holonomic function stays holonomic under an algebraic change of variable, so one equation in $x$ is
found again under each such map. When no factor makes them polynomials, as for $\arsinh$, the logit, the probit, $\tan$
and, with a Gaussian in $y$, the logarithm, the certificate exists only in the transported variable and the direct search
in $x$ correctly finds nothing. The change of variable then enlarges the class of functions that the method describes
beyond the holonomic ones: $\tan x$ is returned as the solution of $y\,g' - g = 0$ in $y = \tan x$.

\paragraph{A fitted power.} When the bank gives nothing and the data lie on $x > 0$, the exponent of $y = x^{\rho}$ is
estimated from the data. Let $s(\rho)$ be the smallest singular value of the first-order weak matrix of degree 1 for the
data transported by $x^{\rho}$. Near an exponent $\rho^*$ at which such an operator exists, $s$ is proportional to
$|\rho - \rho^*|$. A scan over $\rho = 2^{k/6}$, $k = -12, \ldots, 15$, locates the dips, a secant iteration on $s$ refines
each one to the working precision, and the identification step of Section~\ref{sec:exact} then makes the exponent exact.
A fitted power has cost 2. In the regression cases of Section~\ref{sec:cases} the exponents $17/10$, $23/10$, $5/4$ and
$\pi$ were recovered in this way, each with an error near $\tento{-34}$ before identification.

\paragraph{What is and is not learned.} The map is selected from the bank, from the caller's maps or from the power family.
A free search over all maps is ill-posed without a measure of complexity: the probit of the distribution function turns
every density into a Gaussian. The costs in~\eqref{eq:cost} are a crude measure of this kind, and
Section~\ref{sec:limits} lists the choice of map among the limits.

\subsection{Closed forms}\label{sec:solver}

The solver receives the exact operator and nothing else. It builds candidate solutions by the rules below, which are
applied in groups in this order. After each group the data are compared with the candidates found so far, and a later
group is built only when they are not yet matched.
\begin{enumerate}[leftmargin=*,itemsep=1pt,topsep=3pt]
\item \emph{First order}: $\exp\!\bigl(-\!\int p_0 / p_1\bigr)$, with logarithms written as powers. No other rule is applied
to an operator of order 1.
\item \emph{Constant coefficients}: $x^k \ee^{\lambda x}$ for the characteristic roots $\lambda$, in real form. No other
rule is applied to such an operator.
\item \emph{Symmetric square.} If $u$ and $v$ solve $u'' + a u' + b u = 0$, then $y = u v$ solves
$y''' + 3 a y'' + (a' + 2 a^2 + 4 b)\,y' + (2 b' + 4 a b)\,y = 0$. A third-order operator of this form is recognised from
its coefficients, and the products of the solutions of the second-order operator are the candidates: $J_0^2$, $\Ai^2$,
$\Ai \Bi$, $I_0 K_0$.
\item \emph{Series and named functions.} The candidates of the following rules form one group, which is passed to
SymPy's \code{hyperexpand} \cite{roach1996hypergeometric,meurer2017sympy}; it names hypergeometric series where it can
($K$, $E$, erf, incomplete gamma functions and others).
  \begin{itemize}[leftmargin=1.2em,itemsep=1pt,topsep=2pt]
  \item \emph{Airy}: $\Ai$ and $\Bi$ of a linear argument for $f'' = (\alpha x + \beta) f$.
  \item \emph{Euler and two-term operators.} About the origin and about each rational singular point the operator is
  written in $\theta$-form, with $\theta = t\,\dd/\dd t$ in the distance $t$ from that point. With one term,
  $P(\theta) f = 0$, the solutions are $t^{\lambda} \log^k t$ for the roots $\lambda$ of $P$. With two terms whose powers
  of $t$ differ by $m$, put $z = t^m$ and $\vartheta = z\,\dd/\dd z = \theta / m$. The operator then reads
  \begin{equation}\label{eq:twoterm}
    \prod_{j=1}^{q} (\vartheta - b_j)\,f = \mu\,z \prod_{j=1}^{p} (\vartheta - \alpha_j)\,f .
  \end{equation}
  For $q \ge p$ its Frobenius series are
  $z^{b}\,\pFq{p}{q-1}\bigl(b - \alpha_1, \ldots, b - \alpha_p;\ 1 + b - b_j\ \text{for}\ b_j \ne b;\ \mu z\bigr)$, one for
  each exponent $b$ among the $b_j$ whose series has no logarithm, and for $q \le p$ the same rule gives the series in
  $1/z$. For $p = 0$ and $q = 2$ Bessel forms are written directly, with $Y$ and $K$ for the solutions that contain a
  logarithm.
  \item \emph{Normal form}: for order 2 the first-derivative term is removed by $f = W u$, and the two rules above run on
  the operator of $u$.
  \item \emph{Primitive operator.} When the solutions of the operator are the derivatives of the solutions of another
  operator, that operator is computed and solved in its place. This covers the derivative of a Bessel function, whose
  equation has an apparent singular point.
  \item \emph{Antiderivatives.} When the operator ends in $D$, $L = N D$, its solutions are the constants and the
  antiderivatives of the solutions of $N$; this is the case of $f + c$ for a constant $c$. They are taken from the
  primitive operator of $N$ when there is one, and otherwise by integrating the solutions of $N$. For a first-order $N$
  the \emph{binomial primitive} is also tried: the antiderivative of $t^{a} (t^m - u)^{b} (t^m - v)^{c}$ is $t^{a+1}$
  times Appell's $F_1$ \cite{appell1926}, which is Lauricella's $F_D$ in two variables \cite{lauricella1893}, or a
  $\pFq{2}{1}$ when there is one binomial. Incomplete elliptic integrals, the incomplete beta function and the distance
  integrals of cosmology are of this kind.
  \end{itemize}
\item \emph{Polynomial solutions}, of the operator and of the operator after a \emph{gauge}: factors
$\exp(a x^2 + b x) \prod_j q_j^{e_j}$, with $q_j$ a linear or quadratic factor of the leading coefficient and $e_j$ one of
its local exponents, are divided out of the solution.
\item \emph{Reduction of order}, for order 2, from an elementary solution that the previous group found.
\item \emph{Two-term operators after a gauge}: the Euler and two-term rule on the gauged operators, again followed by
\code{hyperexpand}. The parabolic cylinder functions are $\ee^{-x^2/4}$ times two $\pFq{1}{1}$ series.
\item \emph{Meijer $G$.} For the two-term form~\eqref{eq:twoterm} of the operator and of its first two gauged operators,
every split of the parameters into the four groups of $G^{\,k,n}_{\,p,q}$ gives a solution, with upper parameters
$a_j = \alpha_j + 1$, lower parameters $b_j$ and argument $(-1)^{p-k-n} \mu z$ \cite{beals2013meijer,roach1997meijer}.
These include the solutions with logarithms that no Frobenius series gives, such as $G^{\,q,0}_{\,0,q}$, the density of a
product of $q$ normal variables.
\end{enumerate}
The first version of the solver had the rules of groups 1, 2 and 4 to 7 without the primitive operator and the binomial
primitive, and a gauge with exponential and power factors only. Five additions were made after the first comparison with
symbolic regression (Section~\ref{sec:history}): the symmetric square, the primitive operator, the Gaussian and quadratic
factors of the gauge, the binomial primitive and the Meijer $G$ rule.

A candidate is kept only if it satisfies the operator numerically, to a relative $\tento{-12}$ at three points of the
interval. The data then select the smallest set
of candidates, up to four, whose span contains them at nine points of the interval; among sets of equal size the one built
from the simplest expressions wins. The fitted constants go through the identification step, extended by $\Gamma$ at the
rational parameters of the candidates and by exponentials of the operator's own parameters. The result is mapped back
through $\varphi$ by~\eqref{eq:representation} and compared with the data at five other points. The certificate also
records whether SymPy reduces the operator applied to the closed form to zero, which is then a symbolic proof that the
closed form solves the operator.

When the operator of a callable has fitted parameters, they are refitted at twice the working precision before the solver
runs and replaced by exact numbers when the identification step then finds one for every parameter. A fitted parameter
that is known to fewer than twelve digits, as from a float64 table, is re-estimated through the solutions themselves: the
combination $\sum_k C_k f_k(y;\,a)$ of the candidates is matched to the table at 24 points of the fit windows, with the
constants $C_k$ solved at each trial value of $a$, which fixes $a$ to the accuracy of the values where window moments fix
it far less well. The refined operator must pass the hold-out test again. Section~\ref{sec:fresnel} shows a case.

The solver is a set of rules. It has no general algorithm for Liouvillian solutions \cite{kovacic1986}, for second-order
operators that need a change of variable to reach a hypergeometric form \cite{imamoglu2017hypergeometric}, or for operators
with three or more terms that no gauge reduces, the Heun type \cite{ronveaux1995heun}. When it fails the certificate still
holds the exact operator, which the next section uses.

\subsection{Continuation of a certified operator}\label{sec:continuation}

An operator of order $r$ and $r$ numbers determine a function. For a certificate without a closed form, the values
of the solution and of its first $r - 1$ derivatives at a point near the middle of the interval are fitted by least squares
to the data at 24 points. The Taylor coefficients of that solution about any centre follow from the operator by a linear
recurrence. The series (240 terms) converges up to the nearest root of the leading coefficient, and the values and
derivatives at a point inside that radius start the series about the next centre
\cite{vanderhoeven1999holonomic,mezzarobba2010numgfun}. A real singular point on the way is passed on a half circle in the
complex plane, and the function is continued beyond it when the values that arrive are real, as at an apparent singular
point; at a branch point the function ends. Nothing enters except the operator and the data, and the result can be
evaluated inside the interval of the data and beyond it. The operator found for the Voigt profile on $[0.1,\,3]$, continued
to $x = 5$, agrees with the profile there to $3.5 \timestento{-41}$. In the benchmarks we count a target that is reproduced
in this way separately from one with a closed form.

\subsection{Draws, sequences and moments}\label{sec:other}

\paragraph{Draws.} For independent draws from a density, three quarters of the sample fit the operator and one quarter
tests it. The windows are the whole interval between the 0.001 and 0.999 quantiles, its lower 60\% and its upper 60\%; wide
windows keep the sampling error of the moments small. Sampling error is far from uniform over the rows and columns of the
weak matrix, so the rows are divided by their rms error and the columns are recombined so that the noise is the same in
every coefficient direction. The singular values are then signal-to-noise ratios, and the smallest singular vector is an
errors-in-variables estimate of the null direction. Entries that stay numerical after the first identification are
estimated again by generalised least squares given the exact ones, and tried again as fractions. The search is limited to
order 2 and degree 3 in $x$ and to order 1 and degree 2 in a transported variable, because the sampling error of a moment
grows quickly with its order.

The hold-out test is Hotelling's $T^2$ \cite{hotelling1931} of the weak equations $M c$ on the held-out quarter against
their covariance over batches, passed at $p > 0.001$. Table~\ref{tab:calibration} shows its calibration: with the operator
fixed in advance, the true operator was rejected in 0 of 80 runs and an operator with one coefficient changed by 2\% or
2.5\% in 80 of 80. A closed form is fitted to the first six moments of each window. When it is a single solution times a
constant, the constant is replaced by the exact normalisation over the interval between the singular points that enclose
the draws, provided the two agree within the sampling error. The density is then tested against the held-out draws by
$\chi^2$ on 15 bins between the $\tento{-4}$ and $1 - \tento{-4}$ quantiles, which reach beyond the windows of the operator
fit, and a certificate whose closed form fails that test is rejected. Results from draws are graded \code{conjecture} when
all coefficients are exact and \code{fit} otherwise.

\paragraph{Sequences.} For exact terms $a_0, a_1, \ldots$ the method looks for a recurrence
$\sum_{i=0}^{r} p_i(n)\,a_{n+i} = 0$ with polynomial coefficients, up to order 3 and degree 3. The null space is computed in
exact rational arithmetic, so no noise model is needed. The last six equations are left out of the computation and must
then hold exactly. A closed form is returned when SymPy's solver for hypergeometric terms finds one that reproduces every
term.

\paragraph{Moments.} For exact moments $m_n = \int x^n f(x)\,\dd x$ the recurrence of the sequence $m_n$ is found in the
same way and turned into a differential operator for the density. Since $m_{n+i} = \int x^n\,(x^i f)\,\dd x$ and, by
integration by parts with vanishing boundary terms, $n\,m_n = \int x^n \bigl(-(\theta + 1) f\bigr)\,\dd x$,
\begin{equation}\label{eq:moment-operator}
  \sum_i p_i(n)\,m_{n+i} = \int x^n\,(L f)\,\dd x , \qquad L = \sum_i p_i(-\theta - 1)\,x^i .
\end{equation}
A density whose moments obey the recurrence satisfies $L f = 0$ between roots of the leading coefficient of $L$ and can
start or stop only at such a root. Each interval between two real roots (or infinity) is tried as the support, shortest
first. On it the solutions of $L$ are integrated against $x^n$, the combination that reproduces moments 0 to 5 is
selected, and it is tested on moments 6 to 14. From the Catalan numbers as moments this gives $x (x - 4) f' - 2 f = 0$ and
the density $\sqrt{4 - x} / (2 \pi \sqrt{x})$ on $(0, 4)$.

\subsection{Float64 tables in several variables}\label{sec:relations}

The operator search reads window moments of a function of one variable. A second entry point takes float64 points in any
number of variables, scattered or on a grid, and returns one expression. It adds what the operator search cannot read
(several variables) or reads poorly (elementary functions with a scale inside, short tables), and it ends in the operator
search for what remains. The rows are split into 70\% for fitting and 30\% for testing: the 30\% with the largest $x$ for
one variable, so that the test is an extrapolation, and a random 30\% for several. A relation of steps 1 to 3 is fitted on
the first part only. A closed form from step 4 comes from the operator search on the whole table, which holds out a window
of its own, and is then compared with the same 30\% of the rows.
\begin{enumerate}[leftmargin=*,itemsep=1pt,topsep=3pt]
\item \emph{Power factors.} Variables that enter only as $x_i^{a}$ are found by regression of $\log |f|$ on the logarithms of
the variables and divided out, with $a$ identified as a fraction.
\item \emph{Relations.} For a map $v = \psi(f)$ of the values (the identity, $f^2$, $\log f$, $\ee^{f}$, $\sin f$, $\cos f$,
$\tan f$, $\sinh f$, $\tanh f$ and $f \ee^{f}$) and a set of atoms $u$ (the variables, and the cosine and sine, exponential,
logarithm or square root of a variable or of a power of it), the search looks for polynomials $P_0$ and $P_1$ with
\begin{equation}\label{eq:relation}
  v\,P_1(u) + P_0(u) = 0
\end{equation}
on the fit rows, which is again a null-vector problem. A dense search takes all monomials up to a pair of degrees. It
accepts a null space of up to four dimensions, which arises when a common factor fits within the degrees, and takes from it
the vector with zeros at the pivot columns of the others. For one variable a sparse search scores every set of two to four
monomials (to five under noise) and solves the twelve best of each size; it works where the monomials are dependent to
rounding error and under noise. The entries of the null vector are identified as in Section~\ref{sec:exact}, and at most
three may stay numerical (six for a table with noise). That limit refuses a rational approximant of a smooth function,
which has many unidentified coefficients. For one variable the atoms $\exp(c x)$ and $\exp(c / x)$, and the cosine and
sine of $c x$ and of $c / x$, are also tried, with the scale $c$ fitted. When none of these gives a relation with exact
constants, the search also looks for $f$ as a root of $P_2 f^2 + P_1 f + P_0 = 0$ with polynomials $P_k$ in the atoms.
\item \emph{Scaling symmetries.} If $f(\lambda^{w_1} x_1, \ldots, \lambda^{w_n} x_n) = \lambda^{c} f(x)$ for a weight vector
$w$, then $f$ is a power product of the variables times a function of fewer variables, each a monomial of the $x_i$. The
weights are null vectors of $\sum_i w_i\,x_i\,\partial_i f = c f$, with the derivatives taken from a polynomial fit. One
remaining variable gives a table for step 4; several go back to step 2. This is the dimensional reduction that AI~Feynman
obtains from the units of the variables \cite{udrescu2020aifeynman}, found here from the numbers alone.
\item \emph{Operators.} One variable without a relation with exact constants goes to the operator search of
Sections~\ref{sec:measure} to~\ref{sec:solver}, for $f$ and then for $1/f$, $\log f$ and $\sqrt{f}$.
\end{enumerate}
An expression is accepted when it matches the test rows to $\tento{-8}$, relative to $|f|$ plus the rms of the values. It
is \code{verified} when in addition every constant in it is exact, the operator certificate behind a closed form of step
4 is itself \code{verified}, and the match is to $\tento{-12}$; otherwise it is a \code{conjecture}. For one variable a
noise level is estimated from divided differences of order 6 over neighbouring points, which remove the smooth part of
the data and leave the noise. With noise, an expression whose rms residual on the test rows is within 1.35 times that
level is returned, as a \code{conjecture} when all its constants are exact and as a \code{fit} otherwise. Noise is not
handled for several variables.

\subsection{Functions of two variables}\label{sec:bivariate}

For a function $f(x, y)$ given as a callable at $N_p$ digits on a box without singular points, the measurements are the
moments of $f$ over sub-boxes, and the weak form of $x^i y^j \partial_x^a \partial_y^b f$ against a product
$B_s(u) B_t(v)$ of test functions is the product of the two one-variable maps of Section~\ref{sec:measure} applied to
them. The data are not differentiated here either. The fit uses three of the four quadrants of the box and the fourth is
held out.
\begin{enumerate}[leftmargin=*,itemsep=1pt,topsep=3pt]
\item \emph{System.} All operators of order at most 2 with polynomial coefficients of total degree at most $d$ that
annihilate $f$ on the three quadrants. They form a vector space, usually of several dimensions. A singular value counts as
zero when it is below $\kappa$ times its noise, and the first one that does not must exceed that level by a factor
$\tento{6}$. The reduced row echelon form of the null space is unique, and its entries are identified as fractions. A
reduction then keeps the rows that are not linear combinations of left multiples of simpler rows, where only the multiples
that stay within order 2 and degree $d$ are formed. The degree $d$ is the least from 1 to 3 at which two rows are kept,
and otherwise the largest at which one is. The kept rows generate every operator that was found in the search space. They
need not be a minimal set of generators of the annihilating ideal: a kept row can still be a combination of left multiples
of the others that passes through operators of higher order or degree.
\item \emph{Series.} The term $x^i y^j \partial_x^a \partial_y^b$ moves the exponent of a monomial by $(i - a,\,j - b)$. An
operator whose terms move it by $w$ or by $w + s$ only has the form $x^{w} \bigl(P(\theta) + x^{s} Q(\theta)\bigr)$ in
multi-index notation, with $\theta = (x \partial_x,\,y \partial_y)$, and fixes the ratio
\begin{equation}\label{eq:horn}
  \frac{A(k + s)}{A(k)} = -\frac{Q(k)}{P(k + s)}
\end{equation}
of the coefficients of a series $\sum_k A(k)\,x^{k_1} y^{k_2}$. Two operators of this kind with independent shifts $s_1$,
$s_2$ determine a double hypergeometric series up to its leading exponent
\cite{erdelyi1953htf,srivastava1985multiple}. When the factors of the two ratios are those of Appell's $F_1$ to $F_4$
\cite{appell1926} the series is returned under that name, and otherwise as a general double series, which covers Humbert's
confluent functions.
\item \emph{Solution.} The combination of the series at the admissible leading exponents that matches $f$ at 14 points of
the fit quadrants, with its constants identified. It is tested at points of the fourth quadrant and beyond the box.
\item \emph{Changes of variable.} Steps 1 to 3 run in $(x, y)$ and, when that gives no series, with $x / (1 - x)$ or
$1 - x$ applied to both variables.
\end{enumerate}
The result is \code{verified} when every kept operator of the system and both operators of the series have a residual on
the held-out quadrant of at most $\kappa$ times its noise level, at $N_p$ digits. The length scale of
Section~\ref{sec:holdout} is not applied here, and the test is not repeated at a second precision. A function with one
annihilating operator and no second one, such as a function of $x + 2 y$ alone, returns that operator and no series.

\section{A worked example: the Fresnel integral}\label{sec:fresnel}

This section follows the method on $C(x) = \int_0^x \cos(\pi t^2 / 2)\,\dd t$ on the interval $[0.2,\,2]$, first from a
callable and then from the float64 table of Figure~\ref{fig:fresnel_fits}. The intermediate values quoted here were
recorded by calling the learner's own functions in the order in which it calls them.

\subsection{From a callable at 40 digits}

\paragraph{Measure.} The three fit windows are $[0.272,\,0.74]$, $[0.632,\,1.1]$ and $[0.992,\,1.46]$, and the held-out
window is $[1.46,\,1.928]$ (Figure~\ref{fig:fresnel_trace}a). On each window the learner computes the moments
\eqref{eq:moments} of $C$ by Gauss--Legendre quadrature and keeps the difference between the last two quadrature levels as
a realisation of their error.

\paragraph{Search.} The search visits 22 supports (Figure~\ref{fig:fresnel_trace}b). At order 1 the smallest singular value
of the scaled weak matrix falls steadily from $\tento{-0.05}$ at degree 0 to $\tento{-13.4}$ at degree 8, as it does for
any smooth function when columns are added, while its noise level stays near $\tento{-37}$: no first-order operator
annihilates $C$. At order 2 it falls from $\tento{-0.9}$ to $\tento{-24.1}$ against a noise level near $\tento{-35}$, and
again there is no null direction. At order 3 the values at degrees 0, 1 and 2 are $\tento{-1.6}$, $\tento{-4.9}$ and
$\tento{-9.2}$. At degree 3, where the matrix has 48 rows and 16 columns, the smallest singular value is $\tento{-32.5}$
with a noise level of $\tento{-32.2}$, a drop of 23.3 orders of magnitude in one step, and the second smallest stays at
$\tento{-11.8}$. The conditions of~\eqref{eq:accept} hold and the search stops.

\paragraph{Make exact.} Divided by the entry of $x D^3$, the null vector has three entries of order one
(Figure~\ref{fig:fresnel_trace}c),
\begin{equation*}
  x^3 D:\ \ 9.86960440108935861883\ldots, \qquad D^2:\ \ -1.0000000000000000000000374, \qquad x D^3:\ \ 1 ,
\end{equation*}
and 13 entries of size at most $6.4 \timestento{-22}$. The identification step replaces the second by $-1$ and the 13
small ones by 0. The first has no fraction of acceptable height within its error. It is named by the list of forms
$q\,\pi^k$ of Section~\ref{sec:exact}, before any integer-relation search runs: it is $\pi^2$, from which the measured value
differs by a relative $6 \timestento{-24}$. The operator is
\begin{equation*}
  L = x\,D^3 - D^2 + \pi^2 x^3 D ,
\end{equation*}
which is equation~\eqref{eq:fresnel}, with an evidence of 327.5 digits.

\paragraph{Test.} On the held-out window the residual of the exact operator is $9.7 \timestento{-33}$ against a noise
level of $9.2 \timestento{-33}$ at 40 digits, and $8.3 \timestento{-73}$ against $9.9 \timestento{-73}$ at 80 digits. The
residual follows the noise down by forty orders of magnitude when the precision is doubled, which an operator that was
only approximately right would not do. The certificate is graded \code{verified}.

\paragraph{Solve.} The solutions of $L$ are spanned by 1, $C(x)$ and the Fresnel sine integral $S(x)$. The solver arrives
at them through the two-term rule: in $z = x^4$ the operator has the form~\eqref{eq:twoterm}, with the series
$x\,\pFq{1}{2}\bigl(\tfrac14;\,\tfrac12, \tfrac54;\,-\pi^2 x^4/16\bigr)$ and
$x^3\,\pFq{1}{2}\bigl(\tfrac34;\,\tfrac32, \tfrac74;\,-\pi^2 x^4/16\bigr)$, which the expansion step names as $C(x)$ and
$6 S(x)/\pi$. The solver produced 11 candidates in all. The data at nine points select $C(x)$ alone, with the constant 1,
and at the five check points the closed form and the data agree in every digit. The whole run took 10.8\,s.

\subsection{From the float64 table}

The same steps run on the 1101 float64 values of the benchmark table, with the moments taken from a local interpolant
whose estimated relative error is $4.0 \timestento{-17}$. The search stops at the same support, order 3 and degree 3,
where the smallest singular value is $2.2 \timestento{-14}$ against a noise level of $2.4 \timestento{-14}$ and the next
is $1.8 \timestento{-12}$. The gap is now a factor of 80 where the callable had more than twenty orders of magnitude. The
exact entries $-1$, 1 and 0 are still identified, with an evidence of 37.7 digits, but the remaining entry is known only
as $a_1 = 9.8693 \pm 0.0021$, which identifies nothing. The operator at this stage is
$x\,D^3 - D^2 + a_1 x^3 D$ with a fitted parameter.

The solver handles the parameter as a symbol, and its candidates are Fresnel integrals of a scaled argument. The parameter
is then re-estimated through those solutions (Section~\ref{sec:solver}): matching their combination to the table at 24
points of the fit windows fixes $a_1$ to the accuracy of the values in place of the accuracy of the window moments, and at
that accuracy the identification step returns $\pi^2$. The exact operator passes the hold-out test with a residual of
$3.2 \timestento{-15}$ against a noise level of $2.7 \timestento{-15}$, the noise is below $\tento{-8}$ and the evidence
above 10 digits, so the grade is \code{verified}, and the closed form $C(x)$ agrees with the table to
$8 \timestento{-17}$. This run of the operator search took 18.2\,s. In the benchmark of Section~\ref{sec:benchmarks} the
same table goes through the entry point of Section~\ref{sec:relations}, which searches for elementary relations first and
took 57.2\,s in all.

The Fresnel integral was one of 31 special functions of a pilot set that was in view while the method was developed
(Section~\ref{sec:protocol}), and the re-estimation step was added because an earlier version of the learner returned this
table as a conjecture with $a_1 = 9.8693$ left numerical. The example shows how the method works. The generated set of
Section~\ref{sec:generated} measures how it does on functions that it was not developed on.

\subsection{What the other programs returned}

Table~\ref{tab:fresnel} lists the result of each program for the same 1101 points, and Appendix~\ref{app:fresnel} prints
the seven formulas. Operon's formula, of 302 characters, follows $C(x)$ to 0.014\% inside the training range, and RILS-ROLS
returned the shortest one,
\begin{equation*}
  1.055034\,\sin(x) \cos(x^2) - 0.689068\,\cos(x^2) + 0.672111 ,
\end{equation*}
with an error of 0.90\%. Beyond the range, up to $x = 2.9$, the errors are between 27\% and 167\%. Six of the seven
formulas fit the points to within 3.4\%. Each is the best formula that its program found in its budget, and the space that
these programs search does not contain $C(x)$.

\begin{table}[t]
\centering\small
\caption{Results for 1101 float64 points of the Fresnel integral $C(x)$ on $[0.2,\,2]$. The error is the largest value of
$|g - C| / (|C| + \rms)$ at 400 random points inside the training range and at about 200 points beyond it, up to
$x = 2.9$, in percent. The five programs with a time setting ran for 60\,s on one thread; Flash-ANSR, E2E, TPSR and
\VEL{} have none, and TPSR was stopped at 600\,s.}
\label{tab:fresnel}
\begin{tabular}{@{}llccc@{}}
\toprule
 & & Error inside & Error beyond it, & \\
Method & Returned & the data (\%) & to $x = 2.9$ (\%) & Seconds \\
\midrule
\VEL{} & $C(x)$ & \phantom{0}0\phantom{.000} & \phantom{00}0 & \phantom{0}57.2 \\
Operon & formula of 302 characters & \phantom{0}0.014 & \phantom{0}65 & \phantom{0}60.0 \\
RILS-ROLS & formula of 60 characters & \phantom{0}0.90\phantom{0} & \phantom{0}30 & \phantom{0}61.0 \\
PySR & formula of 183 characters & \phantom{0}1.1\phantom{00} & \phantom{0}37 & \phantom{0}60.3 \\
Bingo & formula of 250 characters & \phantom{0}1.2\phantom{00} & \phantom{0}90 & \phantom{0}64.5 \\
Flash-ANSR & formula of 74 characters & \phantom{0}2.0\phantom{00} & 167 & 180.0 \\
E2E & formula of 451 characters & \phantom{0}3.4\phantom{00} & 132 & 263.7 \\
gplearn & formula of 81 characters & 33\phantom{.000} & \phantom{0}27 & \phantom{0}60.2 \\
TPSR & no formula in 600 s & -- & -- & 600.0 \\
\bottomrule
\end{tabular}

\end{table}

\section{Examples}\label{sec:examples}

Every example in this section is a recorded run with the default settings. The callables are evaluated with $N_p = 40$, and
the times are wall-clock seconds of one process; several such processes ran side by side on one machine.

\subsection{Densities and changes of variable}\label{sec:densities}

\paragraph{Lognormal.} For the lognormal density with parameters $\mu = 1/2$ and $\sigma = 3/4$ on $[0.2,\,6]$ the search
in $x$ finds nothing, as it should. Under $y = \log x$ with weight $k = 0$ the learner returns
(Figure~\ref{fig:transport}a,\,b)
\begin{equation*}
  9\,g' + (16 y + 1)\,g = 0 , \qquad
  f(x) = \frac{2 \sqrt{2}\;\ee^{-2/9}}{3 \sqrt{\pi}}\;x^{-1/9}\,\ee^{-8 \log^2 x / 9} ,
\end{equation*}
with the constant identified exactly, in 45.3\,s.

\paragraph{Rice.} The Rice density $f(x) = x\,\ee^{-(x^2 + 1)/2} I_0(x)$ is the distribution of the envelope of a sine
wave in Gaussian noise \cite{rice1944noise,rice1945noise}. On $[0.1,\,4]$ it has an operator in $x$, and the learner ranks
first the cheaper one of its pushforward under $y = x^2$ (Figure~\ref{fig:transport}c,\,d):
\begin{equation*}
  4 y\,g'' + 4 (y + 1)\,g' + (y + 1)\,g = 0, \qquad\text{in $x$:}\quad
  x^2 f'' + x (2 x^2 - 1) f' + (x^4 - x^2 + 1) f = 0 .
\end{equation*}
The residual on the held-out window is $4.0 \timestento{-38}$ against a noise level of $1.6 \timestento{-37}$ at 40
digits, and $1.9 \timestento{-78}$ against $1.4 \timestento{-77}$ at 80. The closed form returned is
$x\,\ee^{-x^2/2 - 1/2} I_0(x)$, in 22.6\,s.

\paragraph{A parameter without a closed form.} Several regression cases (Section~\ref{sec:cases}) contain the number
$a = 0.7390851332151606\ldots$, the root of $\cos a = a$, which none of the identification searches can name. For
$f(x) = \ee^{-a/x} / x^2$ the learner returns the map $y = 1/x$ with $k = 1$, the operator $g' + a_1 g = 0$ and the closed
form $\ee^{-a_1/x}/x^2$, with $a_1$ kept as a symbol. Its fitted value at 80 digits differs from $a$ by
$9.0 \timestento{-82}$.

\paragraph{Fitted exponents.} For $\pi\,x^{\pi - 1} \ee^{-x^{\pi}}$ no map of the bank gives an operator. The fitted
power of Section~\ref{sec:transport} finds the exponent, identifies it as $\pi$ and returns $g' + g = 0$ under
$y = x^{\pi}$ with $k = 1$, in 47.8\,s. The density $\tfrac{17}{10}\,x^{7/10} \ee^{-x^{17/10}}$ comes back in the same
way, with the exponent $17/10$ and $g' + g = 0$. For the function $x^{2} \ee^{-3 x^{23/10}/2}$ the exponent $23/10$ is
identified, and the first certificate is the plain substitution $y = x^{23/10}$ with $k = 0$ and the operator
$46 y\,g' + (69 y - 40)\,g = 0$.

\subsection{Wavefunctions of exactly solvable systems}\label{sec:quantum}

\begin{figure}[t]
\centering
\includegraphics[width=\textwidth]{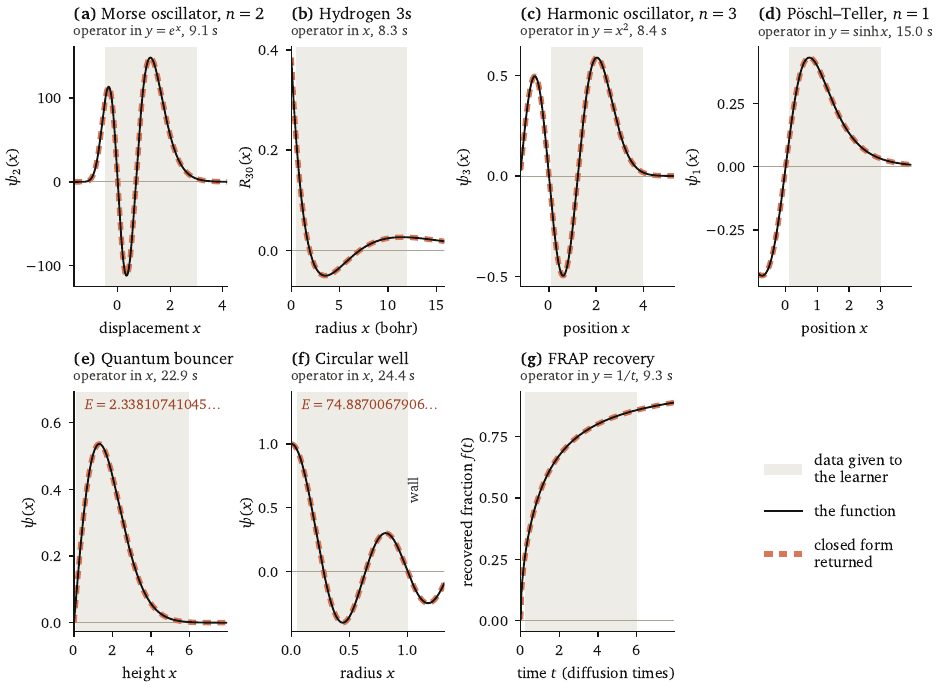}
\caption{Seven of the worked examples. Grey band: the interval of the data given to the learner. Black: the function.
Dashed: the closed form returned, drawn beyond the data as well. Under each title, the variable in which the operator was
found and the time. Panels (e) and (f) give the energy that the learner returned as the fitted coefficient of the
Schr\"odinger equation; panel (f) marks the wall of the well.}
\label{fig:quantum}
\end{figure}

A few exactly solvable systems carry much of quantum chemistry and physics: the harmonic oscillator for small vibrations,
the Morse oscillator for the stretching of a bond \cite{morse1929}, the hydrogen atom for orbitals, and the
P\"oschl--Teller well \cite{poschl1933,landau1977qm}. For eight such systems we gave the learner a routine that returns one
wavefunction on an interval and nothing else: no potential, no energy and no boundary condition.
Table~\ref{tab:examples} lists what it returned for them and for the recovery curve of Section~\ref{sec:frap},
Table~\ref{tab:examples_checks} the tests, and Figure~\ref{fig:quantum} shows seven of the closed forms continued beyond
the data.

\begin{table}[p]
\centering\small
\caption{Nine worked examples from callables at 40 digits. For each: the system, the interval of the data, the variable in
which the operator was found (with the weight $k$ of equation~\eqref{eq:transport} when it is 1), and the time; on the
indented lines, the equation returned, the closed form returned and any coefficient that stayed numerical, with what it
is. In a transported variable $y$ the equation is for $g$ of equation~\eqref{eq:representation}.}
\label{tab:examples}
\begin{tabular}{@{}L{66mm}C{22mm}L{40mm}R{14mm}@{}}
\toprule
System & Interval & Variable & Seconds \\
\multicolumn{4}{@{}l@{}}{\hspace{4mm}\parbox[t]{150mm}{\raggedright Equation returned; closed form returned; fitted parameter\strut}} \\
\midrule
Morse oscillator, $n = 2$ & $[{-}0.5,\,3]$ & $y = e^{x}$ & 9.1 \\
\multicolumn{4}{@{}l@{}}{\hspace{4mm}\parbox[t]{150mm}{\raggedright $2 y^{2} \left(y - 2\right) \left(y - 1\right)\,g' + \left(7 y^{3} - 39 y^{2} + 58 y - 24\right)\,g = 0$\strut}} \\
\multicolumn{4}{@{}l@{}}{\hspace{4mm}\parbox[t]{150mm}{\raggedright $\psi(x) = \nobreak 124416 \sqrt{3}\,\left(e^{x} - 2\right) \left(e^{x} - 1\right) e^{-11 x/2 - 6 e^{-x}}$\strut}} \\
\addlinespace
Hydrogen $3s$ & $[0.5,\,12]$ & $r$ & 8.3 \\
\multicolumn{4}{@{}l@{}}{\hspace{4mm}\parbox[t]{150mm}{\raggedright $3 \left(2 r^{2} - 18 r + 27\right)\,R' + \left(2 r^{2} - 30 r + 81\right)\,R = 0$\strut}} \\
\multicolumn{4}{@{}l@{}}{\hspace{4mm}\parbox[t]{150mm}{\raggedright $R(r) = \nobreak \frac{2 \sqrt{3}}{243}\,\left(2 r^{2} - 18 r + 27\right) e^{- r/3}$\strut}} \\
\addlinespace
Harmonic oscillator, $n = 3$ & $[0.1,\,4]$ & $y = x^{2}$, weight $k = 1$ & 8.4 \\
\multicolumn{4}{@{}l@{}}{\hspace{4mm}\parbox[t]{150mm}{\raggedright $2 \left(2 y - 3\right)\,g' + \left(2 y - 7\right)\,g = 0$\strut}} \\
\multicolumn{4}{@{}l@{}}{\hspace{4mm}\parbox[t]{150mm}{\raggedright $\psi(x) = \nobreak \frac{\sqrt{3}}{3 \pi^{1/4}}\,x \left(2 x^{2} - 3\right) e^{- x^{2}/2}$\strut}} \\
\addlinespace
P\"oschl--Teller well, $n = 1$ & $[0.1,\,3]$ & $y = \sinh x$ & 15.0 \\
\multicolumn{4}{@{}l@{}}{\hspace{4mm}\parbox[t]{150mm}{\raggedright $2 y \left(y^{2} + 1\right)\,g' + \left(3 y^{2} - 2\right)\,g = 0$\strut}} \\
\multicolumn{4}{@{}l@{}}{\hspace{4mm}\parbox[t]{150mm}{\raggedright $\psi(x) = \nobreak \sinh x \,/ \cosh^{5/2} x$\strut}} \\
\addlinespace
Quantum bouncer & $[0.1,\,6]$ & $x$ & 22.9 \\
\multicolumn{4}{@{}l@{}}{\hspace{4mm}\parbox[t]{150mm}{\raggedright $\psi'' + \left(a_{1} - x\right)\,\psi = 0$\strut}} \\
\multicolumn{4}{@{}l@{}}{\hspace{4mm}\parbox[t]{150mm}{\raggedright $\psi(x) = \nobreak \Ai\!\left(x - a_{1}\right)$\strut}} \\
\multicolumn{4}{@{}l@{}}{\hspace{4mm}\parbox[t]{150mm}{\raggedright $a_{1} = 2.3381074104597670384\ldots$, the energy $E$, minus the first zero of $\Ai$\strut}} \\
\addlinespace
Circular well & $[0.05,\,1]$ & $r$ & 24.4 \\
\multicolumn{4}{@{}l@{}}{\hspace{4mm}\parbox[t]{150mm}{\raggedright $r\,\psi'' + \psi' + a_{1} r\,\psi = 0$\strut}} \\
\multicolumn{4}{@{}l@{}}{\hspace{4mm}\parbox[t]{150mm}{\raggedright $\psi(r) = \nobreak J_{0}\!\left(\sqrt{a_{1}} r\right)$\strut}} \\
\multicolumn{4}{@{}l@{}}{\hspace{4mm}\parbox[t]{150mm}{\raggedright $a_{1} = 74.887006790695183444\ldots$, the energy $E$, the square of the third zero of $J_{0}$\strut}} \\
\addlinespace
Oscillator with a wall & $[{-}0.4,\,3]$ & $x$ & 158.9 \\
\multicolumn{4}{@{}l@{}}{\hspace{4mm}\parbox[t]{150mm}{\raggedright $\psi'' + \left(a_{1} - x^{2}\right)\,\psi = 0$\strut}} \\
\multicolumn{4}{@{}l@{}}{\hspace{4mm}\parbox[t]{150mm}{\raggedright $\psi(x) = \nobreak e^{- x^{2}/2} \bigl(0.9066057102\ldots x \,\pFq{1}{1}\!\left(3/4 - a_{1}/4;\,3/2;\,x^{2}\right) + 0.5498413068\ldots \,\pFq{1}{1}\!\left(1/4 - a_{1}/4;\,1/2;\,x^{2}\right)\bigr)$\strut}} \\
\multicolumn{4}{@{}l@{}}{\hspace{4mm}\parbox[t]{150mm}{\raggedright $a_{1} = 2.0607657916305730122\ldots$, the energy $E$ of the ground state with a wall at $x = -1/2$\strut}} \\
\addlinespace
Coulomb wave & $[0.5,\,12]$ & $\rho$ & 10.3 \\
\multicolumn{4}{@{}l@{}}{\hspace{4mm}\parbox[t]{150mm}{\raggedright $\rho\,F'' + \left(\rho + 1\right)\,F = 0$\strut}} \\
\multicolumn{4}{@{}l@{}}{\hspace{4mm}\parbox[t]{150mm}{\raggedright none\strut}} \\
\addlinespace
Fluorescence recovery after photobleaching & $[0.2,\,6]$ & $y = 1/t$ & 9.3 \\
\multicolumn{4}{@{}l@{}}{\hspace{4mm}\parbox[t]{150mm}{\raggedright $y\,g'' + 2 \left(2 y + 1\right)\,g' + 2\,g = 0$\strut}} \\
\multicolumn{4}{@{}l@{}}{\hspace{4mm}\parbox[t]{150mm}{\raggedright $f(t) = \nobreak \bigl(I_{0}\!\left(2/t\right) + I_{1}\!\left(2/t\right)\bigr) e^{-2/t}$\strut}} \\
\bottomrule
\end{tabular}

\end{table}

\begin{table}[t]
\centering\small
\caption{Tests of the nine certificates of Table~\ref{tab:examples}, all graded \code{verified}. Residual and noise level
of the exact operator on the held-out window at 40 and at 80 digits (for the three operators with a fitted parameter,
after the parameter is estimated again on that window); the evidence for the exact coefficients; the largest relative
difference between the closed form and the data at five check points; and the number of certificates returned, which
differ in the change of variable or in the weight $k$.}
\label{tab:examples_checks}
\begin{tabular}{@{}L{40mm}ccccccc@{}}
\toprule
 & & \multicolumn{2}{c}{Hold-out check} & Evidence & Error of the & \\
\cmidrule(lr){3-4}
System & Digits & residual & noise & (digits) & closed form & Certificates \\
\midrule
\multirow[t]{2}{=}{Morse oscillator, $n = 2$} & 40 & $2.0\timestento{-38}$ & $2.0\timestento{-37}$ & 318.3 & $2.0\timestento{-40}$ & 2 \\
 & 80 & $2.4\timestento{-78}$ & $1.4\timestento{-77}$ &  &  &  \\
\addlinespace
\multirow[t]{2}{=}{Hydrogen $3s$} & 40 & $9.4\timestento{-40}$ & $1.0\timestento{-37}$ & 210.3 & $2.4\timestento{-39}$ & 3 \\
 & 80 & $4.0\timestento{-80}$ & $1.0\timestento{-77}$ &  &  &  \\
\addlinespace
\multirow[t]{2}{=}{Harmonic oscillator, $n = 3$} & 40 & $3.2\timestento{-40}$ & $1.0\timestento{-37}$ & 142.9 & $3.2\timestento{-41}$ & 2 \\
 & 80 & $5.6\timestento{-80}$ & $1.0\timestento{-77}$ &  &  &  \\
\addlinespace
\multirow[t]{2}{=}{P\"oschl--Teller well, $n = 1$} & 40 & $2.8\timestento{-39}$ & $1.2\timestento{-37}$ & 247.0 & $2.1\timestento{-41}$ & 6 \\
 & 80 & $6.7\timestento{-79}$ & $1.2\timestento{-77}$ &  &  &  \\
\addlinespace
\multirow[t]{2}{=}{Quantum bouncer} & 40 & $3.1\timestento{-38}$ & $1.6\timestento{-37}$ & 171.3 & $8.6\timestento{-40}$ & 1 \\
 & 80 & $3.8\timestento{-78}$ & $1.4\timestento{-77}$ &  &  &  \\
\addlinespace
\multirow[t]{2}{=}{Circular well} & 40 & $8.8\timestento{-39}$ & $1.3\timestento{-37}$ & 174.0 & $0$ & 9 \\
 & 80 & $1.7\timestento{-78}$ & $1.3\timestento{-77}$ &  &  &  \\
\addlinespace
\multirow[t]{2}{=}{Oscillator with a wall} & 40 & $1.5\timestento{-37}$ & $1.6\timestento{-36}$ & 243.5 & $3.3\timestento{-30}$ & 1 \\
 & 80 & $3.8\timestento{-77}$ & $1.1\timestento{-76}$ &  &  &  \\
\addlinespace
\multirow[t]{2}{=}{Coulomb wave} & 40 & $6.2\timestento{-39}$ & $1.6\timestento{-37}$ & 215.7 & -- & 1 \\
 & 80 & $3.1\timestento{-78}$ & $1.3\timestento{-77}$ &  &  &  \\
\addlinespace
\multirow[t]{2}{=}{Fluorescence recovery after photobleaching} & 40 & $3.1\timestento{-38}$ & $1.3\timestento{-37}$ & 199.6 & $0$ & 1 \\
 & 80 & $8.9\timestento{-79}$ & $1.3\timestento{-77}$ &  &  &  \\
\bottomrule
\end{tabular}

\end{table}

\paragraph{States that are a polynomial times an exponential.} For four of the systems the learner returns a first-order
operator. This is correct and is the least order: the logarithmic derivative of such a state is a rational function of
the right variable, and the search stops at the first order that has a null direction. The Schr\"odinger equation, of
second order, holds as well and is not reported.

The second excited state of the Morse oscillator $-\psi'' + 36\,(\ee^{-2x} - 2\,\ee^{-x})\,\psi = E \psi$ contains the factor
$\exp(-6\,\ee^{-x})$, which satisfies no equation of the form~\eqref{eq:operator} in $x$; the Schr\"odinger equation itself
has coefficients that are not polynomials in $x$. The learner found the variable $y = \ee^{x}$ and returned, in 9.1\,s,
\begin{equation*}
  2 y^2 (y - 2)(y - 1)\,g' + (7 y^3 - 39 y^2 + 58 y - 24)\,g = 0 , \qquad
  \psi(x) = 124416 \sqrt{3}\;(\ee^{x} - 2)(\ee^{x} - 1)\;\ee^{-11 x / 2 - 6\,\ee^{-x}} ,
\end{equation*}
where the constant is that of the routine we supplied, $z^{7/2} \ee^{-z/2} L_2^{(7)}(z)$ with $z = 12\,\ee^{-x}$. For the
hydrogen $3s$ radial function it returned the polynomial, the decay and the normalisation,
$R_{30}(r) = \tfrac{2 \sqrt{3}}{243}\,(2 r^2 - 18 r + 27)\,\ee^{-r/3}$. The harmonic oscillator state $n = 3$ came back as
the pushforward under $y = x^2$ with the constant $\sqrt{3} / (3 \pi^{1/4})$, and the first excited state of the
P\"oschl--Teller well $-\psi'' - \tfrac{35}{4} \sech^2 x\,\psi = E \psi$ under $y = \sinh x$ as
$\sinh x / \cosh^{5/2} x$, a function with no polynomial operator in $x$.

\subsection{Energies as fitted coefficients}\label{sec:energies}

When the state is not of that kind, the operator of least order is the Schr\"odinger equation itself, and the energy
appears in it as a coefficient that has no closed form. The method then returns the equation with exact coefficients and
one numerical parameter.

A particle above a hard floor in a uniform field, the quantum bouncer \cite{gibbs1975bouncer}, obeys
$-\psi'' + x \psi = E \psi$. From values of its ground state on $[0.1,\,6]$ the learner returned
\begin{equation*}
  \psi'' + (a_1 - x)\,\psi = 0, \qquad \psi(x) = \Ai(x - a_1), \qquad a_1 = 2.33810741045976703848919725245 ,
\end{equation*}
in 22.9\,s, with an estimated error of $2.5 \timestento{-36}$ for $a_1$. The energy of the ground state is minus the first
zero of the Airy function, and the 30 digits recorded for $a_1$ agree with it in every digit.

For the third $s$ state of a particle in a circular well of unit radius, the model of an electron in a quantum corral
\cite{crommie1993corral}, the learner returned Bessel's equation and the energy,
\begin{equation*}
  r\,\psi'' + \psi' + a_1 r\,\psi = 0, \qquad \psi(r) = J_0\bigl(\sqrt{a_1}\,r\bigr), \qquad
  a_1 = 74.8870067906951834448890413101 ,
\end{equation*}
in 24.4\,s; $a_1$ is the square of the third zero of $J_0$, again in all 30 recorded digits.

The ground state of a harmonic oscillator with a hard wall at $x = -1/2$ is a parabolic cylinder function
$D_{\nu}(\sqrt{2}\,x)$ of non-integer index. The learner returned $\psi'' + (a_1 - x^2)\,\psi = 0$ with
$a_1 = 2.0607657916305730122\ldots$, which is the energy $E = 2 \nu + 1$, and through the gauge rule a closed form
$\ee^{-x^2/2}$ times a combination of two $\pFq{1}{1}$ series whose parameters contain $a_1$. The two constants of the
combination stayed numerical, the closed form agrees with the data to $3.3 \timestento{-30}$, and the run took 158.9\,s,
the longest of the nine.

For the continuum $s$ wave of hydrogen, the regular Coulomb function $F_0(\eta, \rho)$ with $\eta = -1/2$, the learner
returned $\rho\,F'' + (\rho + 1)\,F = 0$ with exact coefficients in 10.3\,s and no closed form. The standard form of the
solution is a confluent hypergeometric function with complex parameters, and the solver's rules work with real exponents
and rates only. The certificate still determines the function (Section~\ref{sec:continuation}).

\subsection{A recovery curve from cell biology}\label{sec:frap}

In fluorescence recovery after photobleaching, a spot of fluorescent molecules in a membrane is bleached by a laser pulse
and the fluorescence returns as unbleached molecules diffuse in; the recovery curve gives the diffusion coefficient
\cite{axelrod1976frap}. For a uniform circular spot the fraction recovered is commonly quoted in the closed form due to
Soumpasis \cite{soumpasis1983frap},
\begin{equation}\label{eq:frap}
  f(t) = \ee^{-2/t}\,\bigl[I_0(2/t) + I_1(2/t)\bigr] ,
\end{equation}
with $t$ in units of the characteristic diffusion time. From values of $f$ on $[0.2,\,6]$ the learner found the change of
variable $y = 1/t$, the operator and the closed form~\eqref{eq:frap} in 9.3\,s (Figures~\ref{fig:transport}e,\,f
and~\ref{fig:quantum}g):
\begin{equation*}
  y\,g'' + 2 (2 y + 1)\,g' + 2\,g = 0, \qquad\text{in $t$:}\quad t^3 f'' - 4 t f' + 2 f = 0 .
\end{equation*}

\subsection{A Meijer \texorpdfstring{$G$}{G} density from a float64 table}\label{sec:hard}

\begin{figure}[t]
\centering
\includegraphics[width=\textwidth]{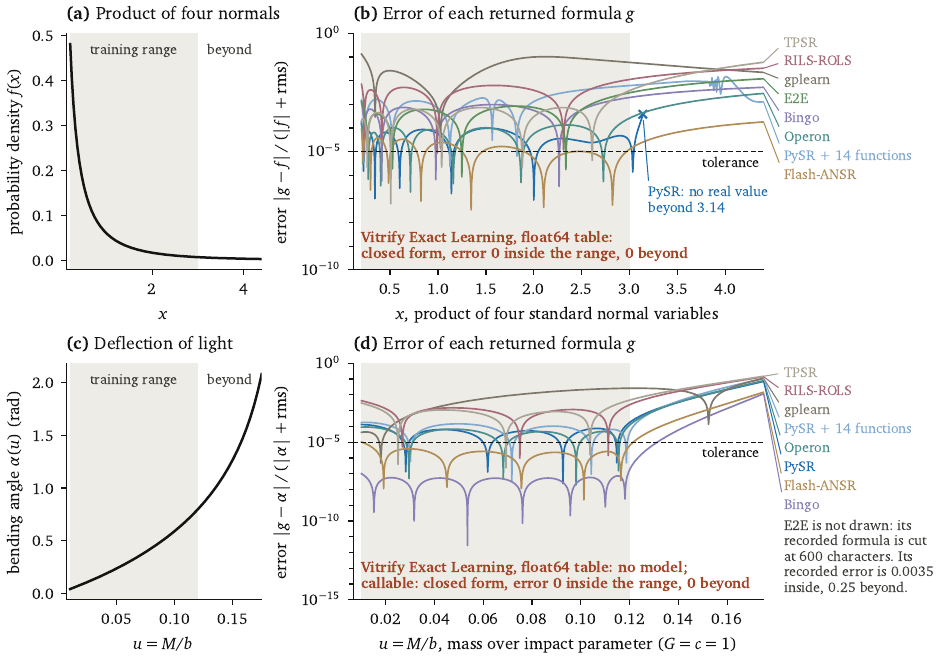}
\caption{Two targets from the physical set of Section~\ref{sec:physical}. (a)~The density of the product of four standard
normal variables, with the training range shaded. (b)~The error of the formula that each other program returned for 1101
float64 points; the dashed line is the tolerance $\tento{-5}$. From the same points \VEL{} returned the closed
form~\eqref{eq:fournormals}. PySR's curve stops at $x = 3.14$, where its formula has no real value. (c)~The bending angle
of light by a Schwarzschild mass against $u = M/b$. (d)~As (b). Here the float64 table gave \VEL{} no model, and the
callable gave the closed form~\eqref{eq:deflection}. E2E is not drawn because its recorded formula is cut at 600
characters; its recorded errors are 0.0035 inside and 0.25 beyond.}
\label{fig:hard_cases}
\end{figure}

The density of the product of $q$ independent standard normal variables is a Meijer $G$-function
\cite{springer1970products}. For $q = 4$, from 1101 float64 values on $[0.2,\,3]$, the learner returned
\begin{equation}\label{eq:fournormals}
  x^3 f'''' + 6 x^2 f''' + 7 x f'' + f' - x f = 0, \qquad
  f(x) = \frac{1}{4 \pi^2}\;\MeijerG{4}{0}{0}{4}{\frac{x^2}{16}\;\middle|\;\begin{matrix}\text{--}\\ 0, 0, 0, 0\end{matrix}} ,
\end{equation}
in 141.0\,s (10.7\,s from the callable), with recorded errors of 0 inside the training range and 0 beyond it
(Figure~\ref{fig:hard_cases}a,\,b). In $\theta$-form the equation is $\theta^4 f = x^2 f$. Its four local exponents at the
origin coincide, so three of its four solutions contain powers of $\log x$ and no hypergeometric series reaches the
density; this is the case for which the Meijer $G$ rule of Section~\ref{sec:solver} was written. Of the eight other
programs, and PySR with 14 special functions added to its operations (Section~\ref{sec:protocol}), the closest is
Flash-ANSR, with errors of $8.5 \timestento{-5}$ inside and $1.8 \timestento{-4}$ beyond.

\subsection{The deflection of light by a black hole}\label{sec:deflection}

A light ray that passes a Schwarzschild mass $M$ at impact parameter $b$ is bent by an angle $\alpha$ that is an elliptic
integral over the roots of a cubic; Darwin analysed the orbits in this field \cite{darwin1959}. We computed $\alpha$ by
quadrature as a function of $u = M/b$ (with $G = c = 1$) on $[0.01,\,0.12]$. From the callable the learner returned, in 10.6\,s, a fourth-order equation and its
solution as a sum of two hypergeometric functions:
\begin{gather}
  u\,(27 u^2 - 1)\,\alpha'''' + 3\,(9 u - 1)(9 u + 1)\,\alpha''' + 501\,u\,\alpha'' + 192\,\alpha' = 0 , \notag\\[2pt]
  \alpha(u) = 4 u\;\pFq{3}{2}\!\left(\tfrac23, 1, \tfrac43;\ \tfrac32, \tfrac32;\ 27 u^2\right)
    + \pi \left[\pFq{2}{1}\!\left(\tfrac16, \tfrac56;\ 1;\ 27 u^2\right) - 1\right] . \label{eq:deflection}
\end{gather}
The series of~\eqref{eq:deflection} begins $4 u + \tfrac{15 \pi}{4} u^2 + \tfrac{128}{3} u^3 + \tfrac{3465 \pi}{64} u^4$;
its first term is Einstein's value $4 M / b$. The singular point $27 u^2 = 1$ of the equation is the critical impact
parameter $b = 3 \sqrt{3}\,M$ below which the light is captured. The returned closed form has recorded errors of 0 inside
the range and 0 beyond it, up to $u = 0.175$.

From 1101 float64 values of the same function the learner returned no model, after 912.3\,s: sixteen digits do not
determine an operator of order 4 and degree 3 on this interval. The closest other program is Bingo, whose formula is
within $1.5 \timestento{-7}$ inside the range and off by $1.2 \timestento{-2}$ beyond it
(Figure~\ref{fig:hard_cases}c,\,d).

\subsection{Elliptic integrals in two of their arguments}\label{sec:pi}

The complete elliptic integral of the third kind,
\begin{equation*}
  \Pi(n \mid m) = \int_0^{\pi/2} \frac{\dd\theta}{(1 - n \sin^2\theta)\,\sqrt{1 - m \sin^2\theta}} ,
\end{equation*}
depends on two parameters. Given as a callable on the box $[0.05,\,0.55]^2$, it gave a system from which the reduction of
Section~\ref{sec:bivariate} kept three operators, among them
\begin{align*}
  2 (m - n)\,\partial_n \partial_m f + \partial_n f - 2\,\partial_m f &= 0 , \\
  4 n (n - 1)\,\partial_n^2 f + 2 (4 n - 1)\,\partial_n f - 4 m (m - 1)\,\partial_m^2 f - 4 m\,\partial_m f + f &= 0 ,
\end{align*}
each with exact coefficients and a residual on the held-out quadrant at its noise level, near $\tento{-37}$
(Appendix~\ref{app:bivariate}). Two further operators found in the same measurements have the two-term form that
fixes the ratios~\eqref{eq:horn}, and the series they determine is Appell's first function:
\begin{equation}\label{eq:pi}
  \Pi(n \mid m) = \frac{\pi}{2}\;F_1\!\left(\tfrac12;\ \tfrac12, 1;\ 1;\ m, n\right) .
\end{equation}
The run took 123.8\,s, and the recorded errors of~\eqref{eq:pi} are 0 inside the box and 0 beyond it.

Carlson's symmetric integral $R_F(x, y, 1) = \tfrac12 \int_0^\infty \bigl[(t + x)(t + y)(t + 1)\bigr]^{-1/2}\,\dd t$
\cite{carlson1977special,carlson1995numerical} gives the capacitance $C$ of a conducting ellipsoid with semi-axes $a$,
$b$ and $c$ through $R_F(a^2, b^2, c^2) = 4 \pi \varepsilon_0 / C$. On the box $[0.3,\,0.8] \times [0.35,\,0.9]$ the search in $(x, y)$ finds
operators and no series. After the change of variable $u = 1 - x$, $v = 1 - y$ it returns, in 221.1\,s,
\begin{equation}\label{eq:rf}
  R_F(x, y, 1) = F_1\!\left(\tfrac12;\ \tfrac12, \tfrac12;\ \tfrac32;\ 1 - y,\ 1 - x\right) ,
\end{equation}
which is Lauricella's function $F_D$ of two variables \cite{lauricella1893}. One of the operators kept is
$2 (u - v)\,\partial_u \partial_v f - \partial_u f + \partial_v f = 0$.

\begin{figure}[t]
\centering
\includegraphics[width=\textwidth]{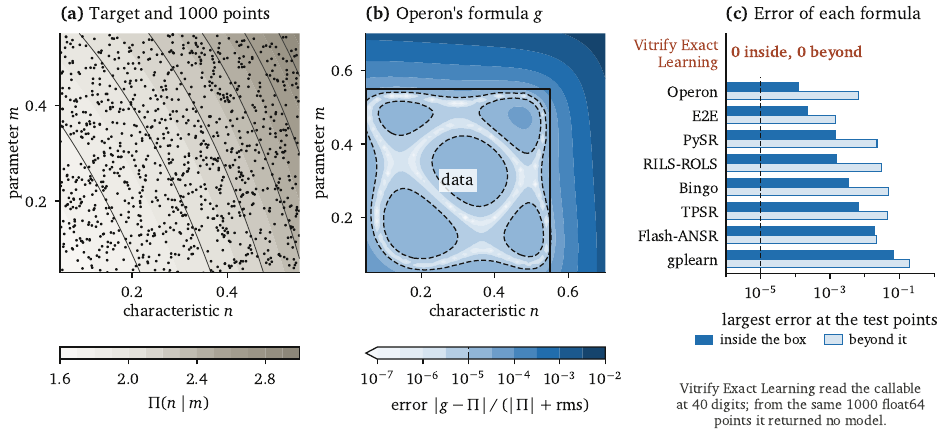}
\caption{The complete elliptic integral $\Pi(n \mid m)$. (a)~Contours of the target on the box $[0.05,\,0.55]^2$ with the
1000 float64 points given to the other programs. (b)~The error of Operon's formula, the closest inside the box, on a wider
region; the box is outlined. (c)~The largest error of each program's formula inside the box and beyond it. \VEL{}
returned~\eqref{eq:pi} from the callable at 40 digits; from the same 1000 float64 points it returned no model.}
\label{fig:elliptic_pi}
\end{figure}

The eight symbolic-regression programs, and PySR with 14 special functions, were given 1000 float64 points of $\Pi$ drawn
from the box (Figure~\ref{fig:elliptic_pi}). The closest inside the box is Operon, with an error of $1.2 \timestento{-4}$ there and
$6.4 \timestento{-3}$ beyond; E2E has $2.3 \timestento{-4}$ and $1.5 \timestento{-3}$. None reaches the tolerance. This
comparison is not like for like, and we return to it in Section~\ref{sec:two}: from the same 1000 float64 points the
learner itself returns no model, and what the example shows is that values to 40 digits suffice to identify the function.

\section{Benchmarks}\label{sec:benchmarks}

This section compares \VEL{} with eight open symbolic-regression programs on the same float64 tables and with the published
results of SRBench. Every number is a count over recorded runs: a program was run on generated data and the function it
returned was compared with the exact target. Figure~\ref{fig:benchmarks} gives the overview.

\subsection{Protocol}\label{sec:protocol}

\paragraph{Scoring.} A program receives training points and returns one expression $g$. Its error against the target $f$ is
\begin{equation}\label{eq:error}
  \max_x\ \frac{|g(x) - f(x)|}{|f(x)| + \rms} ,
\end{equation}
where rms is the root mean square of the training values. The maximum is taken twice: over 400 random points inside the
training range, and over points beyond it. For the published suites the points beyond are those of 600 random points, drawn
up to half a range-width beyond the upper ends, that have a coordinate above its range: about 200 in one variable (164 to
234) and 320 to 590 in several, and fewer where the target has a singular point there. The generated, physical and
two-variable sets use fixed grids of 200 points, in one variable up to half a range-width beyond the upper end and in two
variables up to 30\% of a box width beyond both upper ends. A target is \emph{recovered} when both errors are at most
$\tento{-5}$, or at most the noise level when noise was added and is larger. We also count the recoveries that hold to $\tento{-12}$. The rule does not compare
expression trees. An approximant that fits the training range fails beyond it, and a correct form with fitted constants
passes. Because the rule rewards extrapolation it favours exact forms, so we also report how many expressions match inside
the range only.

\paragraph{Programs.} Table~\ref{tab:methods} lists what was run; Appendix~\ref{app:software} gives the sources. Each
program ran once per target with one seed. The five programs with a time setting ran for 60\,s on one thread.

\begin{table}[t]
\centering\small
\caption{The programs that were run. All received the same float64 training points.}
\label{tab:methods}
\begin{tabular}{@{}L{27mm}L{128mm}@{}}
\toprule
Program & What ran \\
\midrule
\VEL{} & One frozen version of the learner. On a float64 table, the entry point of Section~\ref{sec:relations}, which ends
in the operator search for one variable. On a callable, the operator search at 40 digits, or the two-variable learner,
which was written after that version was frozen. It has no time setting. A run on a float64 table is stopped after
900\,s (600\,s on SRBench), and a run on a callable after 900\,s on the generated set, 3600\,s on the physical set and
3000\,s in two variables. On the tables of the physical set the entry point for tables (900\,s) is followed by the
operator search on the table (3600\,s). \\
\addlinespace[2pt]
PySR \cite{cranmer2023pysr} & Version 2.7.2 with its default search, maximum size 30, the operations $+$, $-$, $\times$,
$\div$, sin, cos, exp, log and sqrt, in float64, 60\,s. \qq{PySR with SRBench's limits} adds the limits that SRBench puts on
the search (no nested trigonometric or exponential functions, among others). \qq{PySR + 14 functions} adds erf, $J_0$,
$J_1$, $Y_0$, $I_0$, $K_0$, Ai, Bi, $K$, $E$, Ei, Si, Ci and $\Gamma$ as unary operations. \\
\addlinespace[2pt]
Operon \cite{burlacu2020operon} & pyoperon 0.6.2 with SRBench's settings, 60\,s. It evaluates in float32. \\
Bingo \cite{randall2022bingo} & bingo-nasa 0.5.7 with SRBench's settings, 60\,s. \\
gplearn \cite{stephens2026gplearn} & Version 0.4.3 with SRBench's settings and exp added, 60\,s. It does not optimise
constants. \\
RILS-ROLS \cite{kartelj2023rilsrols} & Version 1.6.7 with SRBench's settings, 60\,s. \\
Flash-ANSR \cite{saegert2026flashansr} & Version 0.19.1 with its pretrained model: 1024 sampled expressions with refined
constants. No time setting. \\
E2E \cite{kamienny2022e2e} & The authors' code and checkpoint with SRBench's settings: bags of 200 points and 100 candidates
refined by BFGS. No time setting. \\
TPSR \cite{shojaee2023tpsr} & The authors' code through the wrapper of SRBench 2025. No time setting; a run is stopped
at 600\,s where the limit could interrupt it and then counts as no expression. 23 runs returned an expression after 608
to 2545\,s, none of them a recovery. \\
\bottomrule
\end{tabular}
\end{table}

\paragraph{Published suites.} Five suites come from the literature.
\begin{itemize}[leftmargin=*,itemsep=1pt,topsep=3pt]
\item \emph{Classic sets at their published points}: the one-variable functions of the benchmark collection distributed
with deep symbolic regression \cite{petersen2021dsr}, which gathers the Nguyen, Keijzer, Koza, Livermore, R and other sets
\cite{uy2011nguyen,keijzer2003}; 53 distinct targets, 40 of them with 20 training points and the rest with 21 to 1000.
\item \emph{Classic sets at 1101 points}: the same functions at 1101 evenly spaced points (51 targets).
\item \emph{Feynman equations, one variable at a time}: each of the 120 AI Feynman equations \cite{udrescu2020aifeynman}
as a function of one of its variables at 1101 points, with the others fixed inside their ranges. The 468 slices have 254
distinct shapes, and slices that differ by a constant factor are run once.
\item \emph{Feynman equations in all variables}: the 120 equations in their 1 to 9 variables at 1000 uniform random points.
\item \emph{Special functions, pilot set}: 31 Bessel, Airy, elliptic, hypergeometric and related functions at 1101 points,
written for a first comparison.
\end{itemize}
These suites were in view while the method was developed: the entry point of Section~\ref{sec:relations} was written with
the classic and Feynman targets at hand. They are a development set, and Section~\ref{sec:generated} describes the test
that was made after the code was frozen.

\subsection{Published suites without noise}\label{sec:suites}

\begin{table}[t]
\centering\small
\caption{Targets recovered from noise-free float64 tables; in brackets, the number of those that hold to $\tento{-12}$.}
\label{tab:suites}
\begin{tabular}{@{}lccccc@{}}
\toprule
 & \multicolumn{2}{c}{Classic sets} & \multicolumn{2}{c}{Feynman equations} & Special functions \\
\cmidrule(lr){2-3}\cmidrule(lr){4-5}\cmidrule(l){6-6}
 & published & 1101 & one variable & all & \\
 & points & points & at a time & variables & pilot set \\
Method & of 53 & of 51 & of 254 & of 120 & of 31 \\
\midrule
\VEL{} & 45 (45) & 47 (47) & 251 (244) & 108 (108) & 28 (28) \\
PySR & 33 (27) & 32 (23) & 174 (106) & \phantom{0}64 (61)\phantom{0} & \phantom{0}0\phantom{ (00)} \\
PySR with SRBench's limits & 30 (21) & 26 (20) & 164 (96)\phantom{0} & \phantom{0}60 (54)\phantom{0} & \phantom{0}0\phantom{ (00)} \\
Bingo & 26 (23) & 15 (14) & 184 (141) & \phantom{0}48 (39)\phantom{0} & \phantom{0}0\phantom{ (00)} \\
Flash-ANSR & 19 (18) & 21 (20) & 168 (59)\phantom{0} & \phantom{0}69 (64)\phantom{0} & \phantom{0}0\phantom{ (00)} \\
RILS-ROLS & 25 (25) & 19 (18) & 103 (101) & \phantom{0}67 (67)\phantom{0} & \phantom{0}0\phantom{ (00)} \\
Operon & 13 (0)\phantom{0} & \phantom{0}5 (0)\phantom{0} & 104 (0)\phantom{00} & \phantom{0}54 (0)\phantom{00} & \phantom{0}0\phantom{ (00)} \\
E2E & \phantom{0}9 (1)\phantom{0} & \phantom{0}7 (0)\phantom{0} & \phantom{0}92 (13)\phantom{0} & \phantom{0}43 (1)\phantom{00} & \phantom{0}0\phantom{ (00)} \\
TPSR & \phantom{0}8 (0)\phantom{0} & \phantom{0}8 (0)\phantom{0} & \phantom{0}67 (3)\phantom{00} & \phantom{0}11 (0)\phantom{00} & \phantom{0}0\phantom{ (00)} \\
gplearn & \phantom{0}3 (3)\phantom{0} & \phantom{0}4 (4)\phantom{0} & \phantom{00}0\phantom{ (000)} & \phantom{0}14 (14)\phantom{0} & \phantom{0}0\phantom{ (00)} \\
\bottomrule
\end{tabular}

\end{table}

Table~\ref{tab:suites} gives the counts. \VEL{} recovers 45 of the 53 classic targets at their published points and 47 of 51
at 1101 points, 251 of the 254 Feynman slices and 108 of the 120 Feynman equations in all variables. The best other program
on each recovers 33 (PySR), 32 (PySR), 184 (Bingo) and 69 (Flash-ANSR). On the pilot set of special functions it recovers
28 of 31, and no program of Table~\ref{tab:suites} recovers any. That table has no row for PySR with special functions.
In an earlier run on the pilot set, PySR with erf, $J_0$, $J_1$, Ai, Bi and $K$ added to its operations recovered 8 of the
31: seven compositions of those six functions, all of which \VEL{} also recovers, and the digamma function, by a fitted
approximant with errors of $8 \timestento{-6}$ inside the range and $7 \timestento{-6}$ beyond it.

Some targets are recovered by \VEL{} and by none of the nine other rows of Table~\ref{tab:suites}: 6 classic targets (11
at 1101 points), 35 slices, 28 Feynman equations in all variables and the 28 special functions (21 when the run of PySR
with six special functions is counted). At the published points the six are Nguyen-7, returned as
$\log(x^3 + x^2 + x + 1)$, Constant-8, R1, Livermore-22, returned as $x (x + 1)(x^2 + 1)(x^4 + 1)$ from 20 points, and the
polynomials of degree 9, Livermore-9 and Nonic, which came back as conjectures with one spurious coefficient near
$\tento{-16}$. On the slices they are Gaussians, square roots of quadratics, rational functions, and trigonometric and
hyperbolic forms such as $\tanh$, $1/\cosh$ and $\sin^2(a x - b)/(x - c)^2$, with awkward constants.

In the other direction, eight targets were recovered by another program of Table~\ref{tab:suites} and missed by \VEL{}:
R2a, Neat-6 and Poly-5 (PySR and Bingo) and R3a (Flash-ANSR) among the classic targets at their published points, and the
Feynman equations I.50.26 and test\_17 (PySR and RILS-ROLS), II.35.21 (Flash-ANSR) and test\_19 (Operon). At 1101 points
Neat-6 and Poly-5 were missed again and recovered by PySR or Flash-ANSR. In all, \VEL{} missed 8 classic targets, 4 at 1101
points, 3 slices, 12 Feynman equations in all variables (three of them stopped at 900\,s) and 3 special functions. Two of
the three, the Kelvin function ber and the trilogarithm $\mathrm{Li}_3$, have operators of order 4 that the float64 tables
did not determine; from callables the learner returns both in closed form, ber on $[0.5,\,5]$ in 23.5\,s and
$\mathrm{Li}_3$ on $[0.05,\,0.9]$ in 62.4\,s. The third, the digamma function, is not holonomic.

Summed over its 569 runs on these suites, including those with noise, \VEL{} took 8.9 hours with a median of 8.8\,s per run.
The 60\,s programs took 5.8 to 9.7 hours, Flash-ANSR 26.9, E2E 37.7 and TPSR 62.4.

\subsection{What is returned is exact or absent}\label{sec:absent}

\begin{table}[t]
\centering\small
\caption{The expressions returned for the 389 noise-free one-variable tables (the two classic suites, the Feynman slices
and the pilot set). An expression \qq{matches inside the range only} when its error is at most $\tento{-5}$ inside the
training range and larger beyond it.}
\label{tab:one_variable}
\begin{tabular}{@{}lcccccc@{}}
\toprule
 & & & \multicolumn{4}{c}{Of the expressions returned} \\
\cmidrule(l){4-7}
 & & Expressions & & of these & match inside & \\
Method & Tables & returned & recovered & to $\tento{-12}$ & the range only & neither \\
\midrule
\VEL{} & 389 & 371 & 371 & 364 & \phantom{0}0 & \phantom{00}0 \\
PySR & 389 & 389 & 239 & 156 & 42 & 108 \\
PySR with SRBench's limits & 389 & 389 & 220 & 137 & 47 & 122 \\
Bingo & 389 & 389 & 225 & 178 & 46 & 118 \\
Flash-ANSR & 389 & 389 & 208 & \phantom{0}97 & 42 & 139 \\
RILS-ROLS & 389 & 389 & 147 & 144 & \phantom{0}1 & 241 \\
Operon & 389 & 389 & 122 & \phantom{00}0 & 27 & 240 \\
E2E & 389 & 389 & 108 & \phantom{0}14 & 25 & 256 \\
TPSR & 389 & 341 & \phantom{0}83 & \phantom{00}3 & \phantom{0}9 & 249 \\
gplearn & 389 & 389 & \phantom{00}7 & \phantom{00}7 & \phantom{0}0 & 382 \\
\bottomrule
\end{tabular}

\end{table}

\begin{figure}[t]
\centering
\includegraphics[width=\textwidth]{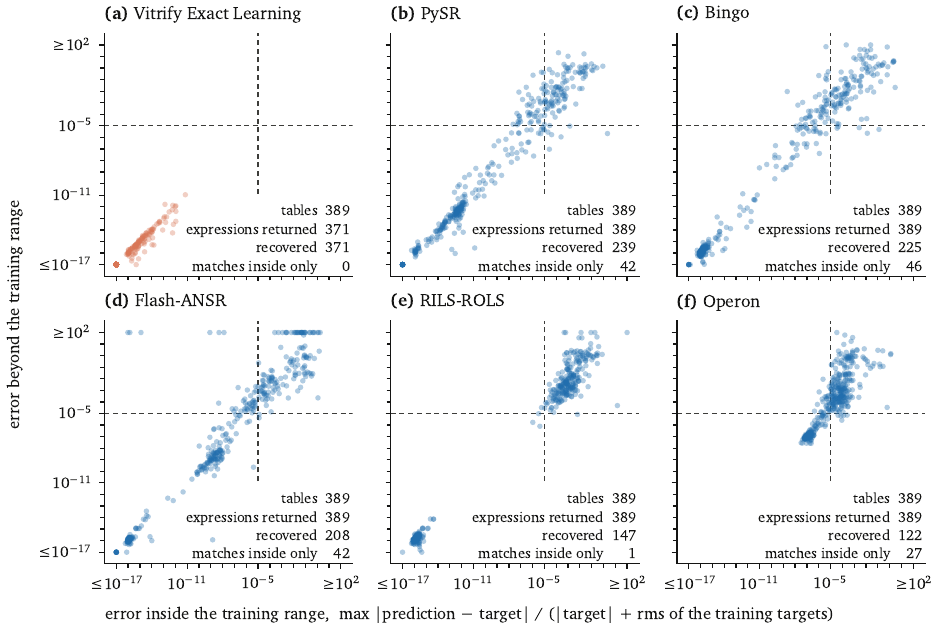}
\caption{Error beyond the training range against error inside it for every expression returned on the 389 noise-free
one-variable tables, for six programs. The dashed lines are the tolerance $\tento{-5}$. Points in the lower left corner
are recovered; points in the upper left quadrant fit the data and fail beyond them.}
\label{fig:exact_or_absent}
\end{figure}

A search program always returns something: its best formula. Table~\ref{tab:one_variable} and
Figure~\ref{fig:exact_or_absent} show what that means on the 389 noise-free tables in one variable. PySR returned 389
expressions, of which 239 are recovered and 42 match inside the training range and fail beyond it. For Bingo the two
numbers are 225 and 46, for Flash-ANSR 208 and 42. Scored inside the range only, PySR would be credited with 281.

\VEL{} returned 371 expressions and no model for the other 18. All 371 are recovered, 364 of them to $\tento{-12}$, and for
250 both recorded errors are exactly zero. On these suites a returned expression was always right on the tested range,
which ends half a range-width beyond the data.

\paragraph{Further from the data.} To see how far that holds, we evaluated the 371 expressions again, without running the
learner, on three bands that the benchmark does not test: from 0.6 to 3 range-widths beyond the upper end, from 4 to 30
range-widths beyond it, and from 0.6 to 3 range-widths below the lower end. Of the 371, 251 are graded \code{verified} and
contain no fitted constant, and 120 are conjectures with at least one float64 constant. Counting the expressions whose
target is real and finite at some point of a band, the error~\eqref{eq:error} stays within $\tento{-9}$ for 363 of 364 in
the first band, 352 of 354 in the second and 327 of 330 in the third. The five expressions that fail are of two kinds.
\begin{itemize}[leftmargin=*,itemsep=1pt,topsep=3pt]
\item \emph{A lost sign in a verified expression.} For the slice
$\bigl(\tfrac{20}{3} - \tfrac{13 x}{9}\bigr) / \sqrt{1 - 49 x^2 / 5184}$ of the Feynman equation I.15.3x on $[1,\,2]$, the
relation search found an exact rational function for $f^2$ and returned its positive square root,
$\sqrt{(-10816 x^2 + 99840 x - 230400) / (49 x^2 - 5184)}$, graded \code{verified} with both recorded errors 0. That is
the absolute value of the target, which changes sign at $x = 60/13$; beyond that point the expression is wrong by up to
1.8. A slice of I.15.1 on $[1,\,2]$, $7 x / \bigl(3 \sqrt{1 - x^2/36}\bigr)$, was returned in the same way as
$14 \sqrt{-x^2 / (x^2 - 36)}$, which has the wrong sign for $x < 0$.
\item \emph{Conjectures with fitted constants.} Two slices of the equation test\_2 were returned as a root of a quadratic
with fitted coefficients, and are wrong by 0.70 and 0.35 below the lower end of the data. A slice of III.9.52,
$552123\,\pi \sin^2(2x/143) / (34 x)$ on $[1,\,3]$, was returned as $54 \pi x\,(1 - 1.304 \timestento{-5}\,x^2)^{4.99976} / 17$,
which follows the target to $\tento{-9}$ up to three range-widths beyond the data and is off by up to
$1.1 \timestento{-4}$ between four and thirty.
\end{itemize}
The two expressions of the first kind are exact on the side of the sign change where the data lie, and nothing in the data
says what happens on the other side. They show what the grade does not cover: \code{verified} is a statement about the
interval of the data.

The property is weaker again on harder inputs, and three further kinds of exception occur in the runs of this paper. On the
generated set of Section~\ref{sec:generated} one \code{conjecture} from a float64 table is wrong beyond the range: the
target B35 of Table~\ref{tab:generated}, from its table of 1101 points (errors of $9.0 \timestento{-12}$ inside and 0.029
beyond) and again from 50 points ($8.8 \timestento{-11}$ and 0.31). On the physical set of Section~\ref{sec:physical} two
float64 tables received a wrong certificate, one of them graded \code{verified}. And under noise most results graded
\code{fit} fail the test (Section~\ref{sec:noise}).

\subsection{SRBench}\label{sec:srbench}

SRBench \cite{lacava2021srbench} publishes results for many methods under one protocol on 130 problems with known
answers: 116 Feynman equations and 14 Strogatz systems \cite{lacava2016strogatz}, as data sets of the Penn Machine Learning
Benchmarks \cite{romano2022pmlb}. We ran \VEL{} under that protocol: a 75/25 split of the rows with SRBench's first three
seeds, Gaussian noise of 0, 0.001, 0.01 and 0.1 times the rms of the target added to the training targets, and SRBench's own
code for cleaning and scoring the returned expression. A run counts as a \emph{symbolic solution} when, after SRBench
rounds the constants to three decimals and simplifies, the expression differs from the true model by a constant or by a
constant factor. The learner read the first 1000 training rows and was stopped at a limit of 600\,s. The limit took effect
late in some runs: 57 of the 69 stopped runs lasted more than 600.5\,s, the longest 1182\,s, and no expression was returned
after 600\,s. The learner has no random part, so a seed changes only the split; we ran three seeds without noise and one
seed at each noise level, 780 runs. SRBench's protocol gives each method ten trials per problem and a budget of $\tento{6}$
evaluations or 8 hours per run \cite{lacava2021srbench}, and the published results hold eight to ten runs per problem and
noise level.

\begin{table}[t]
\centering\small
\caption{SRBench, problems with known answers. Symbolic solution rate at four noise levels, and the share of runs with
$R^2 > 0.999$ on the test rows without noise. The row of \VEL{} is from our runs (three seeds without noise and one seed
at each noise level, a limit of 600\,s). All other rows are the results that SRBench publishes (eight to ten runs per
problem, under a budget of $\tento{6}$ evaluations or 8 hours); the Operon and gplearn rows are those published results
and not the runs of Table~\ref{tab:suites}.}
\label{tab:srbench}
\begin{tabular}{@{}lccccccccc@{}}
\toprule
 & \multicolumn{8}{c}{Symbolic solution rate (\%) at noise level} & $R^2 > 0.999$ (\%) \\
\cmidrule(lr){2-9}\cmidrule(l){10-10}
 & \multicolumn{4}{c}{Feynman} & \multicolumn{4}{c}{Strogatz} & Feynman, \\
\cmidrule(lr){2-5}\cmidrule(lr){6-9}
Method & 0 & 0.001 & 0.01 & 0.1 & 0 & 0.001 & 0.01 & 0.1 & no noise \\
\midrule
\VEL{} & 74.1 & \phantom{0}0.9 & \phantom{0}0.9 & \phantom{0}0.0 & 100.0 & \phantom{0}0.0 & \phantom{0}0.0 & \phantom{0}0.0 & 90.8 \\
AIFeynman & 55.8 & 33.1 & 13.0 & \phantom{0}0.7 & \phantom{0}27.1 & 22.1 & \phantom{0}9.3 & \phantom{0}2.1 & 77.4 \\
AFP\_FE & 27.0 & 21.9 & 20.8 & 13.5 & \phantom{0}20.0 & 15.7 & 13.6 & \phantom{0}2.1 & 57.2 \\
GP-GOMEA & 26.8 & 11.0 & \phantom{0}5.1 & \phantom{0}1.6 & \phantom{0}29.5 & \phantom{0}7.1 & \phantom{0}1.4 & \phantom{0}0.0 & 70.9 \\
ITEA & 22.4 & 14.6 & \phantom{0}7.8 & \phantom{0}1.5 & \phantom{00}7.1 & \phantom{0}7.1 & \phantom{0}6.4 & \phantom{0}1.4 & 27.6 \\
AFP & 21.1 & 19.7 & 16.9 & 13.1 & \phantom{0}15.2 & 13.6 & 11.4 & \phantom{0}4.3 & 41.3 \\
DSR & 19.7 & 19.1 & 19.0 & 16.8 & \phantom{0}19.6 & 19.3 & 18.6 & 15.0 & 22.6 \\
Operon & 16.6 & 13.2 & \phantom{0}2.1 & \phantom{0}0.1 & \phantom{0}11.4 & \phantom{0}5.0 & \phantom{0}0.7 & \phantom{0}0.0 & 85.2 \\
gplearn & 16.3 & 17.0 & 16.9 & 16.6 & \phantom{00}8.9 & \phantom{0}9.3 & \phantom{0}9.3 & \phantom{0}9.3 & 30.7 \\
SBP-GP & 12.7 & \phantom{0}0.8 & \phantom{0}0.0 & \phantom{0}0.0 & \phantom{0}11.6 & \phantom{0}0.0 & \phantom{0}0.0 & \phantom{0}0.0 & 74.2 \\
EPLEX & 12.4 & \phantom{0}9.6 & \phantom{0}8.7 & 10.2 & \phantom{00}8.9 & \phantom{0}9.3 & \phantom{0}4.3 & \phantom{0}2.1 & 45.2 \\
BSR & \phantom{0}2.5 & \phantom{0}0.6 & \phantom{0}0.1 & \phantom{0}0.0 & \phantom{00}0.9 & \phantom{0}0.7 & \phantom{0}0.0 & \phantom{0}0.0 & 11.1 \\
MRGP & \phantom{0}0.0 & \phantom{0}0.0 & \phantom{0}0.0 & \phantom{0}0.0 & \phantom{00}0.0 & \phantom{0}0.0 & \phantom{0}0.0 & \phantom{0}0.0 & 89.9 \\
FEAT & \phantom{0}0.0 & \phantom{0}0.0 & \phantom{0}0.0 & \phantom{0}0.0 & \phantom{00}0.9 & \phantom{0}0.0 & \phantom{0}0.0 & \phantom{0}0.0 & 40.2 \\
FFX & \phantom{0}0.0 & \phantom{0}0.0 & \phantom{0}0.0 & \phantom{0}0.1 & \phantom{00}0.0 & \phantom{0}0.0 & \phantom{0}0.0 & \phantom{0}0.0 & \phantom{0}0.0 \\
\bottomrule
\end{tabular}

\end{table}

Without noise \VEL{} solves 74.1\% of the Feynman problems and all 14 Strogatz problems by SRBench's symbolic test
(Table~\ref{tab:srbench}, Figure~\ref{fig:srbench}). The best published rates are 55.8\% for AI~Feynman on the Feynman
problems and 29.5\% for GP-GOMEA on the Strogatz problems. Of the 116 Feynman problems, 84 are solved in all three runs, 3
in some and 29 in none. The symbolic test undercounts here: on 20 of the Feynman problems the returned expression agrees
with the test rows to $\tento{-9}$ and the symbolic test does not credit it. In 16 of the 20 the returned expression
contains a square root, which the comparison does not reduce to the form of the true model; in the other four it contains
an exact constant such as $3/(4\pi)$, which SRBench's cleaning rounds to three decimals, or is written in an expanded
form. By agreement to $\tento{-9}$ on the test rows the rate is 90.8\%. We used the scorer unchanged and report its
number.

With noise the picture reverses. At noise levels 0.001, 0.01 and 0.1 the Feynman rates of \VEL{} are 0.9\%, 0.9\% and 0.0\%,
and its Strogatz rates are 0. AI~Feynman keeps 33.1\%, 13.0\% and 0.7\%, and methods such as DSR and gplearn are almost
unaffected at their lower level. The cause is plain: all but one of these problems have several variables, and the
method handles noise for one variable only (Section~\ref{sec:relations}). The one problem in one variable, I.6.2a, is the
only one credited under noise, at the levels 0.001 and 0.01. Of the 390 runs with noise, 325 returned no model, 62 were
stopped at the limit and 3, all on that problem, returned a \code{fit}.

The 2025 edition of SRBench \cite{aldeia2025srbench} adds 13 phenomenological and first-principles data sets and 12
black-box data sets, with 4 to 10{,}000 training rows of observations. On these 25 data sets \VEL{} returned no expression in any of
75 runs. Over all 855 runs the statuses are 355 \code{verified}, 3 \code{conjecture}, 3 \code{fit}, 425 \code{no-model}
and 69 stopped at the limit. Every one of the 355 verified expressions agrees with the test rows to better than
$\tento{-13}$.

\subsection{A generated set of special functions}\label{sec:generated}

The suites above were in view during development, so they cannot show how the method does on functions that it was not
written against. For that we wrote a generator of targets and froze the learner before drawing from it.

\paragraph{Targets.} The generator holds 59 function families: Bessel, Airy, Struve, Kelvin and Coulomb functions, complete
and incomplete elliptic integrals, confluent and Gauss hypergeometric functions, Whittaker, parabolic cylinder and Legendre
functions, incomplete gamma and beta functions, the exponential, sine, cosine and Fresnel integrals, the dilogarithm and the
trilogarithm, $\pFq{p}{q}$ up to $\pFq{3}{2}$ and $\pFq{0}{3}$, orthogonal polynomials of non-integer degree, and elementary
factors. A recipe draws one to three members with rational parameters, each composed with one of the maps $x$, $2x$, $x/2$,
$x^2$, $1/x$ and $-x$, and joins them by one of nine shapes. Among the 120 generated targets the shapes are a member plus a
constant (20), a single member (17), a product plus a constant (15), a member times an elementary factor (14), a square
(14), a product (12), a derivative (10), a product plus a third member (9) and a sum (9). For about one recipe in five
outside the elementary tier the result is composed with $\sqrt{x}$ (9 targets), $\log x$ (11) or $\ee^{x}$ (3). The
operator that annihilates each target is computed exactly from the operators of its members by the closure rules of
Section~\ref{sec:holonomic}, and a target is kept only when that operator applied to the function at 40 digits vanishes.

The targets fall into five tiers by the order of that operator (Table~\ref{tab:generated_tiers}): 12 built from elementary
factors only (E), 36 of order 1 or 2 (A), 36 of order 3 or 4 (B), 36 of order 5 to 8 (C), which is beyond the search limits
of the learner, and 12 fixed functions that have no such operator in $x$ (N): $\Gamma(x)$, $1/\Gamma(x)$, $\log \Gamma(x)$,
the digamma function, $\zeta(x)$, Lambert's $W$, the inverse error function, $x^x$, $1/(\ee^x - 1)$, $\tan x$,
$\ee^{\sin x}$ and $\ee^{\ee^{x}/2}$. In each of the tiers A, B and C, 12 targets are compositions of the functions that
PySR + 14 functions is given and 24 need at least one other function. The 120 generated targets use 53 of the 59 families.

The seed of the stream is a string fixed by the frozen version of the learner, so the targets did not exist while the
learner was written. Five other seeds were used during development, and their results are not reported. Every program
received 1101 float64 points of each target on an interval without singular points, and \VEL{} also ran on each target as a
callable. Table~\ref{tab:generated} in Section~\ref{app:generated} lists every target with its formula, its interval and what each program
returned.

\begin{table}[t]
\centering\small
\caption{The generated set: targets recovered in each tier. All programs received the same 1101 float64 points; the two
callable columns of \VEL{} read the target at 40 digits. The column \qq{callable, with continuation} adds the targets for
which the solver found no closed form and the continued solution of the verified operator passes the test.}
\label{tab:generated_tiers}
\begin{tabular}{@{}l@{\hspace{0.8em}}c@{\hspace{0.8em}}c@{\hspace{0.8em}}c@{\hspace{0.8em}}c@{\hspace{0.8em}}c@{\hspace{0.8em}}c@{\hspace{0.8em}}c@{\hspace{0.8em}}c@{\hspace{0.8em}}c@{\hspace{0.8em}}c@{\hspace{0.8em}}c@{\hspace{0.8em}}c@{\hspace{0.8em}}c@{}}
\toprule
 & & \multicolumn{3}{@{}c@{\hspace{0.8em}}}{\VEL} \\
\cmidrule(r{0.8em}){3-5}
Tier & \rotatebox{90}{\strut Targets} & \makebox[2.7em]{\rotatebox{90}{\begin{tabular}{@{}l@{}}callable,\\closed form\end{tabular}}} & \makebox[2.7em]{\rotatebox{90}{\begin{tabular}{@{}l@{}}callable, with\\continuation\end{tabular}}} & \makebox[2.7em]{\rotatebox{90}{\strut float64 table}} & \rotatebox{90}{\begin{tabular}{@{}l@{}}PySR + 14\\functions\end{tabular}} & \rotatebox{90}{\strut PySR} & \rotatebox{90}{\strut Bingo} & \rotatebox{90}{\strut Flash-ANSR} & \rotatebox{90}{\strut RILS-ROLS} & \rotatebox{90}{\strut Operon} & \rotatebox{90}{\strut gplearn} & \rotatebox{90}{\strut E2E} & \rotatebox{90}{\strut TPSR} \\
\midrule
E: elementary & \phantom{0}12 & 11 & 11 & \phantom{0}8 & \phantom{0}3 & \phantom{0}7 & \phantom{0}4 & \phantom{0}2 & \phantom{0}5 & 2 & 1 & 0 & 0 \\
A: operator of order 1 to 2 & \phantom{0}36 & 28 & 35 & 26 & \phantom{0}6 & \phantom{0}0 & \phantom{0}1 & \phantom{0}1 & \phantom{0}0 & 0 & 0 & 0 & 0 \\
B: order 3 to 4 & \phantom{0}36 & 20 & 28 & 10 & \phantom{0}6 & \phantom{0}2 & \phantom{0}3 & \phantom{0}3 & \phantom{0}0 & 1 & 0 & 0 & 0 \\
C: order 5 to 8 & \phantom{0}36 & \phantom{0}3 & \phantom{0}3 & \phantom{0}0 & \phantom{0}2 & \phantom{0}2 & \phantom{0}3 & \phantom{0}2 & \phantom{0}0 & 1 & 0 & 1 & 0 \\
N: no operator in $x$ & \phantom{0}12 & \phantom{0}3 & \phantom{0}3 & \phantom{0}6 & \phantom{0}6 & \phantom{0}5 & \phantom{0}5 & \phantom{0}3 & \phantom{0}5 & 0 & 3 & 0 & 0 \\
\midrule
all & 132 & 65 & 80 & 50 & 23 & 16 & 16 & 11 & 10 & 4 & 4 & 1 & 0 \\
\bottomrule
\end{tabular}

\end{table}

\paragraph{Counts.} From float64 tables \VEL{} recovers 50 of the 132 targets (Table~\ref{tab:generated_tiers},
Figure~\ref{fig:generated_tiers}). PySR + 14 functions recovers 23, PySR and Bingo 16 each, Flash-ANSR 11, RILS-ROLS 10,
Operon and gplearn 4 each, E2E 1 and TPSR none. From callables \VEL{} returns a closed form that passes the test for 65
targets, and for 15 more the continued solution of its verified operator passes, which makes 80.

In tier A the table run recovers 26 of 36 targets and the callable run 35 with continuation, where the best other program
recovers 6. In tier B the numbers are 10 and 28 against 6. In tier C the table run recovers none and the callable run 3,
which are $\operatorname{ker}_{7/5}(-x) - 1$, $\operatorname{bei}_{3/4}(x) - 1$ and
$\pFq{0}{3}\bigl(;\,\tfrac14, \tfrac53, \tfrac95;\,x/2\bigr) - 1$. The operators found for them have order 6, 6 and 5 and
the two-term form~\eqref{eq:banded}, which is what the wide search of Section~\ref{sec:search} looks for. The other
programs recover 0 to 3 targets of tier C.

Tier N holds functions for which the operator search should find nothing in $x$. From tables \VEL{} recovers six of them
through the relation search of Section~\ref{sec:relations}: Lambert's $W$, $x^x$ as $\exp(x \log x)$, and the four built from
exponentials and trigonometric functions. From callables it recovers three, $1/(\ee^x - 1)$ and $\ee^{\ee^x/2}$ under
$y = \ee^x$ and $\tan x$ under $y = \tan x$. For $\Gamma$, $1/\Gamma$, $\log \Gamma$, the digamma function, $\zeta$ and the
inverse error function it returns no model from either input.

\paragraph{Grades.} The table run returned 36 \code{verified} results, 15 conjectures and no model for 81 targets. All 36
verified results and 14 of the conjectures are recovered. The fifteenth is the one expression of this run that fails the
test: for $H_{4/5}(x^2)^2$ on $[-2,\,1]$, with $H_\nu$ the Hermite function, the learner returned
\begin{equation*}
  \bigl[\,1.657822970327457\;x^2\,\pFq{1}{1}\bigl(\tfrac{1}{10};\,\tfrac32;\,x^4\bigr)
    + 0.3243831291665643\;\pFq{1}{1}\bigl(-\tfrac25;\,\tfrac12;\,x^4\bigr)\bigr]^2 ,
\end{equation*}
which agrees with the target to $9 \timestento{-12}$ inside the range and is off by 0.029 beyond it. The form is the right
one: with the two constants replaced by $-2^{9/5} \sqrt{\pi} / \Gamma(-\tfrac25)$ and $2^{4/5} \sqrt{\pi} / \Gamma(\tfrac{1}{10})$,
with which they agree in all 16 digits, it equals the target. The two series grow like $\ee^{x^4}$ and nearly cancel, so
beyond the range the sixteen digits of the fitted constants are used up. The scoring rule counts the result as a failure,
and so do we.

The callable run returned 80 \code{verified} certificates, no model for 51 targets and one error. Each of the 80 operators
annihilates its target to $\tento{-20}$ when applied to it directly, and none has a fitted parameter. The error is the
target $(-x)^{5/4}\,\ee^{-x} I_{5/4}(x) / x^{5/4}$ on $[-4,\,-2]$, whose values are real and whose factors are complex:
the run ended in an exception raised inside the learner's search for a fraction (an ordering comparison of complex
numbers). It counts as a miss, and the learner was left as frozen. The table run recovered the same target.

\paragraph{What kind of answer.} For each recovered expression we also applied the target's own operator to it. In tiers A
to C all 36 expressions that \VEL{} recovered from tables satisfy the operator to $\tento{-8}$: they are the function. Of
the 14 that PySR + 14 functions recovered in these tiers, 9 do, and these are compositions of the functions it was given,
such as $-\Ei(x)\,\erf(x)$, which it returned as $\Ei(x)\,\erf(-1.0000000000002\,x)$. None of the recoveries of the other programs in these tiers does (0 of 4 for PySR, 0 of 7 for
Bingo, 0 of 6 for Flash-ANSR, 0 of 2 for Operon, 0 of 1 for E2E). They are approximants that hold to $\tento{-5}$ over the
test range. They concern seven targets (A21, B06, B12, B13, C12, C34 and C35 of Table~\ref{tab:generated}). Four of them
(B06, B13, C12 and C35) are a function plus the constant 2, and the error~\eqref{eq:error} is relative to $|f|$ plus the
rms of the values, which the constant raises.

\paragraph{Targets found by one side only.} In tiers A to C, 44 targets are recovered by \VEL{} (from a table, or as a
closed form from the callable) and by none of the nine other programs: 28 from tables and 38 from callables. With
continuation the number is 54. Among them are $K_0(x/2)^2$, $\Ei(2x)^2$,
$\pFq{2}{1}\bigl(\tfrac25, \tfrac43;\,\tfrac34;\,x\bigr)$, the elliptic integral $K(2 \log x)$ and $J_1(\ee^{x}/2)$, each
from the table and from the callable, and from the callable $F(\arcsin x \mid \tfrac15)^2$, which was returned as
$x^2\,F_1\bigl(\tfrac12;\,\tfrac12, \tfrac12;\,\tfrac32;\,x^2/5,\,x^2\bigr)^2$.

In the other direction, 13 targets were recovered by another program and missed by the table run of \VEL{}, and for 7 of
them the callable run gave no closed form either. One is elementary: $\log x \cos x + \ee^{-x/2}$, whose operator has order
5, was recovered by PySR and RILS-ROLS. One is a composition of supplied functions: $-\Ei(x)\,\erf(x)$, recovered exactly by
PySR + 14 functions, for which \VEL{} has the verified fourth-order operator from the callable, a continuation that passes,
and no closed form. Two are targets of tier C that other programs fitted with approximants. The last three are
$\Gamma(x)$ and $\log \Gamma(x)$, which PySR + 14 functions returns because $\Gamma$ is among its operations, and the
digamma function, for which it returned the logarithm of a rational function minus a multiple of $1/x$, within the
tolerance ($8 \timestento{-6}$ inside the range and $4 \timestento{-6}$ beyond).

Giving PySR + 14 functions ten times its budget changed little: run for 600\,s on the 24 compositions of its own functions
that it had missed in 60\,s, it recovered 2. Over the 120 targets of tiers E to C, the median time of \VEL{} was 196\,s per
table and 26\,s per callable.

\subsection{Quantities from physics, statistics and geometry}\label{sec:physical}

The generated targets are compositions drawn at random. The third set holds 29 quantities that a calculation produces as
numbers and whose known exact form is a hypergeometric, Meijer $G$, Fox $H$ or Lauricella function: periods of anharmonic
oscillators, generating functions of lattice walks \cite{joyce1973cubic,guttmann2010lattice}, the densities of products of
normal and beta variables \cite{springer1970products,wilks1932}, the $K$ distribution of radar clutter
\cite{jakeman1976kdistribution}, one-sided and symmetric stable laws \cite{penson2010levy,schneider1986stable}, Holtsmark's
distribution of the electric field in a plasma \cite{holtsmark1919,chandrasekhar1943}, fractional relaxation and diffusion
functions \cite{haubold2011mittagleffler,mainardi2010mwright}, the Voigt line profile \cite{armstrong1967voigt}, the
deflection of light of Section~\ref{sec:deflection}, and the elliptic integrals that give the demagnetising factor and the
surface area of an ellipsoid \cite{osborn1945demagnetizing,carlson1995numerical}. Each target is a function of one variable
on an interval, computed by a series or a quadrature. For 26 of the 29 a second, independent definition (a series against a
quadrature, for example, or a Laplace transform) agrees with it to $\tento{-16}$ or better; the period of the quintic
oscillator, the deflection of light and the period of the asymmetric oscillator have no second definition.

These targets were chosen by the authors, and the five additions to the solver that Section~\ref{sec:solver} lists (the
symmetric square, the primitive operator, the extended gauge, the binomial primitive and the Meijer $G$ rule) were made
while working on special-function targets of this kind.
The counts below show which classes of functions the method covers. They are not a blind test.

\begin{table}[t]
\centering\small
\caption{The 29 physical quantities by the class of their known exact form: targets recovered, counting those reproduced
by the continued solution of a certified operator; in brackets, the number recovered as a closed form. All programs but the
first row received 1101 float64 points.}
\label{tab:physical_families}
\begin{tabular}{@{}L{56mm}@{\hspace{0.8em}}c@{\hspace{0.8em}}c@{\hspace{0.8em}}c@{\hspace{0.8em}}c@{\hspace{0.8em}}c@{\hspace{0.8em}}c@{}}
\toprule
 & & Hyper- & & & Appell / & Heun \\
 & All & geometric & Meijer $G$ & Fox $H$ & Lauricella & type \\
Method & of 29 & of 11 & of 5 & of 7 & of 4 & of 2 \\
\midrule
\VEL{}, callable & 29 (23) & 11 (10) & 5 (5) & 7 (7) & 4 (1) & 2 (0) \\
\VEL{}, float64 table & 18 (17) & \phantom{0}8 (7)\phantom{0} & 5 (5) & 4 (4) & 1 (1) & 0\phantom{ (0)} \\
Bingo & \phantom{0}1\phantom{ (00)} & \phantom{0}0\phantom{ (00)} & 0\phantom{ (0)} & 0\phantom{ (0)} & 1\phantom{ (0)} & 0\phantom{ (0)} \\
\hangindent=1em PySR, PySR + 14 functions, Operon, gplearn, Flash-ANSR, RILS-ROLS, E2E, TPSR & \phantom{0}0\phantom{ (00)} & \phantom{0}0\phantom{ (00)} & 0\phantom{ (0)} & 0\phantom{ (0)} & 0\phantom{ (0)} & 0\phantom{ (0)} \\
\bottomrule
\end{tabular}

\end{table}

\paragraph{From callables.} All 29 targets received a \code{verified} operator, in 1.5 to 62.9\,s
(Table~\ref{tab:physical_families}; Tables~\ref{tab:physical} and~\ref{tab:physical_operators} in Section~\ref{app:physical} list every
target, its operator and its closed form). Every operator was found in $x$ itself, has exact coefficients and no fitted
parameter, and applied to the target at 60 digits leaves a residual below $\tento{-57}$ of the size of its terms. For 23
targets the solver returned a closed form that passes the test, three of them with one or two constants left as numbers.
For the other six (the Voigt profile, $\Pi(n \mid \tfrac12)$, the demagnetising factor, the area of the ellipsoid, the
period of the asymmetric oscillator and the simple cubic lattice) it found none, and the continued solution of the operator
reproduces the target inside the range and beyond it.

Some of the results, as returned:
\begin{itemize}[leftmargin=*,itemsep=1pt,topsep=3pt]
\item the period of the oscillator with potential $q^2/2 - q^5/5$ at energy $E$, a hyperelliptic integral, as
$2\pi\,\pFq{4}{3}\bigl(\tfrac{1}{10}, \tfrac{3}{10}, \tfrac{7}{10}, \tfrac{9}{10};\,\tfrac13, \tfrac23, 1;\,1000 E^3/27\bigr)$,
from an operator of order 4;
\item for the density of the symmetric stable law of index $3/2$, the equation
$81 f^{(5)} + 16 x^4 f''' + 240 x^3 f'' + 940 x^2 f' + 880 x f = 0$ and a sum of three hypergeometric series in $x^6$;
Holtsmark's density has an equation of order 5 and degree 9;
\item for the thermonuclear rate integral $\int_0^\infty \ee^{-y - x/\sqrt{y}}\,\dd y$, the equation $x f''' + 2 f = 0$
and the Meijer $G$-function $\pi^{-1/2}\,\MeijerG{3}{0}{0}{3}{x^2/4 \,\middle|\, 0, \tfrac12, 1}$;
\item for the body-centred cubic lattice, the generating function of returns to the origin in the form
$\pFq{2}{1}\bigl(\tfrac14, \tfrac14;\,1;\,x^2\bigr)^2$ that the symmetric-square rule gives, which by Clausen's formula
equals the known form $\pFq{3}{2}\bigl(\tfrac12, \tfrac12, \tfrac12;\,1, 1;\,x^2\bigr)$;
\item for the Voigt profile, the equation $4 f''' + 8 x f'' + (4 x^2 + 13) f' + 8 x f = 0$ and no closed form.
\end{itemize}
The last two classes of Table~\ref{tab:physical_families} show where the solver ends. Three of the four Appell and
Lauricella targets and both targets of Heun type have a verified operator and no closed form. The simple cubic lattice is
the best known of them: Joyce expressed its Green function through the square of a Heun function \cite{joyce1973cubic},
and the solver has no rule for operators of that type \cite{ronveaux1995heun}.

\paragraph{From float64 tables.} From 1101 float64 points \VEL{} recovers 18 of the 29: 17 as closed forms and the Voigt
profile by continuation. The runs took 55 to 1047\,s. Five of the 17 closed forms keep a constant as a float64 number that
the callable run identified, such as the factor of $x^{1/3} K_{2/3}\bigl(2 \sqrt{15 x}/3\bigr)$ in the $K$ distribution.
Nine targets gave no model: the quintic oscillator, the deflection of light, the stable law of index $3/2$, Holtsmark's
density and the five targets without a closed form other than the Voigt profile. Their operators have order 2 to 5 and
degree 3 to 9, which sixteen digits on these intervals do not determine.

Two tables received a wrong certificate, and both carry fitted parameters.
\begin{itemize}[leftmargin=*,itemsep=1pt,topsep=3pt]
\item The relaxation function $E_{1/3}(-t^{1/3})$ was returned as a \code{conjecture}. Its operator has the right form with
the coefficient $2/9$ left as a fitted number, $0.22116$, and the closed form built on that number is off by 0.32 inside
the range.
\item The body-centred cubic lattice was returned as \code{verified}: an operator of order 3 and degree 9 in $y = 1/x$ with
two fitted parameters, which passed both hold-out tests. It is not the operator of the target. Applied to the target it
leaves a residual of $1.6 \timestento{-4}$, and its continued solution is off by $1.1 \timestento{-6}$ inside the range and
by $2.3 \timestento{-4}$ beyond it. No closed form came with it, so it does not count as a recovery, but the certificate
is false. It is the one \code{verified} certificate in the runs of this paper whose operator is wrong on the interval of
its own data. The two expressions of Section~\ref{sec:absent} that lose a sign are a different failure: they are exact
on that interval and wrong beyond a point that the data do not reach.
\end{itemize}

\paragraph{Other programs.} Of the nine other programs, one recovered one target: Bingo returned a rational function with
a sine term for the demagnetising factor, within $1.2 \timestento{-8}$ inside the range and $2.6 \timestento{-6}$ beyond
it. Table~\ref{tab:physical} gives, for each target, the closest formula of any other program and its two errors.
Over the 29 targets its error is between $1.2 \timestento{-8}$ and $1.5 \timestento{-3}$ inside the range and between
$2.6 \timestento{-6}$ and $1.6 \timestento{-2}$ beyond it.

\subsection{Functions of two variables}\label{sec:two}

The fourth set holds eight functions of two variables whose exact form is a double hypergeometric series, and two
controls (Table~\ref{tab:bivariate}): the elliptic integrals of Section~\ref{sec:pi}, Carlson's integral for the
demagnetising factor of an ellipsoid, the Green function of an anisotropic square lattice, an integral of two confluent
hypergeometric functions of the kind that Gordon evaluated for the matrix elements of hydrogen \cite{gordon1929}, an
elliptic integral with an exponential weight, and the Voigt function in both of its arguments. \VEL{} read each as a
callable at 40 digits on a box, and the other programs received 1000 float64 points drawn from the box.

\begin{table}[p]
\centering\small
\caption{Functions of two variables. For each target: what \VEL{} returned from the callable at 40 digits, the number of
operators that the reduction of Section~\ref{sec:bivariate} kept from the system that it found (column \qq{gen.}; the kept
set need not be minimal, and for the Voigt function two of the four are independent), the time, and the closest formula
that any other program returned for 1000 float64 points, with its largest error inside the box and beyond it. $F_1$, $F_2$ and $F_4$ are Appell's functions and
$\Phi_1$ and $\Psi_2$ Humbert's confluent functions \cite{appell1926,erdelyi1953htf,srivastava1985multiple}. All eight
results are graded \code{verified}, with recorded errors of 0 inside the box and beyond it.}
\label{tab:bivariate}
\begin{tabular}{@{}L{47.5mm}@{\hspace{7pt}}L{52mm}@{\hspace{7pt}}c@{\hspace{7pt}}r@{\hspace{7pt}}L{27.5mm}@{}}
\toprule
target, definition and box & returned & gen. & time (s) & closest other program; error inside, beyond \\
\midrule
elliptic Pi, $\Pi(n\,|\,m)$ \newline $\int_0^{\pi/2}\!\dfrac{\dd\theta}{(1-n\sin^2\theta)\sqrt{1-m\sin^2\theta}}$\rule[-2.1ex]{0pt}{5.1ex} \newline on $[0.05,\,0.55]\times[0.05,\,0.55]$ & $\frac{\pi}{2}\,F_{1}\!\left(1/2;\,1/2, 1;\,1;\,m, n\right)$ & 3 & 123.8 & E2E \newline $2\timestento{-4}$, $0.001$ \\
\addlinespace[3pt]
pendulum time, $F(\arcsin s\,|\,m)$ \newline $\int_0^{s}\!\dfrac{\dd t}{\sqrt{(1-t^2)(1-m t^2)}}$\rule[-2.1ex]{0pt}{5.1ex} \newline on $[0.1,\,0.6]\times[0.1,\,0.7]$ & $s F_{1}\!\left(1/2;\,1/2, 1/2;\,3/2;\,s^{2}, m s^{2}\right)$ & 1 & 111.5 & Bingo \newline $7\timestento{-4}$, $0.005$ \\
\addlinespace[3pt]
ellipsoid capacitance, $R_F(x,y,1)$ \newline $\dfrac{1}{2}\int_0^\infty\!\dfrac{\dd t}{\sqrt{(t+x)(t+y)(t+1)}}$\rule[-2.1ex]{0pt}{5.1ex} \newline on $[0.3,\,0.8]\times[0.35,\,0.9]$ & $F_{1}\!\left(1/2;\,1/2, 1/2;\,3/2;\,1 - y, 1 - x\right)$ & 3 & 221.1 & Operon \newline $4\timestento{-5}$, $7\timestento{-5}$ \\
\addlinespace[3pt]
ellipsoid demagnetisation, $R_D(x,y,1)$ \newline $\dfrac{3}{2}\int_0^\infty\!\dfrac{\dd t}{\sqrt{(t+x)(t+y)}\,(t+1)^{3/2}}$\rule[-2.1ex]{0pt}{5.1ex} \newline on $[0.3,\,0.8]\times[0.35,\,0.9]$ & $F_{1}\!\left(3/2;\,1/2, 1/2;\,5/2;\,1 - y, 1 - x\right)$ & 3 & 226.1 & Operon \newline $4\timestento{-5}$, $3\timestento{-4}$ \\
\addlinespace[3pt]
anisotropic square lattice \newline $\int_0^\infty \ee^{-t}\,I_0(at)\,I_0(bt)\,\dd t$ \newline on $[0.05,\,0.35]\times[0.05,\,0.3]$ & $F_{4}\!\left(1, 1/2;\,1, 1;\,b^{2}, a^{2}\right)$ & 2 & 64.7 & Operon \newline $5\timestento{-6}$, $3\timestento{-4}$ \\
\addlinespace[3pt]
Gordon integral \newline $\int_0^\infty \ee^{-t}\,t^{3/2}\,\pFq{1}{1}(1/2;1;xt)$ \newline \hspace*{2em}$\times\,\pFq{1}{1}(1/3;2;yt)\,\dd t$ \newline on $[0.05,\,0.35]\times[0.05,\,0.35]$ & $\frac{3 \sqrt{\pi}}{4}\,F_{2}\!\left(5/2;\,1/3, 1/2;\,2, 1;\,y, x\right)$ & 2 & 73.6 & Operon \newline $3\timestento{-5}$, $0.008$ \\
\addlinespace[3pt]
weighted elliptic K \newline $\dfrac{2}{\pi}\int_0^{\pi/2}\!\dfrac{\exp(y\sin^2\theta)}{\sqrt{1-x\sin^2\theta}}\,\dd\theta$\rule[-2.1ex]{0pt}{5.1ex} \newline on $[0.05,\,0.55]\times[0.1,\,1.1]$ & $\Phi_{1}\!\left(1/2, 1/2;\,1;\,x, y\right)$ & 2 & 110.7 & Operon \newline $7\timestento{-5}$, $0.003$ \\
\addlinespace[3pt]
Voigt function, $\operatorname{Re}\,w(v+\mathrm{i}y)$ \newline $w(z)=\ee^{-z^2}\erfc(-\mathrm{i}z)$ \newline on $[0.2,\,1.7]\times[0.3,\,1.5]$ & $\Psi_{2}\!\left(1/2;\,1/2, 1/2;\,y^{2}, - v^{2}\right) - \frac{2}{\sqrt{\pi}}\,y \Psi_{2}\!\left(1;\,3/2, 1/2;\,y^{2}, - v^{2}\right)$ & 4 & 51.2 & Operon \newline $4\timestento{-4}$, $0.01$ \\
\midrule
control, $\Gamma(x+2y)$ \newline on $[0.5,\,1.5]\times[0.5,\,1.5]$ & the operator $2\partial_{x} - \partial_{y}$ and no series & 1 & 107.9 & PySR + 14 functions, exact \\
\addlinespace[3pt]
control, $\Gamma(x)\,\Gamma(y)$ \newline on $[0.5,\,1.5]\times[0.5,\,1.5]$ & no model & -- & 124.5 & PySR + 14 functions, exact \\
\bottomrule
\end{tabular}

\end{table}

All eight targets came back \code{verified}, in 51 to 226\,s, with recorded errors of 0 inside the box and beyond it. Six
are Appell functions, two of them after the change of variable $1 - x$. The weighted elliptic integral is Humbert's
$\Phi_1$, and the Voigt function is a sum of two $\Psi_2$ series. The reduction kept one to four operators from each
system, and the 36 recorded tests on the held-out quadrant have residuals between $2.0 \timestento{-38}$ and
$3.1 \timestento{-36}$, with noise levels between $1.3 \timestento{-37}$ and $6.3 \timestento{-36}$
(Table~\ref{tab:bivariate_operators}). The kept operators are not minimal sets of generators. For the Voigt function the
four are $A = 2 v\,\partial_v - \partial_y^2 + 2 y\,\partial_y + 2$, $B = \partial_v^2 + 2 v\,\partial_v + 2 y\,\partial_y + 2$
and $y\,(\partial_v^2 + \partial_y^2)$ and $v\,(\partial_v^2 + \partial_y^2)$; since $\partial_v^2 + \partial_y^2 = B - A$,
two of them are independent. For $\Pi$, $R_F$ and $R_D$ a separate check by exact linear algebra, with multipliers of degree up
to 2 and order up to 1, found one of the three kept operators to be a combination of left multiples of the other two. The controls behave as
they should. For
$\Gamma(x + 2y)$ the only operator kept is $2\partial_x - \partial_y$, and the learner returns it and no series.
$\Gamma(x)\,\Gamma(y)$ returns no model.

None of the nine other programs recovers any of the eight targets. The closest result is Operon's formula for the
capacitance integral $R_F$, with errors of $4 \timestento{-5}$ inside the box and $7 \timestento{-5}$ beyond it. PySR + 14
functions recovers both controls exactly, since $\Gamma$ is among its operations.

This comparison is not like for like. From the same 1000 float64 points the table entry point of \VEL{} returns no model
for all eight targets and both controls. The two-variable learner itself needs more than sixteen digits: given the
callables of the anisotropic lattice and of $\Pi(n \mid m)$ with their values rounded to float64, it returned no model for
either (after 97.9 and 124.2\,s). What the set shows is that values to 40 digits are enough to identify these functions,
on targets that the authors chose and that were in view while the two-variable learner was written, after the rest of the
learner had been frozen.

\subsection{Noise and sparse data}\label{sec:noise}

The method was built for accurate values. Two experiments measure what is left when the values are noisy or few. Both are
in one variable, where the entry point of Section~\ref{sec:relations} estimates a noise level; for several variables the
SRBench results of Section~\ref{sec:srbench} are the measure.

\begin{figure}[t]
\centering
\includegraphics[width=\textwidth]{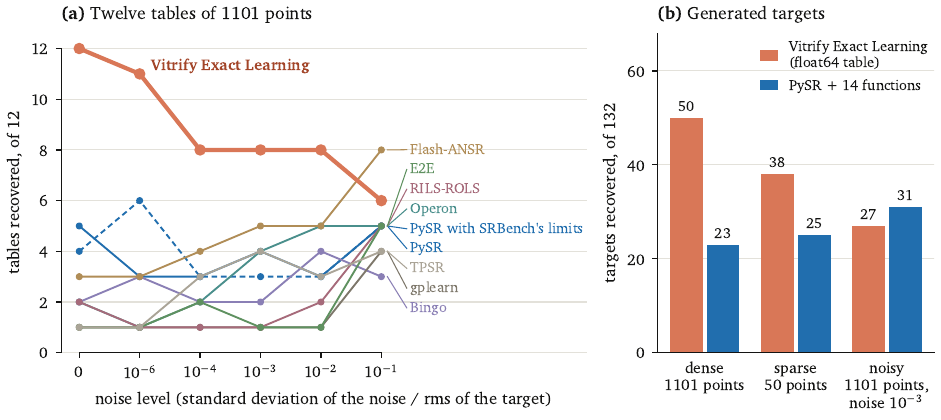}
\caption{Noise and sparse data in one variable. (a)~Twelve tables of 1101 points with Gaussian noise added to the values:
tables recovered by each program against the noise level, at the tolerance $\max(\tento{-5},\,\text{noise level})$.
(b)~The 132 generated targets of Section~\ref{sec:generated} from 1101 points, from 50 points and from 1101 points with
noise of $\tento{-3}$: targets recovered by \VEL{} and by PySR + 14 functions.}
\label{fig:noise}
\end{figure}

\paragraph{Twelve tables.} Gaussian noise with a standard deviation of 0, $\tento{-6}$, $\tento{-4}$, $\tento{-3}$,
$\tento{-2}$ and $\tento{-1}$ times the rms of the target was added to twelve tables of 1101 points: seven classic targets
(Nguyen-1, 5, 7 and 8, Keijzer-9, R1 and Livermore-7), two Feynman slices, and the special functions $\Bi(x)$, $J_0(x)^2$
and $\operatorname{Ci}(x)$. A table counts as recovered when both errors are within the larger of $\tento{-5}$ and the
noise level (Figure~\ref{fig:noise}a). \VEL{} recovers 12, 11, 8, 8, 8 and 6 of the twelve at the six levels. PySR recovers
5, 3, 3, 4, 3 and 5, Operon 2, 1, 2, 4, 5 and 5, and Flash-ANSR 3, 3, 4, 5, 5 and 8. At the highest level the tolerance is
a tenth of the target's rms, which a rough fit passes; most programs recover more tables there than at the lower levels,
and Flash-ANSR recovers more than \VEL{}.

With noise the learner grades nothing \code{verified}. At the level $\tento{-6}$ it returned 10 conjectures and one
\code{fit}, all recovered. At the level $\tento{-1}$ it returned 5 conjectures, one of which fails the test, 4 results
graded \code{fit}, two of which fail, and no model for 3 tables.

\paragraph{The generated set.} The table runs of Section~\ref{sec:generated} were repeated with 50 points in place of
1101, and with 1101 points and noise of $\tento{-3}$ times the rms, for \VEL{} and for PySR + 14 functions
(Figure~\ref{fig:noise}b). From 1101 points, from 50 points and from noisy points \VEL{} recovers 50, 38 and 27 of the 132
targets, and PySR + 14 functions 23, 25 and 31. With 50 points the learner returned 30 \code{verified} results and 9
conjectures, of which one fails beyond the range: the squared Hermite function again. With noise it returned 10
conjectures, all recovered, 77 results graded \code{fit}, of which 60 fail the test at the tolerance $\tento{-3}$, and no
model for 45 targets.

Two conclusions follow. With noise of $\tento{-3}$ on special-function targets the search program is ahead. And the grade
\code{fit} is not a reliable signal: 60 of the 77 expressions with that grade match the noisy rows and do not extrapolate,
which is the behaviour that Section~\ref{sec:absent} described for the search programs.

\subsection{Regression cases, draws and calibration}\label{sec:cases}

The benchmarks above use two of the five kinds of input. The others, and the inputs for which the right answer is no
model, are covered by 125 cases with expected answers that were written with the code. Table~\ref{tab:cases} in the
Section~\ref{app:cases} lists every case with what it returned.

\paragraph{The cases.} The groups are: functions with an operator in $x$ (13 cases); functions of a wider class, such as
the fitted exponents of Section~\ref{sec:densities} (7); changes of variable (9); controls (7); densities of distances in
geometry (4); six densities that an earlier version of the learner fitted by name; float64 and 45-digit tables (9); draws
(7); sequences (7); moments (9); and two later sets of mixed kinds (25 and 22). The frozen learner returns the expected
grade for all 125: 96 \code{verified}, 13 \code{conjecture} and 16 \code{no-model}. For the first ten groups this
agreement followed changes made while looking at failures, so it shows that the code still does what it did and is not a
test. The two later sets were written after the code that they test and were run once before any change. In those first
runs 22 of 25 and 21 of 22 cases returned the expected result. Of the four that did not, two expected the operator in $x$
to be ranked first for $\arctan x$ and for $\log^2 x$, where the learner ranked the cheaper first-order operators under
$y = \arctan x$ and $y = \log x$ first; the expectation was changed. The other two were defects: the integer-relation
routine raised an exception on an entry that rounded to zero, which left Struve's $H_0$ without a model, and Laplace
draws, which have no model in the searched class, returned a conjecture under $y = \sinh x$ whose one solution did not
match the draws. Both were corrected.

The 16 inputs graded \code{no-model} are correct refusals: $\Gamma(x)$, $\zeta(x)$ and $x^x$; $\ee^{\sin x}$; $\tan x$
and a Gaussian in $\arsinh x$ when only $x$ is searched; $1/(1 + x^2)$ when only $\arsinh x$ is searched; a table of $J_0$
with noise of $\tento{-3}$, a table of white noise and a float64 table of $\Gamma$; logistic draws searched in $x$ only,
and Laplace draws; the primes, the partition numbers, Euler's totient function and the Bell numbers. The three longest
cases of the set are $\Gamma$, $x^x$ and $\zeta$, at 239 to 272\,s, for which every search runs to its end. The median
case takes 21.5\,s.

Of the 109 certificates, 101 carry a closed form, and for 93 of those SymPy reduces the operator applied to the closed
form to zero. The eight without one are recurrences. Seven belong to sequences that are not hypergeometric terms (the
Motzkin, harmonic, Ap\'ery, derangement, Schr\"oder, central trinomial and central Delannoy numbers), and one to
$\binom{2n}{n}^2$, whose first-order recurrence the solver that we call did not solve. Some of the results:
\begin{itemize}[leftmargin=*,itemsep=1pt,topsep=3pt]
\item from the Ap\'ery numbers, the recurrence
$(n + 2)^3 a_{n+2} - (2n + 3)(17 n^2 + 51 n + 39)\,a_{n+1} + (n + 1)^3 a_n = 0$;
\item from two million Cauchy draws, the operator $g' = 0$ for the pushforward under $y = \arctan x$, which is uniform,
and the density $1 / (\pi (1 + x^2))$;
\item from two million draws of a beta(2, 3) variable, $x (x - 1) f' + (1 - 3x) f = 0$ and the density $12 x (1 - x)^2$;
\item for the density of the distance between two random points of a unit disk, the operator
$y (y - 4)\,g'' - 2\,g' = 0$ of its pushforward under $y = x^2$;
\item for Struve's function $H_0$, the equation $x f''' + 2 f'' + x f' + f = 0$ and
$\tfrac{2x}{\pi}\,\pFq{1}{2}\bigl(1;\,\tfrac32, \tfrac32;\,-x^2/4\bigr)$.
\end{itemize}
The first certificate of every case was the same under three values of Python's hash seed, on which the order of SymPy's
simplifications depends.

\begin{table}[t]
\centering\small
\caption{Draws: results of ten runs with seeds that no version of the code had seen, for 14 distributions at two sample
sizes. \qq{Expected result} is the exact operator of the density, graded \code{conjecture}, or no certificate for the two
controls. \qq{Numerical fit} is an operator that passes its test with coefficients left numerical, graded \code{fit}.
\qq{Other exact operator} would be an operator with exact coefficients that is not the expected one.}
\label{tab:draws}
\begin{tabular}{@{}l@{\hspace{1.2em}}l@{\hspace{1.6em}}c@{\hspace{0.7em}}c@{\hspace{0.7em}}c@{\hspace{0.7em}}c@{\hspace{1.6em}}c@{\hspace{0.7em}}c@{\hspace{0.7em}}c@{\hspace{0.7em}}c@{}}
\toprule
 & & \multicolumn{4}{@{\hspace{1.6em}}c}{200,000 draws} & \multicolumn{4}{@{\hspace{1.6em}}c@{}}{2,000,000 draws} \\
\cmidrule(l{1.6em}){3-6}\cmidrule(l{1.6em}){7-10}
Distribution & Expected result & \makebox[2.2em]{\rotatebox{90}{\strut expected result}} & \makebox[2.2em]{\rotatebox{90}{\strut no certificate}} & \makebox[2.2em]{\rotatebox{90}{\strut numerical fit}} & \makebox[2.2em]{\rotatebox{90}{\strut other exact operator}} & \makebox[2.2em]{\rotatebox{90}{\strut expected result}} & \makebox[2.2em]{\rotatebox{90}{\strut no certificate}} & \makebox[2.2em]{\rotatebox{90}{\strut numerical fit}} & \makebox[2.2em]{\rotatebox{90}{\strut other exact operator}} \\
\midrule
exponential(1) & exact operator & 10 & \phantom{0}0 & \phantom{0}0 & \phantom{0}0 & 10 & \phantom{0}0 & \phantom{0}0 & \phantom{0}0 \\
gamma(3, 1) & exact operator & 10 & \phantom{0}0 & \phantom{0}0 & \phantom{0}0 & 10 & \phantom{0}0 & \phantom{0}0 & \phantom{0}0 \\
normal(0, 1) & exact operator & 10 & \phantom{0}0 & \phantom{0}0 & \phantom{0}0 & 10 & \phantom{0}0 & \phantom{0}0 & \phantom{0}0 \\
beta(2, 3) & exact operator & \phantom{0}4 & \phantom{0}4 & \phantom{0}2 & \phantom{0}0 & 10 & \phantom{0}0 & \phantom{0}0 & \phantom{0}0 \\
lognormal(0, 1/2) & exact operator & 10 & \phantom{0}0 & \phantom{0}0 & \phantom{0}0 & 10 & \phantom{0}0 & \phantom{0}0 & \phantom{0}0 \\
Weibull(2) & exact operator & 10 & \phantom{0}0 & \phantom{0}0 & \phantom{0}0 & 10 & \phantom{0}0 & \phantom{0}0 & \phantom{0}0 \\
logistic, searched in $x$ only & no certificate & \phantom{0}4 & \phantom{0}0 & \phantom{0}6 & \phantom{0}0 & 10 & \phantom{0}0 & \phantom{0}0 & \phantom{0}0 \\
Rayleigh(1) & exact operator & 10 & \phantom{0}0 & \phantom{0}0 & \phantom{0}0 & 10 & \phantom{0}0 & \phantom{0}0 & \phantom{0}0 \\
chi-square(4) & exact operator & 10 & \phantom{0}0 & \phantom{0}0 & \phantom{0}0 & 10 & \phantom{0}0 & \phantom{0}0 & \phantom{0}0 \\
Student $t$(5) & exact operator & \phantom{0}4 & \phantom{0}0 & \phantom{0}6 & \phantom{0}0 & 10 & \phantom{0}0 & \phantom{0}0 & \phantom{0}0 \\
Cauchy & exact operator & 10 & \phantom{0}0 & \phantom{0}0 & \phantom{0}0 & 10 & \phantom{0}0 & \phantom{0}0 & \phantom{0}0 \\
arcsine & exact operator & 10 & \phantom{0}0 & \phantom{0}0 & \phantom{0}0 & 10 & \phantom{0}0 & \phantom{0}0 & \phantom{0}0 \\
beta(3, 2) & exact operator & \phantom{0}1 & \phantom{0}0 & \phantom{0}9 & \phantom{0}0 & \phantom{0}7 & \phantom{0}1 & \phantom{0}2 & \phantom{0}0 \\
Laplace & no certificate & 10 & \phantom{0}0 & \phantom{0}0 & \phantom{0}0 & 10 & \phantom{0}0 & \phantom{0}0 & \phantom{0}0 \\
\midrule
all 140 runs & & 113 & \phantom{0}\phantom{0}4 & \phantom{0}23 & \phantom{0}\phantom{0}0 & 137 & \phantom{0}\phantom{0}1 & \phantom{0}\phantom{0}2 & \phantom{0}\phantom{0}0 \\
\bottomrule
\end{tabular}

\end{table}

\paragraph{Draws at other seeds and sizes.} The seven cases with draws use one seed each. Table~\ref{tab:draws} repeats
14 distributions over ten seeds that no version of the code had seen, at two sample sizes. With 2{,}000{,}000 draws, 137
of the 140 runs return the expected result; one returns no certificate and two return a \code{fit}. With 200{,}000 draws
113 do, 4 return no certificate and 23 a \code{fit}. The densities with two shape parameters need the larger sample: at
200{,}000 draws the expected operator comes back in 1 of 10 runs for beta(3, 2) and in 4 of 10 for beta(2, 3) and for
Student's $t$ with 5 degrees of freedom. Logistic draws searched in $x$ only, where no model is the right answer, return a
\code{fit} in 6 of 10 runs at the smaller size. No run at either size returned an exact operator other than the
expected one. Together with the calibration of the two statistical decisions in Table~\ref{tab:calibration}, this is the
evidence for the grades given to results from draws.

\subsection{Development history and faults found}\label{sec:history}

All results above come from one frozen version of the learner. How that version came about bears on how they should be
read, and the comparison exposed faults in our code and in the code of others.

\paragraph{The first comparison.} An earlier version of the learner, which had the operator search and no entry point
for tables in several variables, was first run on the three one-variable suites from float64 tables. It recovered 21 of
the 53 classic targets, 34 of the 51 at 1101 points and 167 of the 254 Feynman slices, where PySR recovers 33, 32 and 174.
The run exposed two defects.
\begin{itemize}[leftmargin=*,itemsep=1pt,topsep=3pt]
\item \emph{Five false certificates.} Tables of $\log x$, $\sqrt{x}$ and $\arsinh x$ on intervals up to $x = 100$ were
graded \code{verified} with the operator $g' = 0$ under $y = \ee^{x}$, and the harmonic number with a second-order
operator with two fitted parameters, at two table sizes. The residual of the hold-out test was divided by the size of the
operator's terms, with no length scale. Under $y = \ee^{x}$ on an interval of width 100 the derivative column of the weak
matrix is $\tento{-44}$ times the data column, so $g' = 0$ passed for any data. None of the five came with a closed form,
so none was counted as a recovery, but the certificate is the main claim of the method.
\item \emph{Right operators rejected.} The same test rejected right operators that carry a fitted parameter, because the
noise level of the held-out window did not contain the error of a parameter fitted on the other windows. This accounted
for 39 of the 87 slices that the run missed.
\end{itemize}
The length scale and the two tests for fitted parameters of Section~\ref{sec:holdout} are the corrections. They were made
after looking at these targets.

\paragraph{What was in view.} The entry point of Section~\ref{sec:relations} was then written with the classic and
Feynman targets at hand, and the counts of Table~\ref{tab:suites} are from the learner that includes it. They are results
on a development set, and the first-run counts above, behind PySR on the classic targets at their published points and
about level with it on the other two suites, are the measure of the operator search on targets that it had not seen. The
five additions to the solver that Section~\ref{sec:solver} lists were made with special-function targets in view, among
them those of the physical set (Section~\ref{sec:physical}). The two-variable learner was written after the rest of
the learner was frozen, with its eight targets in view. Only the generated set of Section~\ref{sec:generated} was drawn
after the learner was frozen, and it tests the one-variable learner only.

\paragraph{Faults in the other programs' code.} Three faults affected the baselines, and we report the numbers obtained
after working around them.
\begin{itemize}[leftmargin=*,itemsep=1pt,topsep=3pt]
\item SRBench's wrappers for E2E and TPSR pass the targets as an array of shape $(n)$. The refinement step of E2E then
subtracts predictions of shape $(n, 1)$ from them, its objective becomes a mean over an $n \times n$ array, and no refined
candidate is ever selected. With that shape E2E and TPSR recovered none of the 509 noise-free targets of the published
suites. The rows that we report pass shape $(n, 1)$, as the authors' own evaluation script does. We did not check whether
the results that SRBench publishes for these two programs are affected.
\item The E2E code needed two patches to run: a loop that never ends when a list of candidates is shorter than expected,
and a mask type that current versions of PyTorch reject. TPSR carries its own copy of that code, which ran unpatched; 134
of its 569 runs on the published suites and 36 of its 132 on the generated set ended at the limit of 600\,s. We did not
measure how many of those are that loop.
\item For PySR + 14 functions, Julia's $K_0$ raised an overflow error at a subnormal argument, so the functions that we
supplied are defined on bounded domains.
\end{itemize}
Two pools of E2E and TPSR runs stopped when their machines ran out of memory, and the 42 runs outstanding were repeated.
The first run of the 60\,s programs on the generated set was overwritten by a copying error and was repeated; the reported
rows are the repeat.

\paragraph{Programs that were not run.} LaSR, which guides PySR with a library of concepts written by a language model,
reports 72 of the 100 Feynman equations with GPT-3.5 against 59 for PySR in the same paper \cite{grayeli2024lasr}. The
benchmark LLM-SRBench has 239 problems built to prevent the recitation of known formulas, and the best method on it
reaches a symbolic accuracy of 31.5\% \cite{shojaee2025llmsrbench}. We ran neither. Two further programs of SRBench's 2025
edition, uDSR and Brush, could not be installed in our environment.

\section{Limits}\label{sec:limits}

This section collects the limits of the method and of the evidence for it. Each item names the section that holds the
numbers.

\paragraph{Noise, and few or observed points.} The operator search needs accurate values. The entry point for tables
handles noise for one variable only (Section~\ref{sec:relations}). Under SRBench's protocol the symbolic solution rate
of \VEL{} on the Feynman problems is 74.1\% without noise and 0.9\%, 0.9\% and 0.0\% at the three noise levels, and on
the 25 data sets of observations in SRBench's 2025 edition none of its 75 runs returned an expression
(Section~\ref{sec:srbench}). In one variable, with noise of $\tento{-3}$ on the generated set, PySR + 14 functions
recovers 31 targets and \VEL{} 27, and 60 of the 77 results that the learner graded \code{fit} fail the test
(Section~\ref{sec:noise}). A result graded \code{fit} matches the rows it was compared with and should be read as a fit.

\paragraph{Precision of the data.} A float64 table determines operators of order 1 and 2 and some of order 3 and 4. In
tier B of the generated set the table run recovers 10 of the 36 targets, and the callable run returns a closed form for
20 (Section~\ref{sec:generated}). The Kelvin function ber and the trilogarithm, which have operators of order 4, gave no
model from tables and closed forms from callables (Section~\ref{sec:suites}). In two variables sixteen digits were not
enough for any target: from 1000 float64 points the table entry point returned no model for the eight targets, and the
two-variable learner returned no model for the two targets that we gave it with values rounded to float64
(Section~\ref{sec:two}). The two-variable results of this paper are results for values to 40 digits. Draws need large
samples: the expected operator came back in 137 of 140 runs with two million draws and in 113 of 140 with 200{,}000
(Section~\ref{sec:cases}).

\paragraph{Search bounds.} The search covers operators up to order 4 with coefficients up to degree 8 in $x$, and up to
order 2 and degree 4 in a transported variable; two-term operators up to order 6; order 2 and degree 3 for draws;
recurrences up to order 3 and degree 3; and order 2 with total degree 3 in two variables. At most two coefficients of an
operator may stay numerical. Targets whose operator has order 5 to 8 are mostly out of reach: of the 36 of tier C, three
are recovered from callables, all through two-term operators, and none from tables (Section~\ref{sec:generated}).

\paragraph{The class of functions.} The method describes the functions that satisfy an equation of
the form~\eqref{eq:operator} in $x$ or in a transported variable. For $\Gamma(x)$, $\zeta(x)$ and $x^x$ given as callables it
returns no model. That is the right answer of this search, and it leaves the user of such a function without a result.
The change of variable is selected from a bank of twenty maps, from a fitted power and from maps that the caller
supplies. It is not searched for freely, and the costs in~\eqref{eq:cost} that rank the maps were set by hand
(Section~\ref{sec:transport}). Constants are identified within the small basis of Section~\ref{sec:exact}, and any other
constant stays a number, as in three closed forms of the physical set (Section~\ref{sec:physical}).

\paragraph{The interval.} The caller chooses the interval of the data, and it must contain no singular point of the
function. The learner treats one smooth branch on one interval. It does not choose the interval, split it, or join
results across a singular point.

\paragraph{Closed forms.} The solver is a set of rules (Section~\ref{sec:solver}). It has none for operators of Heun
type, for second-order operators that need a change of variable to reach a hypergeometric form, or for solutions with
complex exponents. For 15 of the 80 verified operators of the generated set from callables, and for 6 of the 29 of the
physical set, it returned no closed form that passes the test, and the function is then available only by continuing the
operator (Section~\ref{sec:continuation}). The Green function of the simple cubic lattice, whose known form is the square
of a Heun function \cite{joyce1973cubic}, is one of the six. Wavefunctions of anharmonic wells that are an exponential
factor times a Heun function \cite{ronveaux1995heun} are a further case of this kind. The Schr\"odinger equation of a
polynomial potential of degree up to 8 is within the bounds of the operator search, with the energy as a coefficient
that stays numerical as in Section~\ref{sec:energies}. What is missing is a rule of the solver for its solutions, and we
leave this case for future work.

\paragraph{What the grade {\normalfont\code{verified}} means.} The grade is the outcome of a numerical test on a part of the interval
that the fit did not read. It is weaker than a proof, and it says nothing about the function outside the interval. The
runs of this paper contain the following failures.
\begin{itemize}[leftmargin=*,itemsep=1pt,topsep=3pt]
\item \emph{One false certificate.} The float64 table of the generating function of the body-centred cubic lattice was
graded \code{verified} with an operator of order 3 and degree 9 in $y = 1/x$ that has two fitted parameters and is not the
operator of the target. Its continued solution is off by $1.1 \timestento{-6}$ inside the range
(Section~\ref{sec:physical}).
\item \emph{Two verified expressions that are wrong beyond a sign change.} Slices of the Feynman equations I.15.3x and
I.15.1 were returned as the positive square root of an exact rational function, which is the absolute value of the
target. Both are exact on the interval of the data and on the tested range (Section~\ref{sec:absent}).
\item \emph{Wrong conjectures.} From noise-free float64 tables, two conjectures fail the test of the benchmark: the
squared Hermite function B35 of the generated set, from 1101 points and again from 50, and the relaxation function
$E_{1/3}(-t^{1/3})$ of the physical set. Three conjectures on Feynman slices pass that test and fail further from the
data (Section~\ref{sec:absent}). All five contain fitted constants. With noise of $\tento{-1}$ one of the five
conjectures returned for the twelve tables of Section~\ref{sec:noise} fails as well.
\item \emph{An earlier version} of the learner returned five false \code{verified} certificates. They led to the length
scale of the hold-out test (Section~\ref{sec:history}).
\end{itemize}
The identification of constants has an error rate of its own. Of 300 random numbers known to a relative $\tento{-6}$,
4 received an exact form, and 20 of 300 at $\tento{-3}$ (Table~\ref{tab:calibration}).

\paragraph{Two variables.} The two-variable learner was tried on eight targets and two controls, all chosen by the
author, and it was written with those targets in view (Section~\ref{sec:two}). It tries the variables as given and two
changes of variable. Its hold-out test has no length scale and is not repeated at a second precision
(Section~\ref{sec:bivariate}). The operators that it keeps generate the operators found within the search bounds and need
not be a minimal set of generators of the annihilating ideal: for the Voigt function two of the four kept operators are
independent.

\paragraph{The comparison.} The benchmarks of Section~\ref{sec:benchmarks} have limits of their own.
\begin{itemize}[leftmargin=*,itemsep=1pt,topsep=3pt]
\item Each of the other programs ran once per target with one seed, and those with a time setting ran for 60\,s on one
thread. Runs of \VEL{} were stopped after 600 to 3600\,s (Table~\ref{tab:methods}); its median time was 8.8\,s per run on
the published suites and 196\,s per table on the generated set. Ten times the budget changed little for PySR + 14
functions on the generated set (Section~\ref{sec:generated}). We did not make that test for the other programs.
\item The scoring rule requires agreement beyond the training range, which favours exact forms.
Table~\ref{tab:one_variable} gives the counts inside the range only.
\item The published suites were in view during development. Their counts come from the entry point for tables, whose
relation search over a fixed set of elementary functions was written with the classic and Feynman targets at hand. In
the first comparison the operator search alone recovered 21, 34 and 167 targets of the three one-variable suites, where
PySR recovers 33, 32 and 174 (Section~\ref{sec:history}).
\item The physical and two-variable sets were chosen by the authors. Only the generated set was drawn after the learner
was frozen. Five development seeds of the same generator preceded it, and the generator, its families and its shapes are
our own.
\item On SRBench our runs are compared with published results that were obtained with other budgets, on other machines
and with more runs per problem (Section~\ref{sec:srbench}).
\item Programs that guess holonomic equations \cite{salvy1994gfun,kauers2015ore} need exact terms of a sequence or of a
series, which the function inputs of the benchmarks do not provide. On the sequence and moment inputs of
Section~\ref{sec:cases} the guesser of ore\_algebra returned the same recurrences as the learner
(Section~\ref{sec:orecheck}). For a callable that accepts complex arguments, Taylor coefficients could be computed by
sampling it on a circle and passed to a guesser; we did not run that comparison. The statement of
Section~\ref{sec:related} about what is new rests on a limited search of the literature.
\end{itemize}

\paragraph{Implementation.} One run of the generated set ended in an exception (target A17, an ordering comparison of
complex numbers); it is counted as a miss, and the code was left as frozen (Section~\ref{sec:generated}). Time limits
inside the solver use Unix signals, so the learner has to run in the main thread of a process on a Unix system.

\section{A check with ore\_algebra}\label{sec:orecheck}

The solver of Section~\ref{sec:solver} is a set of rules, and for 21 certified operators it returned no closed form: 15 of
the generated set (from callables) and 6 of the physical set (the Voigt profile, $\Pi(n \mid \tfrac12)$, the demagnetising
factor and the area of an ellipsoid, the period of the asymmetric oscillator and the simple cubic lattice). To see whether a
general computer-algebra package does better on the same operators, we ran ore\_algebra \cite{kauers2015ore} (source of
September 2026 on passagemath 10.8, without its compiled extensions) on each exact operator, in the variable in which the
learner found it. Each call ran in its own process with a limit of 300\,s. The learner was not changed, and no count of
the main text depends on this section.

\paragraph{Closed forms.} We called four functions on each operator.
\begin{itemize}[leftmargin=*,itemsep=1pt,topsep=3pt]
\item \code{is\_fuchsian}: the five physical operators other than the Voigt profile are Fuchsian; the Voigt operator and
all 15 generated operators have an irregular singular point.
\item \code{hyperexponential\_solutions}, the first-order right factors: for six generated targets of the form
$f + c$ it returned the factor $\partial_y$, the constant solution, which the learner's antiderivative rule already uses; for
the Voigt operator it returned $\partial_y + y \pm \tfrac{\mathrm{i}}{2}$, whose solutions $\ee^{-y^2/2 \mp \mathrm{i} y/2}$
are the factor $\ee^{-z^2}$ of the Faddeeva function $w(z) = \ee^{-z^2} \erfc(-\mathrm{i} z)$ that defines the profile, and not
the profile; for $\Pi(n \mid \tfrac12)$ it returned one first-order factor, whose solution $\sqrt{y / ((1 - y)(1 - 2 y))}$
is not the target; for the other 13 operators it returned none.
\item \code{factor}, the symbolic-numeric factorisation of Chyzak, Goyer and Mezzarobba \cite{chyzak2022factorization}: it
did not finish within 300\,s for the 15 generated operators or for the Voigt operator, none of which is Fuchsian, a case in
which the algorithm is not guaranteed to terminate. It split the operators of the ellipsoid area (1.4\,s) and of the
demagnetising factor (13.8\,s) into a first-order and a second-order factor, and that of $\Pi(n \mid \tfrac12)$ into three
factors of order one (6.2\,s). It returned the asymmetric oscillator and the simple cubic lattice as single factors.
\item \code{is\_provably\_irreducible}: \code{True} for the asymmetric oscillator (0.9\,s) and the simple cubic lattice
(5.2\,s), which supports the Heun type of both; not proven for $\Pi(n \mid \tfrac12)$, for the Voigt operator and for 7 of
the generated operators; an error at an irregular singular point for the other 8; no result within 300\,s for the two
ellipsoid operators.
\end{itemize}
The solutions of the right factor of a product solve the product, so we passed the right factors of the three split
operators to our solver. It found candidates for none of them that match the targets, which also need solutions of the
left factors. The check therefore adds no closed form and changes no count of this paper. It adds a proof that two of the
operators are irreducible.

\paragraph{Continuation.} For the 15 generated operators we compared the function \code{numerical\_solution} of
ore\_algebra, which bounds its error rigorously \cite{mezzarobba2016rigorous}, with the continuation of Section~\ref{sec:continuation}. Both started from
the same initial values at the middle of the interval, computed from the target at 60 digits, and were evaluated at the upper
end of the interval and at the far end of the benchmark's grid, half a range-width beyond it. Against the target, the error
of ore\_algebra is between $1.6 \timestento{-44}$ and $3.1 \timestento{-41}$ in 0.03 to 1.5\,s, and ours between
$2.5 \timestento{-44}$ and $2.7 \timestento{-33}$ in 0.07 to 2.0\,s. The certified radius of ore\_algebra, between
$1.8 \timestento{-44}$ and $8.6 \timestento{-41}$, bounds the continuation from the given initial values and not the error
of those values, which is why the error exceeds the radius at 16 of the 30 points. For two operators with a real singular
point between the centre and the end, the path had to pass the point explicitly on a half circle, as our continuation does.

\paragraph{Guessing.} For the 26 sequence and moment inputs of the regression cases we gave the function \code{guess} of
ore\_algebra the same 30 exact terms that the learner read. For the 22 inputs on which the learner returned a recurrence, \code{guess}
returned the same recurrence up to a constant factor, in at most 0.61\,s. For the four inputs without one (the primes, the
partition numbers, Euler's totient and the Bell numbers) it found none either.

\paragraph{What this means for the method.} ore\_algebra starts where step 2 of the method ends: from an exact operator, or
from exact terms. On those inputs it agrees with the learner, its continuation is more accurate and certified, and it can
prove irreducibility, but on the 21 operators it did not supply a closed form that our rules miss. Its dependence on Sage
makes it a heavy requirement for the learner, so we have not integrated it; it is a natural optional back end for the
continuation of certified operators and for factorisation. The scripts and records of this check are kept with the
source of the paper.

\newcommand{\velcaption}{}

\section{The bank of changes of variable}\label{app:bank}

Table~\ref{tab:bank} lists the maps of Section~\ref{sec:transport}. A map is tried when its domain contains the interval
of the data, with the weights $k = 0$ and $k = 1$ of equation~\eqref{eq:transport}.

\begin{table}[h]
\centering\small
\caption{The bank of twenty changes of variable $y = \varphi(x)$, the domain of $x$ on which each is used, and its cost
in the ranking~\eqref{eq:cost}.}
\label{tab:bank}
\begin{tabular}{@{}llc@{\hspace{2.5em}}llc@{}}
\toprule
Map $y = \varphi(x)$ & Domain of $x$ & Cost & Map $y = \varphi(x)$ & Domain of $x$ & Cost \\
\midrule
$y = x$ & $(-\infty,\,\infty)$ & 0 & $y = x/(1-x)$ & $(-\infty,\,1) \cup (1,\,\infty)$ & 1 \\
$y = \log x$ & $(0,\,\infty)$ & 1 & $y = \logit x = \log\bigl(x/(1-x)\bigr)$ & $(0,\,1)$ & 2 \\
$y = \ee^{x}$ & $(-\infty,\,\infty)$ & 1 & $y = \artanh x$ & $(-1,\,1)$ & 2 \\
$y = x^{2}$ & $(0,\,\infty)$ & 1 & $y = \arsinh x$ & $(-\infty,\,\infty)$ & 2 \\
$y = x^{3}$ & $(0,\,\infty)$ & 1 & $y = \probit x = \sqrt{2}\,\erf^{-1}(2x-1)$ & $(0,\,1)$ & 2 \\
$y = x^{1/2}$ & $(0,\,\infty)$ & 1 & $y = \tan x$ & $(-\pi/2,\,\pi/2)$ & 2 \\
$y = x^{1/3}$ & $(0,\,\infty)$ & 1 & $y = \arctan x$ & $(-\infty,\,\infty)$ & 2 \\
$y = x^{3/2}$ & $(0,\,\infty)$ & 1 & $y = \arcsin x$ & $(-1,\,1)$ & 2 \\
$y = x^{2/3}$ & $(0,\,\infty)$ & 1 & $y = \sinh x$ & $(-\infty,\,\infty)$ & 2 \\
$y = 1/x$ & $(-\infty,\,0) \cup (0,\,\infty)$ & 1 & $y = \tanh x$ & $(-\infty,\,\infty)$ & 2 \\
\bottomrule
\end{tabular}

\end{table}

\section{Software and data}\label{app:software}

\paragraph{Versions.} \VEL{} ran on Python 3.12 with mpmath 1.3.0, SymPy 1.14.0, NumPy 2.4.6 and SciPy 1.18.1. The
other programs were the following.
\begin{itemize}[leftmargin=*,itemsep=1pt,topsep=3pt]
\item PySR 2.7.2, with SymbolicRegression.jl 2.7.2 on Julia 1.13.4 and SpecialFunctions.jl 2.9.2 for the added
functions.
\item pyoperon 0.6.2, gplearn 0.4.3, bingo-nasa 0.5.7 and rils-rols 1.6.7.
\item flash-ansr 0.19.1 with the pretrained model \nolinkurl{psaegert/flash-ansr-v25.0-T8-20M}.
\item E2E: the repository \nolinkurl{facebookresearch/symbolicregression} at commit \code{c4144c9}, with the checkpoint
\code{model1.pt}.
\item TPSR: the repository \nolinkurl{deep-symbolic-mathematics/TPSR} at commit \code{bdff709}.
\end{itemize}
The settings of Operon, Bingo, gplearn, RILS-ROLS and E2E, the wrapper of TPSR, the scoring code of
Section~\ref{sec:srbench} and the published results of Table~\ref{tab:srbench} are those of the repository
\nolinkurl{cavalab/srbench} at commit \code{dc3f6da} of its branch \nolinkurl{srbench_2025}. The neural programs ran on
CPU with PyTorch 2.14.1.

\paragraph{Runs.} All runs were made between 18 and 20 November 2026 on four machines with 12 to 32 cores, with one
thread per process.

\paragraph{Benchmark definitions.} The classic targets are the rows with one variable of \code{benchmarks.csv}, a file
of the deep symbolic optimization repository \cite{petersen2021dsr}. The authors' files of the Feynman equations
\cite{udrescu2020aifeynman} were not reachable when the benchmarks were set up. The 100 equations were taken from a
mirror and compared with the metadata of the Penn Machine Learning Benchmarks \cite{romano2022pmlb}: the formula and the
ranges of 99 equations agree, and II.11.17 has no data set there under that name. The 20 bonus equations were read from
that metadata. Every data set was generated from its formula. The data sets of Section~\ref{sec:srbench} are those of
the Penn Machine Learning Benchmarks.

\section{The formulas returned for the Fresnel integral}\label{app:fresnel}

The seven formulas of Figure~\ref{fig:fresnel_fits} and Table~\ref{tab:fresnel}, as each program printed them for the
1101 float64 points of $C(x)$ on $[0.2,\,2]$. The variable is written \code{X1}, \code{x0}, \code{X\_0}, \code{x1},
\code{x\_0} or \code{X0}.


\noindent\begin{minipage}[t]{\linewidth}
\noindent\textbf{Operon}
\begin{lstlisting}
(0.000000125 + (0.999999762 * ((cos((3.496690989 * X1)) / (cos(((cos((1.513441682 * X1)) / exp((1.282563090 * X1))) - (2.349471331 + (2.207667589 * X1)))) - 4.305118084)) - (((-0.368798584) + ((-0.158383533) * X1)) + (log((42.630218506 * X1)) * (cos(exp((1.091885924 * X1))) / (85.690597534 * X1)))))))
\end{lstlisting}
\end{minipage}\par

\noindent\begin{minipage}[t]{\linewidth}
\noindent\textbf{RILS-ROLS}
\begin{lstlisting}
1.055034*sin(x0)*cos(x0**2) - 0.689068*cos(x0**2) + 0.672111
\end{lstlisting}
\end{minipage}\par

\noindent\begin{minipage}[t]{\linewidth}
\noindent\textbf{PySR}
\begin{lstlisting}
sin(1.293268038178877*sin(sqrt(x0 - cos(cos(2*x0 + log(x0 + 0.15918748681300737))) + 1.7189315973901353)) - 1.293268038178877*cos(sin(2*x0 - 0.3192569350292012)) + 0.3346267909056614)
\end{lstlisting}
\end{minipage}\par

\noindent\begin{minipage}[t]{\linewidth}
\noindent\textbf{Bingo}
\begin{lstlisting}
0.7549593188478704 + (1.9902988474237653)*((-0.619568464883273 + (sin(X_0))*(0.3448095545514621 + X_0 + (-2.113835900246371)*(sin(X_0))))*(0.35096750328067633 + ((X_0)**(-1))*((sin(X_0))*(0.3448095545514621 + X_0 + (-2.113835900246371)*(sin(X_0))))))
\end{lstlisting}
\end{minipage}\par

\noindent\begin{minipage}[t]{\linewidth}
\noindent\textbf{Flash-ANSR}
\begin{lstlisting}
(pow(x1, 2) + 0.12 - 1.4 * (x1 * sin(x1 * (x1 - 0.997)))) / cosh(0.9 * x1)
\end{lstlisting}
\end{minipage}\par

\noindent\begin{minipage}[t]{\linewidth}
\noindent\textbf{E2E}
\begin{lstlisting}
((0.039227331370364646 + (0.02268610950825978 * cos(((-0.476143579171859 + (-1.2658648260338818 * (-2.116951860213973 + (1.9245016911036117 * x_0)))) - (0.43284357917180205 + (0.4056648260340491 * (-2.116951860213973 + (1.9245016911036117 * x_0)))))))) * ((18.776749565117353 + (2.0887116771274212 * (-2.116951860213973 + (1.9245016911036117 * x_0)))) - (4.5332504348826514 + (-2.020311677127389 * (-2.116951860213973 + (1.9245016911036117 * x_0))))))
\end{lstlisting}
\end{minipage}\par

\noindent\begin{minipage}[t]{\linewidth}
\noindent\textbf{gplearn}
\begin{lstlisting}
mul(sin(sqrt(sin(sin(add(add(0.532, X0), add(-0.701, X0)))))), sin(exp(cos(X0))))
\end{lstlisting}
\end{minipage}\par

\velneed{0.5\textheight}
\section{The generated set, target by target}\label{app:generated}

\renewcommand{\velcaption}{The 132 targets of the generated set (Section~\ref{sec:generated}). \emph{Order, degree}: of
the operator that the generator computed for the target from the operators of its members; the learner can return
another operator, as for C18, where it found a two-term operator of order 6. \emph{Callable}: \qq{closed form} when the
closed form returned from the callable passes the test of Section~\ref{sec:protocol}; \qq{continuation} when none does
and the continued solution of the verified operator passes; \qq{error} for the run that ended in an exception.
\emph{Float64 table}: \qq{recovered} when the expression returned from 1101 float64 points passes the test, \qq{wrong}
when it fails. \emph{Other programs}: those that recovered the target from the same points, with P14 for PySR + 14
functions, P for PySR, B for Bingo, F for Flash-ANSR, R for RILS-ROLS, O for Operon, G for gplearn and E for E2E; TPSR
recovered none. In the formulas $K(m)$ and $E(m)$ are complete elliptic integrals of the parameter $m$, $E_\nu$ is the
generalised exponential integral, $H_\nu$ the Hermite function, $\mathbf{H}_\nu$ the Struve function, $D_\nu$ the
parabolic cylinder function, $P_\nu$ and $Q_\nu$ are Legendre functions, $C_\nu^{(\lambda)}$ is the Gegenbauer function, $W_{\kappa,\mu}$ and $M_{\kappa,\mu}$ Whittaker
functions, $F_\ell(\eta, \rho)$ and $G_\ell(\eta, \rho)$ Coulomb functions, $U(a, b, x)$ is the confluent hypergeometric
function of the second kind, $\Gamma(a, x)$ the incomplete gamma function, $C$ the Fresnel integral and $W(x)$
Lambert's function \cite{dlmf}.}
{\footnotesize
\begin{longtable}{@{}l@{\hspace{0.6em}}L{55mm}@{\hspace{0.6em}}l@{\hspace{0.6em}}c@{\hspace{0.6em}}l@{\hspace{0.6em}}l@{\hspace{0.6em}}l@{}}
\caption{\velcaption}\label{tab:generated}\\
\toprule
& & & & \multicolumn{2}{@{}c@{\hspace{0.6em}}}{\VEL{}} & \\
\cmidrule(r{0.6em}){5-6}
& & & Order, & & float64 & Other \\
Target & Formula & Interval & degree & callable & table & programs \\
\midrule
\endfirsthead
\toprule
& & & & \multicolumn{2}{@{}c@{\hspace{0.6em}}}{\VEL{}} & \\
\cmidrule(r{0.6em}){5-6}
& & & Order, & & float64 & Other \\
Target & Formula & Interval & degree & callable & table & programs \\
\midrule
\endhead
\midrule
\multicolumn{7}{r@{}}{continued on the next page} \\
\endfoot
\bottomrule
\endlastfoot
\multicolumn{7}{@{}l}{\textit{Tier E: elementary (12 targets)}} \\*
\addlinespace[2pt]
E01 & $\frac{1}{2} + e^{- x}$ & $[1/4,\,2]$ & 2, 0 & closed form & recovered & P, B, R \\
E02 & $\dfrac{e^{- x^{2}}}{\left(x^{2} + 1\right)^{3/2}}$\rule[-2.0ex]{0pt}{5.0ex} & $[1/10,\,3/5]$ & 1, 3 & closed form & recovered & -- \\
E03 & $\sqrt{x} + \dfrac{e^{- x}}{x}$\rule[-2.0ex]{0pt}{5.0ex} & $[2,\,10]$ & 2, 3 & closed form & recovered & P14, P, B, F \\
E04 & $\sqrt{x} + e^{- x^{2}}$ & $[1/2,\,4]$ & 2, 4 & closed form & recovered & P, R \\
E05 & $e^{- x^{2}/2} \cos\!\left(3 x/2\right)$ & $[1/10,\,3/5]$ & 2, 2 & closed form & recovered & -- \\
E06 & $x^{3/2} e^{- x}$ & $[1/10,\,3/5]$ & 1, 1 & closed form & recovered & -- \\
E07 & $e^{- x} \log\!\left(x\right)$ & $[1/4,\,2]$ & 2, 1 & closed form & recovered & P14, P, B, F, R, O, G \\
E08 & $e^{- 2 x^{2}} + \dfrac{1}{\sqrt{3 x^{2} + 1}}$\rule[-2.0ex]{0pt}{5.0ex} & $[{-}1/2,\,2/5]$ & 2, 6 & closed form & no model & -- \\
E09 & $\dfrac{e^{- x}}{x^{2}/2 + 1} - 1$\rule[-2.0ex]{0pt}{5.0ex} & $[1/10,\,3/5]$ & 2, 4 & closed form & no model & -- \\
E10 & $\log\!\left(x\right) \cos\!\left(x\right) + e^{- x/2}$ & $[2,\,10]$ & 5, 6 & no model & no model & P, R \\
E11 & $\sin\!\left(2 x\right) + \frac{1}{2}$ & $[{-}2,\,1]$ & 3, 0 & closed form & recovered & P14, P, B, R, O \\
E12 & $1 + \dfrac{e^{- x^{2}}}{\left(x^{2}/4 + 1\right)^{3/2}}$\rule[-2.0ex]{0pt}{5.0ex} & $[1/10,\,3/5]$ & 2, 6 & closed form & no model & P \\
\addlinespace
\multicolumn{7}{@{}l}{\textit{Tier A: operator of order 1 to 2 (36 targets)}} \\*
\addlinespace[2pt]
A01 & $E\!\left(2 \log\!\left(x\right)\right) e^{- 2 \log^{2}\!\left(x\right)}$ & $[1/10,\,3/5]$ & 2, 4 & closed form & no model & -- \\
A02 & $\dfrac{J_{1}\!\left(1/x\right)}{\sqrt{x^{2}/4 + 1}}$\rule[-2.0ex]{0pt}{5.0ex} & $[{-}5,\,{-}2]$ & 2, 8 & closed form & no model & -- \\
A03 & $\Ei\!\left(x\right)$ & $[1/10,\,3/5]$ & 2, 1 & closed form & recovered & P14 \\
A04 & $E\!\left(x\right)$ & $[1/10,\,3/5]$ & 2, 3 & closed form & recovered & P14 \\
A05 & $\left(2 x^{2} + 1\right)^{3/2} \Bi\!\left(- x\right)$ & $[1,\,6]$ & 2, 5 & closed form & no model & -- \\
A06 & $K\!\left(2 \log\!\left(x\right)\right)$ & $[1/10,\,3/5]$ & 2, 2 & closed form & recovered & -- \\
A07 & $\Bi\!\left(2 x\right)$ & $[1/4,\,2]$ & 2, 1 & closed form & recovered & P14 \\
A08 & $J_{1}\!\left(x\right)$ & $[1,\,6]$ & 2, 2 & closed form & recovered & P14 \\
A09 & $K\!\left(x\right)$ & $[1/20,\,1/2]$ & 2, 2 & closed form & recovered & P14 \\
A10 & $K_{0}\!\left(x\right)$ & $[1/4,\,2]$ & 2, 1 & closed form & recovered & P14 \\
A11 & $J_{1}\!\left(e^{x}/2\right)$ & $[{-}3,\,2]$ & 2, 2 & closed form & recovered & -- \\
A12 & $\dfrac{K_{0}\!\left(x\right)}{\left(2 x^{2} + 1\right)^{3/2}}$\rule[-2.0ex]{0pt}{5.0ex} & $[1/20,\,1/2]$ & 2, 5 & closed form & no model & -- \\
A13 & $\dfrac{J_{13/5}\!\left(1/x\right)/2 - J_{3/5}\!\left(1/x\right)/2}{x^{2}}$\rule[-2.0ex]{0pt}{5.0ex} & $[1/20,\,1/2]$ & 2, 6 & continuation & no model & -- \\
A14 & $- \frac{1}{4}\,K_{1/6}\!\left(x/2\right) - \frac{1}{4}\,K_{13/6}\!\left(x/2\right)$ & $[1/4,\,2]$ & 2, 4 & closed form & recovered & -- \\
A15 & $W_{7/5,\,4/5}\!\left(1/\log\!\left(x\right)\right)$ & $[2,\,10]$ & 2, 4 & continuation & no model & -- \\
A16 & $\frac{27}{4}\,x \,\pFq{2}{1}\!\left(7/4, 14/5;\,7/5;\,x^{2}\right)$ & $[1/20,\,1/2]$ & 2, 4 & closed form & recovered & -- \\
A17 & $\dfrac{\left(- x\right)^{5/4} I_{5/4}\!\left(x\right) e^{- x}}{x^{5/4}}$\rule[-2.0ex]{0pt}{5.0ex} & $[{-}4,\,{-}2]$ & 2, 2 & error & recovered & -- \\
A18 & $C_{4/3}^{(7/4)}\!\left(- x\right) e^{- x}$ & $[{-}1/2,\,2/5]$ & 2, 2 & closed form & recovered & -- \\
A19 & $\dfrac{J_{7/5}\!\left(1/x\right)/2 - J_{- 3/5}\!\left(1/x\right)/2}{x^{2}}$\rule[-2.0ex]{0pt}{5.0ex} & $[1/10,\,3/5]$ & 2, 6 & continuation & recovered & -- \\
A20 & $\,\pFq{2}{1}\!\left(2/5, 4/3;\,3/4;\,x\right)$ & $[1/10,\,3/5]$ & 2, 2 & closed form & recovered & -- \\
A21 & $- \frac{1}{22}\,\,\pFq{2}{1}\!\left(5/4, 4/3;\,17/6;\,- \log\!\left(x\right)\right)$ & $[2,\,10]$ & 2, 2 & closed form & recovered & B, F \\
A22 & $\dfrac{\operatorname{erfi}\!\left(x\right)}{\left(3 x^{2} + 1\right)^{3/2}}$\rule[-2.0ex]{0pt}{5.0ex} & $[{-}4/5,\,1/5]$ & 2, 5 & closed form & no model & -- \\
A23 & $2\,\sqrt{x} \bigl(Y_{- 3/5}\!\left(x\right)/2 - Y_{7/5}\!\left(x\right)/2\bigr)$ & $[1/10,\,3/5]$ & 2, 8 & continuation & recovered & -- \\
A24 & $J_{5/4}\!\left(x\right)$ & $[1/2,\,4]$ & 2, 2 & closed form & recovered & -- \\
A25 & $H_{3/4}\!\left(- \log\!\left(x\right)\right)$ & $[1/20,\,1/2]$ & 2, 1 & closed form & recovered & -- \\
A26 & $E_{1/4}\!\left(- x\right)$ & $[{-}5,\,{-}2]$ & 2, 1 & closed form & recovered & -- \\
A27 & $\dfrac{J_{9/4}\!\left(1/x\right)/2 - J_{1/4}\!\left(1/x\right)/2}{x^{2}}$\rule[-2.0ex]{0pt}{5.0ex} & $[1/2,\,4]$ & 2, 6 & continuation & no model & -- \\
A28 & $\dfrac{K_{- 1/4}\!\left(1/x\right)/2 + K_{7/4}\!\left(1/x\right)/2}{x^{2}}$\rule[-2.0ex]{0pt}{5.0ex} & $[1/4,\,2]$ & 2, 6 & continuation & recovered & -- \\
A29 & $\sqrt{x} Q_{4/5}\!\left(x/2\right)$ & $[1/10,\,3/5]$ & 2, 4 & closed form & recovered & -- \\
A30 & $\dfrac{Y_{9/4}\!\left(1/x\right)/2 - Y_{1/4}\!\left(1/x\right)/2}{x^{2}}$\rule[-2.0ex]{0pt}{5.0ex} & $[1/20,\,1/2]$ & 2, 6 & continuation & recovered & -- \\
A31 & $I_{1/6}\!\left(x^{2}\right) e^{- x/2}$ & $[1/10,\,3/5]$ & 2, 4 & closed form & no model & -- \\
A32 & $W_{5/6,\,2/5}\!\left(1/x\right)$ & $[1/10,\,3/5]$ & 2, 4 & closed form & recovered & -- \\
A33 & $\dfrac{I_{3/5}\!\left(1/x\right)}{\left(x^{2}/4 + 1\right)^{3/2}}$\rule[-2.0ex]{0pt}{5.0ex} & $[2,\,10]$ & 2, 8 & closed form & no model & -- \\
A34 & $M_{2/5,\,2/3}\!\left(- x\right)$ & $[{-}4,\,{-}2]$ & 2, 2 & closed form & recovered & -- \\
A35 & $\dfrac{\,\pFq{1}{1}\!\left(7/6;\,2/3;\,x/2\right)}{\sqrt{x}}$\rule[-2.0ex]{0pt}{5.0ex} & $[1/4,\,2]$ & 2, 2 & closed form & recovered & -- \\
A36 & $C_{7/4}^{(4/3)}\!\left(x/2\right)$ & $[{-}1/2,\,2/5]$ & 2, 2 & closed form & recovered & -- \\
\addlinespace
\multicolumn{7}{@{}l}{\textit{Tier B: operator of order 3 to 4 (36 targets)}} \\*
\addlinespace[2pt]
B01 & $\operatorname{Si}\!\left(x^{2}\right) e^{x/2}$ & $[{-}4/5,\,1/5]$ & 3, 4 & closed form & no model & P14 \\
B02 & $\Bi\!\left(x\right) \erf\!\left(x^{2}\right)$ & $[{-}4,\,{-}2]$ & 4, 15 & no model & no model & -- \\
B03 & $K\!\left(x\right)^{2}$ & $[1/20,\,1/2]$ & 3, 4 & closed form & no model & P14 \\
B04 & $- \Ei\!\left(x\right) \erf\!\left(x\right)$ & $[1/20,\,1/2]$ & 4, 8 & continuation & no model & P14 \\
B05 & $\Bi\!\left(1/x\right)^{2}$ & $[{-}5,\,{-}2]$ & 3, 6 & closed form & recovered & -- \\
B06 & $J_{0}\!\left(\sqrt{x}\right) + 2$ & $[1/4,\,2]$ & 3, 2 & closed form & no model & P14, B, F \\
B07 & $K_{0}\!\left(x/2\right)^{2}$ & $[1/10,\,3/5]$ & 3, 2 & closed form & recovered & -- \\
B08 & $J_{1}\!\left(2 x\right) K_{0}\!\left(- x\right)$ & $[{-}4,\,{-}2]$ & 4, 6 & continuation & no model & -- \\
B09 & $Y_{0}\!\left(- x\right) + \frac{1}{2}$ & $[{-}5,\,{-}2]$ & 3, 2 & continuation & no model & -- \\
B10 & $I_{0}\!\left(2 \log\!\left(x\right)\right) + \frac{1}{2}$ & $[2,\,10]$ & 3, 2 & no model & no model & -- \\
B11 & $\Ei\!\left(2 x\right)^{2}$ & $[1/4,\,2]$ & 3, 2 & closed form & recovered & -- \\
B12 & $E\!\left(- x\right)^{2}$ & $[1/20,\,1/2]$ & 3, 4 & closed form & recovered & P14, P, B, F \\
B13 & $2 + \,\pFq{1}{1}\!\left(1/3;\,3/5;\,x/2\right)$ & $[1/10,\,3/5]$ & 3, 1 & closed form & no model & P14, P, B, F, O \\
B14 & $F_{1/3}\!\left(7/6, 1/x\right) + 1$ & $[1/4,\,2]$ & 3, 6 & continuation & no model & -- \\
B15 & $G_{4/5}\!\left(5/4, x^{2}\right)^{2}$ & $[{-}4,\,{-}2]$ & 3, 5 & continuation & no model & -- \\
B16 & $\left(3 x^{2} + 1\right)^{3/2} \operatorname{ker}_{1/3}\!\left(1/x\right)$ & $[1/20,\,1/2]$ & 4, 16 & no model & no model & -- \\
B17 & $\dfrac{\operatorname{Chi}\!\left(- \log\!\left(x\right)\right)}{\log^{2}\!\left(x\right)/2 + 1}$\rule[-2.0ex]{0pt}{5.0ex} & $[1/10,\,3/5]$ & 3, 3 & no model & no model & -- \\
B18 & $C_{8/5}^{(1/6)}\!\left(1/\sqrt{x}\right)^{2}$ & $[2,\,10]$ & 3, 7 & closed form & no model & -- \\
B19 & $G_{9/5}\!\left(3/4, 1/x\right) + 2$ & $[1/2,\,4]$ & 3, 6 & continuation & no model & -- \\
B20 & $F_{9/5}\!\left(3/5, x\right) + \frac{1}{2}$ & $[1/2,\,4]$ & 3, 4 & continuation & no model & -- \\
B21 & $F\!\left(\arcsin\!\left(x\right)\,|\,1/5\right)^{2}$ & $[{-}4/5,\,1/5]$ & 3, 4 & closed form & no model & -- \\
B22 & $D_{6/5}\!\left(- \log\!\left(x\right)\right) + \frac{1}{2}$ & $[1/4,\,2]$ & 3, 4 & no model & no model & -- \\
B23 & $\,\pFq{0}{2}\!\left(;\,1/6, 8/5;\,x^{2}\right) - \frac{1}{2}$ & $[{-}2,\,1]$ & 4, 3 & closed form & no model & -- \\
B24 & $F_{5/3}\!\left(6/5, x^{2}\right) - 1$ & $[{-}2,\,1]$ & 3, 8 & no model & no model & -- \\
B25 & $E_{3/4}\!\left(x\right)^{2}$ & $[1/4,\,2]$ & 3, 2 & closed form & recovered & -- \\
B26 & $F_{3/5}\!\left(7/4, - x\right) - 1$ & $[{-}4,\,{-}2]$ & 3, 4 & continuation & no model & -- \\
B27 & $W_{1/6,\,11/6}\!\left(- x\right) + 2$ & $[{-}5,\,{-}2]$ & 3, 4 & closed form & no model & -- \\
B28 & $U\!\left(7/6, 4/5, x/2\right) + 1$ & $[1/2,\,4]$ & 3, 1 & closed form & recovered & -- \\
B29 & $\,\pFq{1}{1}\!\left(1/4;\,1/6;\,1/x\right)^{2}$ & $[{-}5,\,{-}2]$ & 3, 5 & closed form & recovered & -- \\
B30 & $\,\pFq{2}{1}\!\left(1/6, 7/4;\,11/6;\,x^{2}\right) - \frac{1}{2}$ & $[1/10,\,3/5]$ & 3, 4 & closed form & no model & -- \\
B31 & $\frac{9}{154}\,\,\pFq{3}{2}\!\left(7/6, 11/5, 9/4;\,13/6, 17/6;\,x/2\right)$ & $[{-}5,\,{-}2]$ & 3, 3 & closed form & no model & -- \\
B32 & $\dfrac{\sqrt{2} \sqrt{\pi}}{2}\,C\!\left(\sqrt{2} x/\sqrt{\pi}\right)$\rule[-2.0ex]{0pt}{5.0ex} & $[{-}4,\,{-}2]$ & 3, 3 & closed form & recovered & -- \\
B33 & $U\!\left(7/5, 3/5, e^{x}/2\right)^{2}$ & $[{-}5,\,{-}2]$ & 3, 2 & no model & recovered & -- \\
B34 & $\Gamma\!\left(3/5, - x\right)^{2}$ & $[{-}4,\,{-}2]$ & 3, 2 & closed form & recovered & -- \\
B35 & $H_{4/5}\!\left(x^{2}\right)^{2}$ & $[{-}2,\,1]$ & 3, 8 & closed form & wrong & -- \\
B36 & $D_{11/6}\!\left(\log\!\left(x\right)/2\right) - \frac{1}{2}$ & $[1/4,\,2]$ & 3, 4 & no model & no model & -- \\
\addlinespace
\multicolumn{7}{@{}l}{\textit{Tier C: operator of order 5 to 8 (36 targets)}} \\*
\addlinespace[2pt]
C01 & $J_{1}\!\left(1/x\right) E\!\left(x\right) + \erf\!\left(x/2\right)$ & $[1/20,\,1/2]$ & 6, 35 & no model & no model & -- \\
C02 & $\Ai\!\left(1/x\right) \Ai\!\left(x\right) + 1$ & $[{-}4,\,{-}2]$ & 5, 24 & no model & no model & -- \\
C03 & $\Ei\!\left(x\right) + \operatorname{Si}\!\left(x/2\right) Y_{0}\!\left(2 x\right)$ & $[1/20,\,1/2]$ & 8, 15 & no model & no model & -- \\
C04 & $\Ei\!\left(x/2\right) \Ai\!\left(x/2\right) + \Bi\!\left(x^{2}\right)$ & $[1/20,\,1/2]$ & 6, 32 & no model & no model & -- \\
C05 & $\operatorname{Si}\!\left(1/x\right) \Bi\!\left(1/x\right)$ & $[1/2,\,4]$ & 6, 22 & no model & no model & -- \\
C06 & $J_{0}\!\left(\log\!\left(x\right)/2\right) J_{1}\!\left(\log^{2}\!\left(x\right)\right) + 2$ & $[1/10,\,3/5]$ & 5, 20 & no model & no model & -- \\
C07 & $J_{0}\!\left(\sqrt{x}/2\right) \erf\!\left(1/\sqrt{x}\right) + 2$ & $[1/4,\,2]$ & 5, 20 & no model & no model & -- \\
C08 & $- \operatorname{Si}\!\left(\log\!\left(x\right)\right) - \Ai\!\left(\log^{2}\!\left(x\right)\right) J_{1}\!\left(\log\!\left(x\right)\right)$ & $[1/10,\,3/5]$ & 7, 49 & no model & no model & -- \\
C09 & $I_{0}\!\left(x/2\right) E\!\left(- x\right) + 1$ & $[1,\,6]$ & 5, 11 & no model & no model & -- \\
C10 & $\operatorname{Ci}\!\left(\sqrt{x}\right) J_{1}\!\left(\sqrt{x}\right)$ & $[1,\,6]$ & 6, 8 & no model & no model & -- \\
C11 & $\Ei\!\left(1/\sqrt{x}\right) E\!\left(\sqrt{x}/2\right) + 1$ & $[1/20,\,1/2]$ & 5, 14 & no model & no model & -- \\
C12 & $\Ai\!\left(x/2\right) E\!\left(- x\right) + 2$ & $[1/20,\,1/2]$ & 5, 17 & no model & no model & P14, P, B, F, O \\
C13 & $\operatorname{Shi}\!\left(x/2\right) + \,\pFq{2}{1}\!\left(7/5, 11/6;\,4/5;\,x^{2}\right)$ & $[1/20,\,1/2]$ & 5, 12 & no model & no model & -- \\
C14 & $\Ei\!\left(x^{2}\right) \,\pFq{1}{2}\!\left(1/6;\,2/5, 5/3;\,x\right) - \frac{1}{2}$ & $[1/10,\,3/5]$ & 7, 26 & no model & no model & -- \\
C15 & $\operatorname{bei}_{5/3}\!\left(x/2\right) + Y_{0}\!\left(1/x\right)$ & $[1/10,\,3/5]$ & 6, 24 & no model & no model & -- \\
C16 & $\operatorname{bei}_{6/5}\!\left(- x\right) + \operatorname{Li}_{3}\!\left(x\right)$ & $[{-}5,\,{-}2]$ & 8, 27 & no model & no model & -- \\
C17 & $\operatorname{erfi}\!\left(1/x\right) \,\pFq{2}{2}\!\left(5/6, 11/6;\,6/5, 7/4;\,2 x\right)$ & $[{-}4,\,{-}2]$ & 6, 26 & no model & no model & -- \\
C18 & $\operatorname{ker}_{7/5}\!\left(- x\right) - 1$ & $[{-}4,\,{-}2]$ & 5, 8 & closed form & no model & -- \\
C19 & $P_{3/5}\!\left(2 x\right) \,\pFq{0}{2}\!\left(;\,1/3, 4/3;\,1/x\right) - 1$ & $[2,\,10]$ & 7, 36 & no model & no model & -- \\
C20 & $Y_{6/5}\!\left(- x\right) + \,\pFq{0}{3}\!\left(;\,1/5, 7/6, 5/3;\,2 x\right)$ & $[{-}5,\,{-}2]$ & 6, 13 & no model & no model & -- \\
C21 & $I_{3/5}\!\left(x/2\right) + \operatorname{kei}_{5/6}\!\left(x^{2}\right)$ & $[1/4,\,2]$ & 6, 26 & no model & no model & -- \\
C22 & $\operatorname{Chi}\!\left(\sqrt{x}/2\right) + Q_{8/5}\!\left(\sqrt{x}/2\right)$ & $[1/4,\,2]$ & 5, 9 & no model & no model & -- \\
C23 & $\operatorname{Chi}\!\left(1/x\right) \Bi\!\left(x/2\right)$ & $[1/2,\,4]$ & 6, 26 & no model & no model & -- \\
C24 & $\operatorname{bei}_{3/4}\!\left(x\right) - 1$ & $[1/2,\,4]$ & 5, 8 & closed form & no model & -- \\
C25 & $\operatorname{Chi}\!\left(x/2\right) H_{5/3}\!\left(- x\right)$ & $[1/20,\,1/2]$ & 6, 15 & no model & no model & -- \\
C26 & $J_{1}\!\left(e^{x}\right) G_{3/5}\!\left(11/6, e^{- x}\right) + K_{1/6}\!\left(e^{2 x}\right)$ & $[1/4,\,2]$ & 6, 40 & no model & no model & -- \\
C27 & $W_{5/6,\,7/6}\!\left(1/x\right) \,\pFq{0}{2}\!\left(;\,2/5, 7/6;\,- x\right) + 1$ & $[1/4,\,2]$ & 7, 28 & no model & no model & -- \\
C28 & $Y_{2/3}\!\left(x^{2}\right) P_{3/4}\!\left(x/2\right) + 2$ & $[1/20,\,1/2]$ & 5, 32 & no model & no model & -- \\
C29 & $C_{4/3}^{(4/5)}\!\left(- x\right) + \operatorname{kei}_{7/4}\!\left(2 x\right)$ & $[1/20,\,1/2]$ & 6, 22 & no model & no model & -- \\
C30 & $H_{3/5}\!\left(1/x\right) + U\!\left(3/4, 2/3, x^{2}\right) U\!\left(5/4, 1/4, x\right)$ & $[1/20,\,1/2]$ & 6, 31 & no model & no model & -- \\
C31 & $\mathbf{H}_{7/5}\!\left(x/2\right)^{2}$ & $[1/2,\,4]$ & 6, 6 & no model & no model & -- \\
C32 & $\Ei\!\left(- \sqrt{x}\right) D_{5/3}\!\left(2 \sqrt{x}\right) - \frac{1}{2}$ & $[1/10,\,3/5]$ & 5, 16 & no model & no model & -- \\
C33 & $F_{1/3}\!\left(5/6, - x\right) \Gamma\!\left(7/5, - x\right) + \frac{1}{2}$ & $[{-}4,\,{-}2]$ & 5, 10 & no model & no model & -- \\
C34 & $\,\pFq{0}{3}\!\left(;\,1/4, 5/3, 9/5;\,x/2\right) - 1$ & $[1/2,\,4]$ & 5, 3 & closed form & no model & P14, P, B, F, E \\
C35 & $G_{1/3}\!\left(2/3, 1/x\right) \operatorname{erfi}\!\left(1/x\right) + 2$ & $[2,\,10]$ & 5, 20 & no model & no model & B \\
C36 & $\operatorname{Chi}\!\left(1/x\right) E_{8/5}\!\left(x^{2}\right) + Y_{0}\!\left(1/x\right)$ & $[1/4,\,2]$ & 8, 58 & no model & no model & -- \\
\addlinespace
\multicolumn{7}{@{}l}{\textit{Tier N: no operator in $x$ (12 targets)}} \\*
\addlinespace[2pt]
N01 & $\Gamma\!\left(x\right)$ & $[1/2,\,4]$ & -- & no model & no model & P14 \\
N02 & $\dfrac{1}{\Gamma\!\left(x\right)}$\rule[-2.0ex]{0pt}{5.0ex} & $[1/2,\,4]$ & -- & no model & no model & -- \\
N03 & $\log\!\left(\Gamma\!\left(x\right)\right)$ & $[1/2,\,4]$ & -- & no model & no model & P14 \\
N04 & $\psi\!\left(x\right)$ & $[1/2,\,4]$ & -- & no model & no model & P14 \\
N05 & $\zeta\!\left(x\right)$ & $[3/2,\,5]$ & -- & no model & no model & -- \\
N06 & $W\!\left(x\right)$ & $[1/5,\,3]$ & -- & no model & recovered & -- \\
N07 & $\erf^{-1}\!\left(x\right)$ & $[{-}1/2,\,2/5]$ & -- & no model & no model & -- \\
N08 & $x^{x}$ & $[1/5,\,5/2]$ & -- & no model & recovered & P, B, R \\
N09 & $\dfrac{1}{e^{x} - 1}$\rule[-2.0ex]{0pt}{5.0ex} & $[3/10,\,3]$ & -- & closed form & recovered & P, B, F, R \\
N10 & $\dfrac{\sin\!\left(x\right)}{\cos\!\left(x\right)}$\rule[-2.0ex]{0pt}{5.0ex} & $[{-}1,\,3/5]$ & -- & closed form & recovered & P14, P, B, F, R, G \\
N11 & $e^{\sin\!\left(x\right)}$ & $[{-}1,\,5]$ & -- & no model & recovered & P14, P, B, F, R, G \\
N12 & $e^{e^{x}/2}$ & $[{-}2,\,1]$ & -- & closed form & recovered & P14, P, B, R, G \\
\end{longtable}
}

\velneed{0.4\textheight}
\section{The physical set, target by target}\label{app:physical}

\renewcommand{\velcaption}{The 29 quantities of the physical set (Section~\ref{sec:physical}), grouped by the class of
their known exact form. \emph{From the callable}: the order and degree of the verified operator, the result and the time
in seconds. The result is \qq{closed form} when the closed form returned agrees with the target to $\tento{-5}$ inside
the range of the data and beyond it, and \qq{continuation} when only the continued solution of the operator does.
\emph{From the float64 table}: the same for 1101 float64 points; \qq{wrong} marks a returned certificate that fails
that test (Section~\ref{sec:physical}). \emph{Closest other program}: the program whose formula has the smallest error
when the larger of its two errors is taken, with its errors inside the range and beyond it.}
{\footnotesize
\begin{longtable}{@{}L{30mm}@{\hspace{0.45em}}l@{\hspace{0.45em}}c@{\hspace{0.45em}}c@{\hspace{0.45em}}l@{\hspace{0.45em}}r@{\hspace{0.45em}}l@{\hspace{0.45em}}L{16.5mm}@{\hspace{0.45em}}c@{\hspace{0.45em}}c@{}}
\caption{\velcaption}\label{tab:physical}\\
\toprule
 & & \multicolumn{4}{c}{From the callable} & From the & \multicolumn{3}{c}{Closest other program, its error} \\
\cmidrule(r{0.45em}){3-6}\cmidrule{8-10}
Quantity & Interval & Ord. & Deg. & Result & Sec. & float64 table & Program & inside & beyond \\
\midrule
\endfirsthead
\toprule
 & & \multicolumn{4}{c}{From the callable} & From the & \multicolumn{3}{c}{Closest other program, its error} \\
\cmidrule(r{0.45em}){3-6}\cmidrule{8-10}
Quantity & Interval & Ord. & Deg. & Result & Sec. & float64 table & Program & inside & beyond \\
\midrule
\endhead
\bottomrule
\endlastfoot
\multicolumn{10}{@{}l}{\textit{Hypergeometric}} \\*
\hangindent=0.8em\hyphenpenalty=10000\exhyphenpenalty=10000 Pendulum period over its small-angle period, against $k = \sin(\text{amplitude}/2)$ & 0.05--0.6 & 2 & 3 & closed form & 1.7 & closed form & \hangindent=0.8em\hyphenpenalty=10000\exhyphenpenalty=10000 PySR & $1\timestento{-6}$ & $2\timestento{-4}$ \\
\hangindent=0.8em\hyphenpenalty=10000\exhyphenpenalty=10000 Ellipse perimeter over $2\pi a$, against the eccentricity & 0.05--0.6 & 2 & 3 & closed form & 2.4 & closed form & \hangindent=0.8em\hyphenpenalty=10000\exhyphenpenalty=10000 PySR + 14 functions & $1\timestento{-7}$ & $1\timestento{-4}$ \\
\hangindent=0.8em\hyphenpenalty=10000\exhyphenpenalty=10000 Mutual inductance of two equal coaxial loops over $\mu_0 a$, against Maxwell's modulus $k$ & 0.1--0.6 & 2 & 4 & closed form & 3.6 & closed form & \hangindent=0.8em\hyphenpenalty=10000\exhyphenpenalty=10000 Flash-ANSR & $1\timestento{-5}$ & $0.02$ \\
\hangindent=0.8em\hyphenpenalty=10000\exhyphenpenalty=10000 Lemniscate arc length from the node, against the radius & 0.05--0.6 & 2 & 4 & closed form & 1.9 & closed form & \hangindent=0.8em\hyphenpenalty=10000\exhyphenpenalty=10000 Flash-ANSR & $2\timestento{-8}$ & $2\timestento{-4}$ \\
\hangindent=0.8em\hyphenpenalty=10000\exhyphenpenalty=10000 Period in the potential $q^2/2 - q^3/3$, against the energy & 0.01--0.1 & 2 & 2 & closed form & 2.8 & closed form & \hangindent=0.8em\hyphenpenalty=10000\exhyphenpenalty=10000 E2E & $3\timestento{-5}$ & $0.001$ \\
\hangindent=0.8em\hyphenpenalty=10000\exhyphenpenalty=10000 Period in the potential $q^2/2 - q^4/4$, against the energy & 0.01--0.15 & 2 & 2 & closed form & 2.8 & closed form & \hangindent=0.8em\hyphenpenalty=10000\exhyphenpenalty=10000 Flash-ANSR & $2\timestento{-7}$ & $0.002$ \\
\hangindent=0.8em\hyphenpenalty=10000\exhyphenpenalty=10000 Period in the potential $q^2/2 - q^5/5$, against the energy & 0.01--0.18 & 4 & 4 & closed form & 46.0 & no model & \hangindent=0.8em\hyphenpenalty=10000\exhyphenpenalty=10000 PySR & $2\timestento{-6}$ & $3\timestento{-4}$ \\
\hangindent=0.8em\hyphenpenalty=10000\exhyphenpenalty=10000 Comoving distance over $c/H_0$, flat universe with $\Omega_m = 3/10$, against the redshift & 0.05--3 & 2 & 3 & closed form & 4.0 & closed form & \hangindent=0.8em\hyphenpenalty=10000\exhyphenpenalty=10000 Bingo & $4\timestento{-4}$ & $3\timestento{-4}$ \\
\hangindent=0.8em\hyphenpenalty=10000\exhyphenpenalty=10000 Bending angle of light by a Schwarzschild mass $M$, against $M/b$ & 0.01--0.12 & 4 & 3 & closed form & 10.6 & no model & \hangindent=0.8em\hyphenpenalty=10000\exhyphenpenalty=10000 Bingo & $1\timestento{-7}$ & $0.01$ \\
\hangindent=0.8em\hyphenpenalty=10000\exhyphenpenalty=10000 Generating function of returns to the origin, body-centred cubic lattice, against its argument & 0.05--0.6 & 3 & 4 & closed form & 5.2 & wrong & \hangindent=0.8em\hyphenpenalty=10000\exhyphenpenalty=10000 PySR & $5\timestento{-8}$ & $2\timestento{-4}$ \\
\hangindent=0.8em\hyphenpenalty=10000\exhyphenpenalty=10000 Voigt profile: unit Gaussian convolved with a Lorentzian of half-width $1/2$, against its argument & 0.1--3 & 3 & 2 & continuation & 2.4 & continuation & \hangindent=0.8em\hyphenpenalty=10000\exhyphenpenalty=10000 Flash-ANSR & $0.002$ & $0.01$ \\
\addlinespace
\multicolumn{10}{@{}l}{\textit{Meijer $G$}} \\*
\hangindent=0.8em\hyphenpenalty=10000\exhyphenpenalty=10000 Density of the product of three standard normal variables, against the product & 0.2--3 & 3 & 2 & closed form & 6.2 & closed form & \hangindent=0.8em\hyphenpenalty=10000\exhyphenpenalty=10000 Flash-ANSR & $1\timestento{-4}$ & $3\timestento{-4}$ \\
\hangindent=0.8em\hyphenpenalty=10000\exhyphenpenalty=10000 Density of the product of four standard normal variables, against the product & 0.2--3 & 4 & 3 & closed form & 10.7 & closed form & \hangindent=0.8em\hyphenpenalty=10000\exhyphenpenalty=10000 Flash-ANSR & $8\timestento{-5}$ & $2\timestento{-4}$ \\
\hangindent=0.8em\hyphenpenalty=10000\exhyphenpenalty=10000 Thermonuclear rate integral $\int_0^\infty \ee^{-y - z/\sqrt{y}}\,\dd y$, against $z$ & 0.2--4 & 3 & 1 & closed form & 4.3 & closed form & \hangindent=0.8em\hyphenpenalty=10000\exhyphenpenalty=10000 PySR & $5\timestento{-4}$ & $0.002$ \\
\hangindent=0.8em\hyphenpenalty=10000\exhyphenpenalty=10000 Density of Wilks' $\Lambda(3, 6, 3)$, against $\Lambda$ & 0.05--0.6 & 3 & 4 & closed form & 62.9 & closed form & \hangindent=0.8em\hyphenpenalty=10000\exhyphenpenalty=10000 Operon & $3\timestento{-4}$ & $0.004$ \\
\hangindent=0.8em\hyphenpenalty=10000\exhyphenpenalty=10000 K distribution: density of an exponential variable with gamma-distributed mean, shape $5/3$, against the variable & 0.1--3 & 2 & 1 & closed form & 3.4 & closed form & \hangindent=0.8em\hyphenpenalty=10000\exhyphenpenalty=10000 Operon & $2\timestento{-4}$ & $4\timestento{-4}$ \\
\addlinespace
\multicolumn{10}{@{}l}{\textit{Fox $H$}} \\*
\hangindent=0.8em\hyphenpenalty=10000\exhyphenpenalty=10000 Density of the one-sided stable law of index $1/3$, against its variable & 0.05--2 & 2 & 3 & closed form & 1.6 & closed form & \hangindent=0.8em\hyphenpenalty=10000\exhyphenpenalty=10000 Flash-ANSR & $9\timestento{-5}$ & $1\timestento{-4}$ \\
\hangindent=0.8em\hyphenpenalty=10000\exhyphenpenalty=10000 Density of the one-sided stable law of index $2/3$, against its variable & 0.1--2 & 2 & 3 & closed form & 4.0 & closed form & \hangindent=0.8em\hyphenpenalty=10000\exhyphenpenalty=10000 Flash-ANSR & $4\timestento{-4}$ & $0.001$ \\
\hangindent=0.8em\hyphenpenalty=10000\exhyphenpenalty=10000 Density of the symmetric stable law of index $3/2$, against its variable & 0.1--4 & 5 & 4 & closed form & 30.0 & no model & \hangindent=0.8em\hyphenpenalty=10000\exhyphenpenalty=10000 Flash-ANSR & $0.001$ & $0.004$ \\
\hangindent=0.8em\hyphenpenalty=10000\exhyphenpenalty=10000 Holtsmark density of the electric field strength in a plasma, against the field strength & 0.2--4 & 5 & 9 & closed form & 35.0 & no model & \hangindent=0.8em\hyphenpenalty=10000\exhyphenpenalty=10000 Flash-ANSR & $0.001$ & $0.009$ \\
\hangindent=0.8em\hyphenpenalty=10000\exhyphenpenalty=10000 Fractional relaxation function $E_{1/3}(-t^{1/3})$, against $t$ & 0.1--6 & 3 & 2 & closed form & 13.9 & wrong & \hangindent=0.8em\hyphenpenalty=10000\exhyphenpenalty=10000 Flash-ANSR & $8\timestento{-6}$ & $6\timestento{-5}$ \\
\hangindent=0.8em\hyphenpenalty=10000\exhyphenpenalty=10000 Fractional relaxation function $E_{1/2}(-t^{1/2})$, against $t$ & 0.1--6 & 2 & 1 & closed form & 1.5 & closed form & \hangindent=0.8em\hyphenpenalty=10000\exhyphenpenalty=10000 Flash-ANSR & $5\timestento{-5}$ & $2\timestento{-5}$ \\
\hangindent=0.8em\hyphenpenalty=10000\exhyphenpenalty=10000 Mainardi's $M$ function of index $2/3$, against its argument & 0.1--3 & 2 & 2 & closed form & 1.6 & closed form & \hangindent=0.8em\hyphenpenalty=10000\exhyphenpenalty=10000 PySR & $6\timestento{-5}$ & $3\timestento{-5}$ \\
\addlinespace
\multicolumn{10}{@{}l}{\textit{Appell / Lauricella}} \\*
\hangindent=0.8em\hyphenpenalty=10000\exhyphenpenalty=10000 Incomplete elliptic integral $F(\arcsin x \mid 1/2)$, against $x$ & 0.05--0.6 & 2 & 4 & closed form & 6.1 & closed form & \hangindent=0.8em\hyphenpenalty=10000\exhyphenpenalty=10000 PySR & $1\timestento{-7}$ & $3\timestento{-4}$ \\
\hangindent=0.8em\hyphenpenalty=10000\exhyphenpenalty=10000 Complete elliptic integral of the third kind $\Pi(n \mid 1/2)$, against $n$ & 0.05--0.6 & 3 & 3 & continuation & 8.6 & no model & \hangindent=0.8em\hyphenpenalty=10000\exhyphenpenalty=10000 Bingo & $4\timestento{-6}$ & $0.01$ \\
\hangindent=0.8em\hyphenpenalty=10000\exhyphenpenalty=10000 Demagnetising factor along the short axis $c$ of an ellipsoid with $a = 2$, $b = 1$, against $c$ & 0.2--0.7 & 3 & 5 & continuation & 9.7 & no model & \hangindent=0.8em\hyphenpenalty=10000\exhyphenpenalty=10000 Bingo & $1\timestento{-8}$ & $3\timestento{-6}$ \\
\hangindent=0.8em\hyphenpenalty=10000\exhyphenpenalty=10000 Surface area over $4\pi$ of an ellipsoid with $a = 2$, $b = 1$, against the short axis $c$ & 0.2--0.7 & 3 & 6 & continuation & 5.6 & no model & \hangindent=0.8em\hyphenpenalty=10000\exhyphenpenalty=10000 Flash-ANSR & $4\timestento{-7}$ & $4\timestento{-5}$ \\
\addlinespace
\multicolumn{10}{@{}l}{\textit{Heun type}} \\*
\hangindent=0.8em\hyphenpenalty=10000\exhyphenpenalty=10000 Period in the potential $q^2/2 + q^3/3 + q^4/4$, against the energy & 0.05--2 & 2 & 4 & continuation & 13.2 & no model & \hangindent=0.8em\hyphenpenalty=10000\exhyphenpenalty=10000 Flash-ANSR & $4\timestento{-4}$ & $5\timestento{-5}$ \\
\hangindent=0.8em\hyphenpenalty=10000\exhyphenpenalty=10000 Generating function of returns to the origin, simple cubic lattice, against its argument & 0.05--0.6 & 3 & 6 & continuation & 17.1 & no model & \hangindent=0.8em\hyphenpenalty=10000\exhyphenpenalty=10000 Flash-ANSR & $2\timestento{-7}$ & $4\timestento{-5}$ \\
\end{longtable}
}

\begin{landscape}
\renewcommand{\velcaption}{The differential equation that \VEL{} found for each of the 29 quantities of the physical set
from the callable, and the closed form returned. Every equation was found in $x$ itself, has exact coefficients and no
fitted parameter, and applied to the target at 60 digits leaves a residual below $\tento{-57}$ of the size of its terms.
A number followed by dots is a constant of the closed form that stayed numerical. The Meijer $G$-function is written
with its upper parameters above its lower ones, and a dash for an empty list.}
{\footnotesize
\begin{longtable}{@{}L{39mm}@{\hspace{1em}}L{95mm}@{\hspace{1em}}L{104mm}@{}}
\caption{\velcaption}\label{tab:physical_operators}\\
\toprule
Quantity & Differential equation found for $f(x)$ & Closed form returned \\
\midrule
\endfirsthead
\toprule
Quantity & Differential equation found for $f(x)$ & Closed form returned \\
\midrule
\endhead
\bottomrule
\endlastfoot
\multicolumn{3}{@{}l}{\textit{Hypergeometric}} \\*
Pendulum period & \hangindent=0.8em\hyphenpenalty=10000\exhyphenpenalty=10000 $x \left(x - 1\right) \left(x + 1\right)\,f'' + \left(3 x^{2} - 1\right)\,f' + x\,f = 0$ & \hangindent=0.8em\hyphenpenalty=10000\exhyphenpenalty=10000 \lineskiplimit=2pt\lineskip=3pt $\displaystyle \frac{2}{\pi}\,K\!\left(x^{2}\right)$ \\
\addlinespace[3pt]
Ellipse perimeter & \hangindent=0.8em\hyphenpenalty=10000\exhyphenpenalty=10000 $x \left(x - 1\right) \left(x + 1\right)\,f'' + \left(x - 1\right) \left(x + 1\right)\,f' - x\,f = 0$ & \hangindent=0.8em\hyphenpenalty=10000\exhyphenpenalty=10000 \lineskiplimit=2pt\lineskip=3pt $\displaystyle \frac{2}{\pi}\,E\!\left(x^{2}\right)$ \\
\addlinespace[3pt]
Coaxial loops & \hangindent=0.8em\hyphenpenalty=10000\exhyphenpenalty=10000 $x^{2} \left(x - 1\right) \left(x + 1\right)\,f'' + x \left(x^{2} + 1\right)\,f' + 3\,f = 0$ & \hangindent=0.8em\hyphenpenalty=10000\exhyphenpenalty=10000 \lineskiplimit=2pt\lineskip=3pt $\displaystyle \frac{\left(2 - x^{2}\right) K\!\left(x^{2}\right) - 2 E\!\left(x^{2}\right)}{x}$ \\
\addlinespace[3pt]
Lemniscate arc & \hangindent=0.8em\hyphenpenalty=10000\exhyphenpenalty=10000 $\left(x - 1\right) \left(x + 1\right) \left(x^{2} + 1\right)\,f'' + 2 x^{3}\,f' = 0$ & \hangindent=0.8em\hyphenpenalty=10000\exhyphenpenalty=10000 \lineskiplimit=2pt\lineskip=3pt $\displaystyle x \,\pFq{2}{1}\!\left(1/4, 1/2;\,5/4;\,x^{4}\right)$ \\
\addlinespace[3pt]
Cubic oscillator & \hangindent=0.8em\hyphenpenalty=10000\exhyphenpenalty=10000 $6 x \left(6 x - 1\right)\,f'' + 6 \left(12 x - 1\right)\,f' + 5\,f = 0$ & \hangindent=0.8em\hyphenpenalty=10000\exhyphenpenalty=10000 \lineskiplimit=2pt\lineskip=3pt $\displaystyle 2 \pi\,\,\pFq{2}{1}\!\left(1/6, 5/6;\,1;\,6 x\right)$ \\
\addlinespace[3pt]
Quartic oscillator & \hangindent=0.8em\hyphenpenalty=10000\exhyphenpenalty=10000 $4 x \left(4 x - 1\right)\,f'' + 4 \left(8 x - 1\right)\,f' + 3\,f = 0$ & \hangindent=0.8em\hyphenpenalty=10000\exhyphenpenalty=10000 \lineskiplimit=2pt\lineskip=3pt $\displaystyle 2 \pi\,\,\pFq{2}{1}\!\left(1/4, 3/4;\,1;\,4 x\right)$ \\
\addlinespace[3pt]
Quintic oscillator & \hangindent=0.8em\hyphenpenalty=10000\exhyphenpenalty=10000 $10 x \left(10 x - 3\right) \left(100 x^{2} + 30 x + 9\right)\,f^{(4)} + 30 \left(4000 x^{3} - 27\right)\,f''' + 367000 x^{2}\,f'' + 268000 x\,f' + 15309\,f = 0$ & \hangindent=0.8em\hyphenpenalty=10000\exhyphenpenalty=10000 \lineskiplimit=2pt\lineskip=3pt $\displaystyle 2 \pi\,\,\pFq{4}{3}\!\bigl(1/10, 3/10, 7/10, 9/10;\,1/3, 2/3, 1;\,1000 x^{3}/27\bigr)$ \\
\addlinespace[3pt]
$\Lambda$CDM distance & \hangindent=0.8em\hyphenpenalty=10000\exhyphenpenalty=10000 $2 \left(3 x^{3} + 9 x^{2} + 9 x + 10\right)\,f'' + 9 \left(x + 1\right)^{2}\,f' = 0$ & \hangindent=0.8em\hyphenpenalty=10000\exhyphenpenalty=10000 \lineskiplimit=2pt\lineskip=3pt $\displaystyle \frac{\sqrt{70}}{7}\,\left(x + 1\right) \,\pFq{2}{1}\!\left(1/3, 1/2;\,4/3;\,- 3 \left(x + 1\right)^{3}/7\right) - 1.140666875\ldots$ \\
\addlinespace[3pt]
Light deflection & \hangindent=0.8em\hyphenpenalty=10000\exhyphenpenalty=10000 $x \left(27 x^{2} - 1\right)\,f^{(4)} + 3 \left(9 x - 1\right) \left(9 x + 1\right)\,f''' + 501 x\,f'' + 192\,f' = 0$ & \hangindent=0.8em\hyphenpenalty=10000\exhyphenpenalty=10000 \lineskiplimit=2pt\lineskip=3pt $\displaystyle 4\,x \,\pFq{3}{2}\!\left(2/3, 1, 4/3;\,3/2, 3/2;\,27 x^{2}\right) - \pi + \pi\,\,\pFq{2}{1}\!\left(1/6, 5/6;\,1;\,27 x^{2}\right)$ \\
\addlinespace[3pt]
Body-centred cubic lattice & \hangindent=0.8em\hyphenpenalty=10000\exhyphenpenalty=10000 $x^{2} \left(x - 1\right) \left(x + 1\right)\,f''' + 3 x \left(2 x^{2} - 1\right)\,f'' + \left(7 x^{2} - 1\right)\,f' + x\,f = 0$ & \hangindent=0.8em\hyphenpenalty=10000\exhyphenpenalty=10000 \lineskiplimit=2pt\lineskip=3pt $\displaystyle \,\pFq{2}{1}\!\left(1/4, 1/4;\,1;\,x^{2}\right)^{2}$ \\
\addlinespace[3pt]
Voigt profile & \hangindent=0.8em\hyphenpenalty=10000\exhyphenpenalty=10000 $4\,f''' + 8 x\,f'' + \left(4 x^{2} + 13\right)\,f' + 8 x\,f = 0$ & \hangindent=0.8em\hyphenpenalty=10000\exhyphenpenalty=10000 none; the continued operator reproduces the target \\
\addlinespace[3pt]
\addlinespace
\multicolumn{3}{@{}l}{\textit{Meijer $G$}} \\*
Three normals & \hangindent=0.8em\hyphenpenalty=10000\exhyphenpenalty=10000 $x^{2}\,f''' + 3 x\,f'' + f' + x\,f = 0$ & \hangindent=0.8em\hyphenpenalty=10000\exhyphenpenalty=10000 \lineskiplimit=2pt\lineskip=3pt $\displaystyle \frac{\sqrt{2}}{4 \pi^{3/2}}\,\MeijerG{3}{0}{0}{3}{x^{2}/8\,\middle|\,\begin{smallmatrix}\text{--}\\ 0, 0, 0\end{smallmatrix}}$ \\
\addlinespace[3pt]
Four normals & \hangindent=0.8em\hyphenpenalty=10000\exhyphenpenalty=10000 $x^{3}\,f^{(4)} + 6 x^{2}\,f''' + 7 x\,f'' + f' - x\,f = 0$ & \hangindent=0.8em\hyphenpenalty=10000\exhyphenpenalty=10000 \lineskiplimit=2pt\lineskip=3pt $\displaystyle \frac{1}{4 \pi^{2}}\,\MeijerG{4}{0}{0}{4}{x^{2}/16\,\middle|\,\begin{smallmatrix}\text{--}\\ 0, 0, 0, 0\end{smallmatrix}}$ \\
\addlinespace[3pt]
Reaction rate & \hangindent=0.8em\hyphenpenalty=10000\exhyphenpenalty=10000 $x\,f''' + 2\,f = 0$ & \hangindent=0.8em\hyphenpenalty=10000\exhyphenpenalty=10000 \lineskiplimit=2pt\lineskip=3pt $\displaystyle \frac{1}{\sqrt{\pi}}\,\MeijerG{3}{0}{0}{3}{x^{2}/4\,\middle|\,\begin{smallmatrix}\text{--}\\ 0, 1/2, 1\end{smallmatrix}}$ \\
\addlinespace[3pt]
Wilks' $\Lambda$ & \hangindent=0.8em\hyphenpenalty=10000\exhyphenpenalty=10000 $4 x^{3} \left(x - 1\right)\,f''' - 6 x^{2} \left(2 x - 1\right)\,f'' + 3 x \left(9 x - 4\right)\,f' - 6 \left(5 x - 2\right)\,f = 0$ & \hangindent=0.8em\hyphenpenalty=10000\exhyphenpenalty=10000 \lineskiplimit=2pt\lineskip=3pt $\displaystyle \frac{1575 \sqrt{\pi}}{32}\,\MeijerG{3}{0}{3}{3}{x\,\middle|\,\begin{smallmatrix}5/2, 3, 7/2\\ 1, 3/2, 2\end{smallmatrix}}$ \\
\addlinespace[3pt]
K distribution & \hangindent=0.8em\hyphenpenalty=10000\exhyphenpenalty=10000 $3 x\,f'' + f' - 5\,f = 0$ & \hangindent=0.8em\hyphenpenalty=10000\exhyphenpenalty=10000 \lineskiplimit=2pt\lineskip=3pt $\displaystyle \frac{5 \cdot\nobreak 2^{1/6} \cdot\nobreak 3^{11/12} \cdot\nobreak 5^{1/3} \sqrt{\Gamma\!\left(1/6\right)}}{6 \pi^{3/4}}\,x^{1/3} K_{2/3}\!\left(2 \sqrt{15} \sqrt{x}/3\right)$ \\
\addlinespace[3pt]
\addlinespace
\multicolumn{3}{@{}l}{\textit{Fox $H$}} \\*
L\'evy 1/3 & \hangindent=0.8em\hyphenpenalty=10000\exhyphenpenalty=10000 $27 x^{3}\,f'' + 108 x^{2}\,f' + \left(60 x - 1\right)\,f = 0$ & \hangindent=0.8em\hyphenpenalty=10000\exhyphenpenalty=10000 \lineskiplimit=2pt\lineskip=3pt $\displaystyle \frac{\sqrt{3}}{9}\,\frac{I_{- 1/3}\!\left(2 \sqrt{3}/\left(9 \sqrt{x}\right)\right) - I_{1/3}\!\left(2 \sqrt{3}/\left(9 \sqrt{x}\right)\right)}{x^{3/2}}$ \\
\addlinespace[3pt]
L\'evy 2/3 & \hangindent=0.8em\hyphenpenalty=10000\exhyphenpenalty=10000 $27 x^{3}\,f'' + \left(135 x^{2} - 8\right)\,f' + 105 x\,f = 0$ & \hangindent=0.8em\hyphenpenalty=10000\exhyphenpenalty=10000 \lineskiplimit=2pt\lineskip=3pt $\displaystyle \frac{2^{1/6} \cdot\nobreak 3^{1/4}}{9 \pi^{3/4} x^{7/3} \sqrt{\Gamma\!\left(1/6\right)}}\,\bigl(3 \cdot\nobreak 2^{2/3} \sqrt{\pi}\,x^{2/3} \,\pFq{1}{1}\!\left(5/6;\,2/3;\,-4/\left(27 x^{2}\right)\right) + \Gamma\!\left(1/6\right)\,\,\pFq{1}{1}\!\left(7/6;\,4/3;\,-4/\left(27 x^{2}\right)\right)\bigr)$ \\
\addlinespace[3pt]
Stable 3/2 & \hangindent=0.8em\hyphenpenalty=10000\exhyphenpenalty=10000 $81\,f^{(5)} + 16 x^{4}\,f''' + 240 x^{3}\,f'' + 940 x^{2}\,f' + 880 x\,f = 0$ & \hangindent=0.8em\hyphenpenalty=10000\exhyphenpenalty=10000 \lineskiplimit=2pt\lineskip=3pt $\displaystyle 0.02456431415\ldots\,x^{4} \,\pFq{2}{3}\!\left(13/12, 19/12;\,7/6, 3/2, 5/3;\,- 4 x^{6}/729\right) - \frac{1}{3 \pi}\,x^{2} \,\pFq{3}{4}\!\left(3/4, 1, 5/4;\,2/3, 5/6, 7/6, 4/3;\,- 4 x^{6}/729\right) + \frac{2 \cdot\nobreak 2^{5/6} \cdot\nobreak 3^{3/4}}{9 \pi^{1/4} \sqrt{\Gamma\!\left(1/6\right)}}\,\,\pFq{2}{3}\!\left(5/12, 11/12;\,1/3, 1/2, 5/6;\,- 4 x^{6}/729\right)$ \\
\addlinespace[3pt]
Holtsmark & \hangindent=0.8em\hyphenpenalty=10000\exhyphenpenalty=10000 $81 x^{5}\,f^{(5)} - 486 x^{4}\,f^{(4)} + 8 x^{3} \left(2 x^{6} + 243\right)\,f''' + 24 x^{2} \left(10 x^{6} - 243\right)\,f'' + 4 x \left(235 x^{6} + 2916\right)\,f' + 16 \left(55 x^{6} - 729\right)\,f = 0$ & \hangindent=0.8em\hyphenpenalty=10000\exhyphenpenalty=10000 \lineskiplimit=2pt\lineskip=3pt $\displaystyle 0.05203095088\ldots\,x^{6} \,\pFq{2}{3}\!\left(17/12, 23/12;\,4/3, 3/2, 11/6;\,- 4 x^{6}/729\right) - 0.1965145132\ldots\,x^{4} \,\pFq{2}{3}\!\left(13/12, 19/12;\,2/3, 7/6, 3/2;\,- 4 x^{6}/729\right) + \frac{4}{3 \pi}\,x^{2} \,\pFq{3}{4}\!\left(3/4, 1, 5/4;\,1/3, 2/3, 5/6, 7/6;\,- 4 x^{6}/729\right)$ \\
\addlinespace[3pt]
Relaxation 1/3 & \hangindent=0.8em\hyphenpenalty=10000\exhyphenpenalty=10000 $9 x^{2}\,f''' + 9 x \left(x + 2\right)\,f'' + 2 \left(9 x + 1\right)\,f' + 2\,f = 0$ & \hangindent=0.8em\hyphenpenalty=10000\exhyphenpenalty=10000 \lineskiplimit=2pt\lineskip=3pt $\displaystyle \frac{3 \sqrt{3} \Gamma\!\left(1/3\right)}{4 \pi}\,x^{2/3} \,\pFq{1}{1}\!\left(1;\,5/3;\,- x\right) - \frac{3}{\Gamma\!\left(1/3\right)}\,x^{1/3} \,\pFq{1}{1}\!\left(1;\,4/3;\,- x\right) + e^{- x}$ \\
\addlinespace[3pt]
Relaxation 1/2 & \hangindent=0.8em\hyphenpenalty=10000\exhyphenpenalty=10000 $2 x\,f'' - \left(2 x - 1\right)\,f' - f = 0$ & \hangindent=0.8em\hyphenpenalty=10000\exhyphenpenalty=10000 \lineskiplimit=2pt\lineskip=3pt $\displaystyle \left(1 - \erf\!\left(\sqrt{x}\right)\right) e^{x}$ \\
\addlinespace[3pt]
Wright $M$ 2/3 & \hangindent=0.8em\hyphenpenalty=10000\exhyphenpenalty=10000 $9\,f'' + 4 x^{2}\,f' + 10 x\,f = 0$ & \hangindent=0.8em\hyphenpenalty=10000\exhyphenpenalty=10000 \lineskiplimit=2pt\lineskip=3pt $\displaystyle \frac{2^{1/6} \cdot\nobreak 3^{1/4}}{6 \pi^{3/4} \sqrt{\Gamma\!\left(1/6\right)}}\,\bigl(x \Gamma\!\left(1/6\right) \,\pFq{1}{1}\!\left(7/6;\,4/3;\,- 4 x^{3}/27\right) + 3 \cdot\nobreak 2^{2/3} \sqrt{\pi} \,\pFq{1}{1}\!\left(5/6;\,2/3;\,- 4 x^{3}/27\right)\bigr)$ \\
\addlinespace[3pt]
\addlinespace
\multicolumn{3}{@{}l}{\textit{Appell / Lauricella}} \\*
Incomplete elliptic $F$ & \hangindent=0.8em\hyphenpenalty=10000\exhyphenpenalty=10000 $\left(x - 1\right) \left(x + 1\right) \left(x^{2} - 2\right)\,f'' + x \left(2 x^{2} - 3\right)\,f' = 0$ & \hangindent=0.8em\hyphenpenalty=10000\exhyphenpenalty=10000 \lineskiplimit=2pt\lineskip=3pt $\displaystyle x F_{1}\!\left(1/2;\,1/2, 1/2;\,3/2;\,x^{2}/2, x^{2}\right)$ \\
\addlinespace[3pt]
Elliptic $\Pi$ & \hangindent=0.8em\hyphenpenalty=10000\exhyphenpenalty=10000 $2 x \left(x - 1\right) \left(2 x - 1\right)\,f''' + \left(26 x^{2} - 24 x + 3\right)\,f'' + 4 \left(8 x - 3\right)\,f' + 4\,f = 0$ & \hangindent=0.8em\hyphenpenalty=10000\exhyphenpenalty=10000 none; the continued operator reproduces the target \\
\addlinespace[3pt]
Demagnetising factor & \hangindent=0.8em\hyphenpenalty=10000\exhyphenpenalty=10000 $x \left(x - 2\right) \left(x - 1\right) \left(x + 1\right) \left(x + 2\right)\,f''' + \left(11 x^{4} - 25 x^{2} - 4\right)\,f'' + x \left(29 x^{2} - 10\right)\,f' + 2 \left(8 x^{2} + 5\right)\,f = 0$ & \hangindent=0.8em\hyphenpenalty=10000\exhyphenpenalty=10000 none; the continued operator reproduces the target \\
\addlinespace[3pt]
Ellipsoid area & \hangindent=0.8em\hyphenpenalty=10000\exhyphenpenalty=10000 $x^{2} \left(x - 2\right) \left(x - 1\right) \left(x + 1\right) \left(x + 2\right)\,f''' + x \left(5 x^{4} - 5 x^{2} - 12\right)\,f'' + 2 \left(x^{2} + 2\right) \left(x^{2} + 3\right)\,f' - 2 x \left(x^{2} + 5\right)\,f = 0$ & \hangindent=0.8em\hyphenpenalty=10000\exhyphenpenalty=10000 none; the continued operator reproduces the target \\
\addlinespace[3pt]
\addlinespace
\multicolumn{3}{@{}l}{\textit{Heun type}} \\*
Asymmetric oscillator & \hangindent=0.8em\hyphenpenalty=10000\exhyphenpenalty=10000 $12 x \left(24 x + 7\right) \left(144 x^{2} + 12 x + 7\right)\,f'' + 12 \left(6912 x^{3} + 3312 x^{2} + 168 x + 49\right)\,f' + \left(7776 x^{2} + 5124 x - 49\right)\,f = 0$ & \hangindent=0.8em\hyphenpenalty=10000\exhyphenpenalty=10000 none; the continued operator reproduces the target \\
\addlinespace[3pt]
Cubic lattice & \hangindent=0.8em\hyphenpenalty=10000\exhyphenpenalty=10000 $x^{2} \left(x - 3\right) \left(x - 1\right) \left(x + 1\right) \left(x + 3\right)\,f''' + 3 x \left(3 x^{4} - 20 x^{2} + 9\right)\,f'' + 9 \left(2 x^{4} - 8 x^{2} + 1\right)\,f' + 6 x \left(x^{2} - 2\right)\,f = 0$ & \hangindent=0.8em\hyphenpenalty=10000\exhyphenpenalty=10000 none; the continued operator reproduces the target \\
\end{longtable}
}

\velneed{0.45\textheight}
\section{Operators found for the functions of two variables}\label{app:bivariate}

\renewcommand{\velcaption}{Every distinct operator recorded for the eight functions of two variables of
Section~\ref{sec:two} and for the control $\Gamma(x + 2y)$, with its test on the held-out quadrant at 40 digits: the
residual of the exact operator and the noise level, both relative to the size of its terms. An operator passes when the
residual is at most ten times the noise. Role \qq{generator}: an operator that the reduction of
Section~\ref{sec:bivariate} kept from the system; the kept operators need not be a minimal set of generators. Role
\qq{series}: one of the two two-term operators that fix the ratios~\eqref{eq:horn}. An operator can have both roles.
For the Voigt function the reduction kept four rows, of which $y\,(\partial_v^2 + \partial_y^2)$ and
$v\,(\partial_v^2 + \partial_y^2)$ print as one operator once the common factor is removed (\qq{recorded twice}), and
$\partial_v^2 + \partial_y^2$ is the difference of the other two operators listed. The operators of the two ellipsoid
integrals are printed in the variables $u = 1 - x$ and $v = 1 - y$ in which they were found.}
{\footnotesize
\begin{longtable}{@{}L{140mm}@{\hspace{8pt}}L{40mm}@{\hspace{8pt}}l@{\hspace{8pt}}l@{}}
\caption{\velcaption}\label{tab:bivariate_operators}\\
\toprule
Operator & Role & Residual & Noise \\
\midrule
\endfirsthead
\toprule
Operator & Role & Residual & Noise \\
\midrule
\endhead
\bottomrule
\endfoot
\multicolumn{4}{@{}l}{\textit{elliptic Pi}, variables $(n,\,m)$} \\*
\quad $2 \left(m - n\right)\partial_{n}\partial_{m} + \partial_{n} - 2\partial_{m}$ & generator & $2.9\timestento{-37}$ & $3.6\timestento{-37}$ \\*
\quad $4 n \left(n - 1\right)\partial_{n}^{2} + 2 \left(4 n - 1\right)\partial_{n} - 4 m \left(m - 1\right)\partial_{m}^{2} - 4 m\partial_{m} + 1$ & generator & $6.7\timestento{-38}$ & $1.5\timestento{-37}$ \\*
\quad $2 n \left(n - 1\right)\partial_{n}^{2} + 2 m \left(n - 1\right)\partial_{n}\partial_{m} + \left(5 n - 2\right)\partial_{n} + 2 m\partial_{m} + 1$ & series & $4.8\timestento{-37}$ & $5.1\timestento{-37}$ \\*
\quad $4 n \left(m - 1\right)\partial_{n}\partial_{m} + 2 n\partial_{n} + 4 m \left(m - 1\right)\partial_{m}^{2} + 4 \left(2 m - 1\right)\partial_{m} + 1$ & series & $9.4\timestento{-37}$ & $9.4\timestento{-37}$ \\*
\quad $4 m \left(m - 1\right)\partial_{n}\partial_{m} + 2 \left(m + n - 1\right)\partial_{n} + 4 m \left(m - 1\right)\partial_{m}^{2} + 4 m\partial_{m} + 1$ & generator & $1.0\timestento{-36}$ & $9.9\timestento{-37}$ \\
\addlinespace[4pt]
\multicolumn{4}{@{}l}{\textit{pendulum time}, variables $(s,\,m)$} \\*
\quad $2 \left(m s^{2} - 1\right)\partial_{s}\partial_{m} + s^{2}\partial_{s}$ & series & $3.1\timestento{-36}$ & $1.4\timestento{-36}$ \\*
\quad $s \left(s - 1\right) \left(s + 1\right)\partial_{s}^{2} - 2 m \left(s - 1\right) \left(s + 1\right)\partial_{s}\partial_{m} + s^{2}\partial_{s}$ & series & $1.0\timestento{-36}$ & $5.5\timestento{-37}$ \\*
\quad $2 s \left(m - 1\right)\partial_{s}\partial_{m} + s\partial_{s} - 4 m \left(m - 1\right)\partial_{m}^{2} - 4 \left(2 m - 1\right)\partial_{m} - 1$ & generator & $4.6\timestento{-37}$ & $3.4\timestento{-37}$ \\
\addlinespace[4pt]
\multicolumn{4}{@{}l}{\textit{ellipsoid capacitance}, variables $u=1-x$, $v=1-y$} \\*
\quad $2 \left(u - v\right)\partial_{u}\partial_{v} - \partial_{u} + \partial_{v}$ & generator & $9.0\timestento{-37}$ & $1.1\timestento{-36}$ \\*
\quad $4 u \left(v - 1\right)\partial_{u}\partial_{v} + 2 u\partial_{u} + 4 v \left(v - 1\right)\partial_{v}^{2} + 2 \left(4 v - 3\right)\partial_{v} + 1$ & series & $1.1\timestento{-36}$ & $1.3\timestento{-36}$ \\*
\quad $4 u \left(u - 1\right)\partial_{u}^{2} + 4 v \left(u - 1\right)\partial_{u}\partial_{v} + 2 \left(4 u - 3\right)\partial_{u} + 2 v\partial_{v} + 1$ & series & $7.8\timestento{-37}$ & $9.8\timestento{-37}$ \\*
\quad $2 u \left(u - 1\right)\partial_{u}^{2} + \left(3 u - 2\right)\partial_{u} - 2 v \left(v - 1\right)\partial_{v}^{2} - \left(3 v - 2\right)\partial_{v}$ & generator & $1.3\timestento{-37}$ & $2.6\timestento{-37}$ \\*
\quad $4 v \left(v - 1\right)\partial_{u}\partial_{v} + 2 \left(u + v - 1\right)\partial_{u} + 4 v \left(v - 1\right)\partial_{v}^{2} + 2 \left(3 v - 2\right)\partial_{v} + 1$ & generator & $1.2\timestento{-36}$ & $1.4\timestento{-36}$ \\
\addlinespace[4pt]
\multicolumn{4}{@{}l}{\textit{ellipsoid demagnetisation}, variables $u=1-x$, $v=1-y$} \\*
\quad $2 \left(u - v\right)\partial_{u}\partial_{v} - \partial_{u} + \partial_{v}$ & generator & $9.5\timestento{-38}$ & $3.7\timestento{-37}$ \\*
\quad $4 u \left(v - 1\right)\partial_{u}\partial_{v} + 2 u\partial_{u} + 4 v \left(v - 1\right)\partial_{v}^{2} + 2 \left(6 v - 5\right)\partial_{v} + 3$ & series & $4.5\timestento{-38}$ & $3.0\timestento{-37}$ \\*
\quad $4 u \left(u - 1\right)\partial_{u}^{2} + 4 v \left(u - 1\right)\partial_{u}\partial_{v} + 2 \left(6 u - 5\right)\partial_{u} + 2 v\partial_{v} + 3$ & series & $3.7\timestento{-38}$ & $3.4\timestento{-37}$ \\*
\quad $2 u \left(u - 1\right)\partial_{u}^{2} + \left(5 u - 4\right)\partial_{u} - 2 v \left(v - 1\right)\partial_{v}^{2} - \left(5 v - 4\right)\partial_{v}$ & generator & $3.7\timestento{-38}$ & $2.1\timestento{-37}$ \\*
\quad $4 v \left(v - 1\right)\partial_{u}\partial_{v} + 2 \left(u + v - 1\right)\partial_{u} + 4 v \left(v - 1\right)\partial_{v}^{2} + 2 \left(5 v - 4\right)\partial_{v} + 3$ & generator & $5.3\timestento{-38}$ & $2.8\timestento{-37}$ \\
\addlinespace[4pt]
\multicolumn{4}{@{}l}{\textit{anisotropic square lattice}, variables $(a,\,b)$} \\*
\quad $a b\partial_{a}^{2} + b\partial_{a} - a b\partial_{b}^{2} - a\partial_{b}$ & generator & $2.9\timestento{-37}$ & $4.5\timestento{-37}$ \\*
\quad $\left(a^{2} + b^{2} - 1\right)\partial_{a}\partial_{b} + b\partial_{a} + 2 a b\partial_{b}^{2} + 3 a\partial_{b}$ & generator & $1.9\timestento{-36}$ & $6.3\timestento{-36}$ \\*
\quad $a^{2} b\partial_{a}^{2} + 2 a b^{2}\partial_{a}\partial_{b} + 4 a b\partial_{a} + b \left(b - 1\right) \left(b + 1\right)\partial_{b}^{2} + \left(2 b - 1\right) \left(2 b + 1\right)\partial_{b} + 2 b$ & series & $6.1\timestento{-37}$ & $9.4\timestento{-37}$ \\*
\quad $a \left(a - 1\right) \left(a + 1\right)\partial_{a}^{2} + 2 a^{2} b\partial_{a}\partial_{b} + \left(2 a - 1\right) \left(2 a + 1\right)\partial_{a} + a b^{2}\partial_{b}^{2} + 4 a b\partial_{b} + 2 a$ & series & $3.6\timestento{-37}$ & $7.0\timestento{-37}$ \\
\addlinespace[4pt]
\multicolumn{4}{@{}l}{\textit{Gordon integral}, variables $(x,\,y)$} \\*
\quad $6 x y\partial_{x}\partial_{y} + 2 x\partial_{x} + 6 y \left(y - 1\right)\partial_{y}^{2} + \left(23 y - 12\right)\partial_{y} + 5$ & generator, series & $1.4\timestento{-37}$ & $3.1\timestento{-37}$ \\*
\quad $4 x \left(x - 1\right)\partial_{x}^{2} + 4 x y\partial_{x}\partial_{y} + 4 \left(4 x - 1\right)\partial_{x} + 2 y\partial_{y} + 5$ & series & $7.9\timestento{-38}$ & $1.9\timestento{-37}$ \\*
\quad $12 x \left(x - 1\right)\partial_{x}^{2} + 4 \left(11 x - 3\right)\partial_{x} - 12 y \left(y - 1\right)\partial_{y}^{2} - 8 \left(5 y - 3\right)\partial_{y} + 5$ & generator & $3.7\timestento{-38}$ & $1.4\timestento{-37}$ \\
\addlinespace[4pt]
\multicolumn{4}{@{}l}{\textit{weighted elliptic K}, variables $(x,\,y)$} \\*
\quad $2 x\partial_{x}\partial_{y} - 2\partial_{x} + \partial_{y}$ & generator & $4.4\timestento{-37}$ & $2.5\timestento{-37}$ \\*
\quad $2 \left(x - 1\right)\partial_{x} - 2 y\partial_{y}^{2} + \left(2 y - 1\right)\partial_{y} + 1$ & generator & $1.4\timestento{-37}$ & $1.5\timestento{-37}$ \\*
\quad $2 x\partial_{x}\partial_{y} - 2 x\partial_{x} + 2 y\partial_{y}^{2} - 2 \left(y - 1\right)\partial_{y} - 1$ & series & $3.1\timestento{-37}$ & $2.2\timestento{-37}$ \\*
\quad $4 x \left(x - 1\right)\partial_{x}^{2} + 4 y \left(x - 1\right)\partial_{x}\partial_{y} + 4 \left(2 x - 1\right)\partial_{x} + 2 y\partial_{y} + 1$ & series & $3.3\timestento{-37}$ & $2.3\timestento{-37}$ \\
\addlinespace[4pt]
\multicolumn{4}{@{}l}{\textit{Voigt function}, variables $(v,\,y)$} \\*
\quad $\partial_{v}^{2} + \partial_{y}^{2}$ & generator (recorded twice) & $6.1\timestento{-37}$ & $8.8\timestento{-37}$ \\*
\quad $2 v\partial_{v} - \partial_{y}^{2} + 2 y\partial_{y} + 2$ & generator, series & $3.2\timestento{-38}$ & $1.4\timestento{-37}$ \\*
\quad $\partial_{v}^{2} + 2 v\partial_{v} + 2 y\partial_{y} + 2$ & generator, series & $2.0\timestento{-38}$ & $1.3\timestento{-37}$ \\
\addlinespace[4pt]
\multicolumn{4}{@{}l}{\textit{control $\Gamma(x+2y)$}, variables $(x,\,y)$} \\*
\quad $2\partial_{x} - \partial_{y}$ & generator & $1.5\timestento{-38}$ & $1.1\timestento{-37}$ \\
\end{longtable}
}

\velneed{0.45\textheight}
\section{The regression cases}\label{app:cases}

\renewcommand{\velcaption}{The 125 regression cases of Section~\ref{sec:cases} with the first certificate that the
frozen learner returned for each. \emph{Data}: the kind of input. \emph{Change of variable}: the map $y = \varphi(x)$
of the certificate, with ($k = 1$) when the search ran on the pushforward of equation~\eqref{eq:transport}; a dash when
the operator was found in $x$. \emph{Operator}: the equation for $f(x)$, for $g(y)$ of
equation~\eqref{eq:representation} after a change of variable, or the recurrence for a sequence $a_n$; for moments it is
the equation of the density. $a_1$ is a coefficient that stayed numerical. The groups are those of
Section~\ref{sec:cases}; \qq{named densities} are six densities that an earlier version of the learner fitted by name,
and the two later sets were written after the code that they test. Case names are those of the benchmark script.}
{\footnotesize\renewcommand{\arraystretch}{0.97}
\begin{longtable}{@{}L{42mm}L{12.5mm}L{15.5mm}L{28mm}L{58mm}L{53mm}R{11mm}@{}}
\caption{\velcaption}\label{tab:cases}\\
\toprule
case & data & status & change of variable & operator & closed form & seconds \\
\midrule
\endfirsthead
\toprule
case & data & status & change of variable & operator & closed form & seconds \\
\midrule
\endhead
\bottomrule
\endfoot
\multicolumn{7}{@{}l}{\textit{direct} (13 cases)} \\*[1pt]
\code{exp(-x)} & callable & verified & -- & $f' + f = \nobreak 0$ & $e^{- x}$ & 21.5 \\
\addlinespace[2pt]
\code{1/(1-x)} & callable & verified & -- & $\left(x - 1\right)\,f' + f = \nobreak 0$ & $\displaystyle \frac{1}{1 - x}$ & 9.6 \\
\addlinespace[2pt]
\code{exp(-x\textasciicircum{}2)} & callable & verified & -- & $f' + 2 x\,f = \nobreak 0$ & $e^{- x^{2}}$ & 11.0 \\
\code{Ai(x)} & callable & verified & -- & $f'' - x\,f = \nobreak 0$ & $\Ai\!\left(x\right)$ & 12.6 \\
\code{J0(x)} & callable & verified & -- & $x\,f'' + f' + x\,f = \nobreak 0$ & $J_{0}\!\left(x\right)$ & 26.0 \\
\code{erf(x)} & callable & verified & -- & $f'' + 2 x\,f' = \nobreak 0$ & $\erf\!\left(x\right)$ & 9.8 \\
\addlinespace[2pt]
\code{(exp(x)-1)/\allowbreak{}x} & callable & verified & -- & $x\,f'' - \left(x - 2\right)\,f' - f = \nobreak 0$ & $\displaystyle \frac{e^{x} - 1}{x}$ & 23.0 \\
\addlinespace[2pt]
\code{x exp(-x)} & callable & verified & -- & $x\,f' + \left(x - 1\right)\,f = \nobreak 0$ & $x e^{- x}$ & 29.9 \\
\code{exp(x)+cos(x)} & callable & verified & -- & $f''' - f'' + f' - f = \nobreak 0$ & $e^{x} + \cos\!\left(x\right)$ & 12.2 \\
\code{exp(x) sin(x)} & callable & verified & -- & $f'' - 2\,f' + 2\,f = \nobreak 0$ & $e^{x} \sin\!\left(x\right)$ & 22.0 \\
\code{log(1+x)} & callable & verified & -- & $\left(x + 1\right)\,f'' + f' = \nobreak 0$ & $\log\!\left(x + 1\right)$ & 25.3 \\
\addlinespace[2pt]
\code{sin(x)/\allowbreak{}x} & callable & verified & -- & $x\,f'' + 2\,f' + x\,f = \nobreak 0$ & $\displaystyle \frac{\sin\!\left(x\right)}{x}$ & 24.7 \\
\addlinespace[2pt]
\code{constant 7/3} & callable & verified & -- & $f' = \nobreak 0$ & $\frac{7}{3}$ & 13.1 \\
\addlinespace[4pt]
\multicolumn{7}{@{}l}{\textit{wider class} (7 cases)} \\*[1pt]
\code{K(m) complete elliptic} & callable & verified & -- & $4 x \left(x - 1\right)\,f'' + 4 \left(2 x - 1\right)\,f' + f = \nobreak 0$ & $K\!\left(x\right)$ & 38.3 \\
\addlinespace[2pt]
\code{Dawson F(x)} & callable & verified & -- & $f'' + 2 x\,f' + 2\,f = \nobreak 0$ & $\displaystyle \frac{\sqrt{\pi}}{2}\,\operatorname{erfi}\!\left(x\right) e^{- x^{2}}$ & 25.8 \\
\addlinespace[2pt]
\code{Ai(x)\textasciicircum{}2} & callable & verified & -- & $f''' - 4 x\,f' - 2\,f = \nobreak 0$ & $\Ai\!\left(x\right)^{2}$ & 14.2 \\
\code{Rice density} & callable & verified & $y = x^{2}$ ($k = 1$) & $4 y\,g'' + 4 \left(y + 1\right)\,g' + \left(y + 1\right)\,g = \nobreak 0$ & $x I_{0}\!\left(x\right) e^{- x^{2}/2 - 1/2}$ & 22.6 \\
\code{x\textasciicircum{}2 exp(-3 x\textasciicircum{}2.3 / 2)} & callable & verified & $y = x^{23/10}$ & $46 y\,g' + \left(69 y - 40\right)\,g = \nobreak 0$ & $x^{2} e^{- 3 x^{23/10}/2}$ & 61.7 \\
\code{1.7 x\textasciicircum{}0.7 exp(-x\textasciicircum{}1.7)} & callable & verified & $y = x^{17/10}$ ($k = 1$) & $g' + g = \nobreak 0$ & $\frac{17}{10}\,x^{7/10} e^{- x^{17/10}}$ & 63.4 \\
\addlinespace[2pt]
\code{pi x\textasciicircum{}(pi-1) exp(-x\textasciicircum{}pi)} & callable & verified & $y = x^{\pi}$ ($k = 1$) & $g' + g = \nobreak 0$ & $\displaystyle \pi\,\frac{x^{\pi} e^{- x^{\pi}}}{x}$ & 47.8 \\
\addlinespace[4pt]
\multicolumn{7}{@{}l}{\textit{change of variable} (9 cases)} \\*[1pt]
\code{2x exp(-x\textasciicircum{}2)} & callable & verified & $y = x^{2}$ ($k = 1$) & $g' + g = \nobreak 0$ & $2\,x e^{- x^{2}}$ & 19.0 \\
\code{(3/2) x\textasciicircum{}(1/2) exp(-a x\textasciicircum{}(3/2))} & callable & verified & $y = x^{3/2}$ ($k = 1$) & $g' + a_{1}\,g = \nobreak 0$ & $\frac{3}{2}\,\sqrt{x} e^{- a_{1} x^{3/2}}$ & 29.0 \\
\addlinespace[2pt]
\code{exp(-a/x)/\allowbreak{}x\textasciicircum{}2} & callable & verified & $y = 1/x$ ($k = 1$) & $g' + a_{1}\,g = \nobreak 0$ & $\displaystyle \frac{e^{- a_{1}/x}}{x^{2}}$ & 28.6 \\
\addlinespace[2pt]
\code{exp(-a x/(1-x))/\allowbreak{}(1-x)\textasciicircum{}2} & callable & verified & $y = x/(1 - x)$ ($k = 1$) & $g' + a_{1}\,g = \nobreak 0$ & $\displaystyle \frac{e^{a_{1} x/\left(x - 1\right)}}{\left(x - 1\right)^{2}}$ & 42.1 \\
\addlinespace[2pt]
\code{exp(-a asinh(x)\textasciicircum{}2)/\allowbreak{}sqrt(1+x\textasciicircum{}2)} & callable & verified & $y = \arsinh x$ ($k = 1$) & $g' + a_{1} y\,g = \nobreak 0$ & $\displaystyle \frac{e^{- a_{1} \arsinh^{2}\!\left(x\right)/2}}{\sqrt{x^{2} + 1}}$ & 28.6 \\
\addlinespace[2pt]
\code{exp(-a logit(x)\textasciicircum{}2)/\allowbreak{}(x(1-x))} & callable & verified & $y = \logit x$ ($k = 1$) & $g' + a_{1} y\,g = \nobreak 0$ & $\displaystyle \frac{e^{- a_{1} \log^{2}\!\left(- x/\left(x - 1\right)\right)/2}}{x \left(1 - x\right)}$ & 66.3 \\
\addlinespace[2pt]
\code{sqrt(2 pi) exp(-(a-1/2) probit(x)\textasciicircum{}2)} & callable & verified & $y = \probit x$ & $g' + a_{1} y\,g = \nobreak 0$ & $\sqrt{2} \sqrt{\pi} e^{- a_{1} \erf^{-1}\!\left(2 x - 1\right)^{2}}$ & 76.1 \\
\code{tan(x)} & callable & verified & $y = \tan x$ & $y\,g' - g = \nobreak 0$ & $\tan\!\left(x\right)$ & 27.0 \\
\addlinespace[2pt]
\code{exp(-a logit(x/2)\textasciicircum{}2)/\allowbreak{}(x(2-x)), map supplied by the caller} & callable & verified & $y = \logit(x/2)$ ($k = 1$) & $g' + a_{1} y\,g = \nobreak 0$ & $\displaystyle \frac{e^{- a_{1} \log^{2}\!\left(- x/\left(x - 2\right)\right)/2}}{x \left(2 - x\right)}$ & 46.1 \\
\addlinespace[4pt]
\multicolumn{7}{@{}l}{\textit{control} (7 cases)} \\*[1pt]
\code{1/(1+x\textasciicircum{}2), arcsinh only} & callable & no-model & -- & -- & -- & 3.6 \\
\addlinespace[2pt]
\code{1/(pi (1+x\textasciicircum{}2)), full bank} & callable & verified & -- & $\left(x^{2} + 1\right)\,f' + 2 x\,f = \nobreak 0$ & $\displaystyle \frac{1}{\pi \left(x^{2} + 1\right)}$ & 5.9 \\
\addlinespace[2pt]
\code{exp(-a asinh(x)\textasciicircum{}2)/\allowbreak{}sqrt(1+x\textasciicircum{}2), direct only} & callable & no-model & -- & -- & -- & 86.0 \\
\code{tan(x), direct only} & callable & no-model & -- & -- & -- & 81.3 \\
\code{Gamma(x)} & callable & no-model & -- & -- & -- & 238.9 \\
\code{zeta(x)} & callable & no-model & -- & -- & -- & 272.0 \\
\code{x\textasciicircum{}x} & callable & no-model & -- & -- & -- & 245.6 \\
\addlinespace[4pt]
\multicolumn{7}{@{}l}{\textit{geometry} (4 cases)} \\*[3pt]
\code{disk distance density} & callable & verified & $y = x^{2}$ ($k = 1$) & $y \left(y - 4\right)\,g'' - 2\,g' = \nobreak 0$ & $\displaystyle \frac{1}{\pi}\,x \allowbreak\bigl(4 \arcsin\!\left(\sqrt{4 - x^{2}}/2\right) - x \sqrt{4 - x^{2}}\bigr)$ & 38.0 \\
\addlinespace[2pt]
\code{ball distance density} & callable & verified & -- & $x \left(x - 2\right) \left(x + 4\right)\,f' - \left(5 x^{2} + 10 x - 16\right)\,f = \nobreak 0$ & $\frac{1}{16}\,\left(3 x + 12\right) \left(2 x - x^{2}\right)^{2}$ & 24.1 \\
\addlinespace[2pt]
\code{circle chord density} & callable & verified & -- & $\left(x - 2\right) \left(x + 2\right)\,f' + x\,f = \nobreak 0$ & $\displaystyle \frac{2}{\pi \sqrt{4 - x^{2}}}$ & 20.5 \\
\addlinespace[2pt]
\code{segment distance density} & callable & verified & -- & $\left(x - 1\right)\,f' - f = \nobreak 0$ & $2 - 2\,x$ & 27.4 \\
\addlinespace[4pt]
\multicolumn{7}{@{}l}{\textit{named densities} (6 cases)} \\*[3pt]
\code{gamma(5/2, 2/3)} & callable & verified & -- & $2 x\,f' + 3 \left(x - 1\right)\,f = \nobreak 0$ & $\displaystyle \frac{3 \sqrt{6}}{2 \sqrt{\pi}}\,x^{3/2} e^{- 3 x/2}$ & 15.9 \\
\addlinespace[2pt]
\code{GB1(a=2, b=1, p=3/2, q=5/2)} & callable & verified & -- & $x \left(x - 1\right) \left(x + 1\right)\,f' - \left(5 x^{2} - 2\right)\,f = \nobreak 0$ & $\displaystyle \frac{32}{\pi}\,x^{2} \left(1 - x^{2}\right)^{3/2}$ & 31.3 \\
\addlinespace[2pt]
\code{GB2(a=2, b=1, p=1, q=3/2)} & callable & verified & $y = x^{2}$ ($k = 1$) & $2 \left(y + 1\right)\,g' + 5\,g = \nobreak 0$ & $\displaystyle 3\,\frac{x}{\left(x^{2} + 1\right)^{5/2}}$ & 19.5 \\
\addlinespace[2pt]
\code{beta prime(2, 3)} & callable & verified & -- & $x \left(x + 1\right)\,f' + \left(4 x - 1\right)\,f = \nobreak 0$ & $\displaystyle 12\,\frac{x}{\left(x + 1\right)^{5}}$ & 27.2 \\
\addlinespace[2pt]
\code{lognormal(1/2, 3/4)} & callable & verified & $y = \log x$ & $9\,g' + \left(16 y + 1\right)\,g = \nobreak 0$ & $\displaystyle \frac{2 \sqrt{2} e^{- 8 \log^{2}\!\left(x\right)/9}}{3 \sqrt{\pi} x^{1/9} e^{2/9}}$ & 45.3 \\
\addlinespace[2pt]
\code{gamma-gamma(3, 1)} & callable & verified & -- & $x\,f'' - f' - 3\,f = \nobreak 0$ & $9\,x K_{2}\!\left(2 \sqrt{3} \sqrt{x}\right)$ & 133.2 \\
\addlinespace[4pt]
\multicolumn{7}{@{}l}{\textit{tables} (9 cases)} \\*[1pt]
\code{J0, 45-digit table of 551 points} & table & verified & -- & $x\,f'' + f' + x\,f = \nobreak 0$ & $J_{0}\!\left(x\right)$ & 6.4 \\
\code{Ai(x-3), 45-digit table} & table & verified & -- & $f'' - \left(x - 3\right)\,f = \nobreak 0$ & $\Ai\!\left(x - 3\right)$ & 7.9 \\
\code{J0, float64 table of 1101 points} & table & verified & -- & $x\,f'' + f' + x\,f = \nobreak 0$ & $J_{0}\!\left(x\right)$ & 5.0 \\
\code{exp(-x\textasciicircum{}2/2), float64 table} & table & verified & -- & $f' + x\,f = \nobreak 0$ & $e^{- x^{2}/2}$ & 2.2 \\
\addlinespace[2pt]
\code{lognormal(0, 1/2), float64 table, full bank} & table & verified & $y = \log x$ & $g' + \left(4 y + 1\right)\,g = \nobreak 0$ & $\displaystyle \frac{e^{- 2 \log^{2}\!\left(x\right)}}{x}$ & 53.9 \\
\addlinespace[2pt]
\code{J0 + 1e-6 noise, float64 table} & table & conjecture & -- & $x\,f'' + f' + x\,f = \nobreak 0$ & $J_{0}\!\left(x\right)$ & 2.4 \\
\code{J0 + 1e-3 noise, float64 table} & table & no-model & -- & -- & -- & 21.0 \\
\code{white noise table} & table & no-model & -- & -- & -- & 37.7 \\
\code{Gamma(x), float64 table} & table & no-model & -- & -- & -- & 25.6 \\
\addlinespace[4pt]
\multicolumn{7}{@{}l}{\textit{draws} (7 cases)} \\*[1pt]
\code{2e6 draws, exponential(1)} & draws & conjecture & -- & $f' + f = \nobreak 0$ & $e^{- x}$ & 31.9 \\
\code{2e6 draws, gamma(3, 1)} & draws & conjecture & -- & $x\,f' + \left(x - 2\right)\,f = \nobreak 0$ & $\frac{1}{2}\,x^{2} e^{- x}$ & 38.9 \\
\addlinespace[2pt]
\code{2e6 draws, normal(0, 1)} & draws & conjecture & -- & $f' + x\,f = \nobreak 0$ & $\displaystyle \frac{\sqrt{2} e^{- x^{2}/2}}{2 \sqrt{\pi}}$ & 12.9 \\
\addlinespace[2pt]
\code{2e6 draws, beta(2, 3)} & draws & conjecture & -- & $x \left(x - 1\right)\,f' - \left(3 x - 1\right)\,f = \nobreak 0$ & $12\,x \left(1 - x\right)^{2}$ & 45.5 \\
\addlinespace[2pt]
\code{2e6 draws, lognormal(0, 1/2)} & draws & conjecture & $y = \log x$ ($k = 1$) & $g' + 4 y\,g = \nobreak 0$ & $\displaystyle \frac{\sqrt{2} e^{- 2 \log^{2}\!\left(x\right)}}{\sqrt{\pi} x}$ & 40.5 \\
\addlinespace[2pt]
\code{2e6 draws, Weibull(2)} & draws & conjecture & $y = x^{2}$ ($k = 1$) & $g' + g = \nobreak 0$ & $2\,x e^{- x^{2}}$ & 35.7 \\
\code{2e6 draws, logistic, direct only} & draws & no-model & -- & -- & -- & 3.5 \\
\addlinespace[4pt]
\multicolumn{7}{@{}l}{\textit{sequences} (7 cases)} \\*[1pt]
\code{n!} & sequence & verified & -- & $a_{n + 1} - \left(n + 1\right)\,a_{n} = \nobreak 0$ & $n!$ & 0.1 \\
\addlinespace[2pt]
\code{binomial(2n, n)} & sequence & verified & -- & $\left(n + 1\right)\,a_{n + 1} - 2 \left(2 n + 1\right)\,a_{n} = \nobreak 0$ & $\displaystyle \frac{4^{n} \left(1/2\right)_{n}}{n!}$ & 0.1 \\
\addlinespace[2pt]
\code{Catalan} & sequence & verified & -- & $\left(n + 2\right)\,a_{n + 1} - 2 \left(2 n + 1\right)\,a_{n} = \nobreak 0$ & $\displaystyle \frac{4^{n} \left(1/2\right)_{n}}{\left(2\right)_{n}}$ & 0.2 \\
\addlinespace[2pt]
\code{Fibonacci} & sequence & verified & -- & $a_{n + 2} - a_{n + 1} - a_{n} = \nobreak 0$ & $\displaystyle \frac{\sqrt{5}}{5}\,2^{- n} \allowbreak\bigl(\left(1 + \sqrt{5}\right)^{n} - \left(1 - \sqrt{5}\right)^{n}\bigr)$ & 1.3 \\
\addlinespace[2pt]
\code{Motzkin} & sequence & verified & -- & $\left(n + 4\right)\,a_{n + 2} - \left(2 n + 5\right)\,a_{n + 1} - 3 \left(n + 1\right)\,a_{n} = \nobreak 0$ & -- & 0.1 \\
\code{harmonic numbers} & sequence & verified & -- & $\left(n + 2\right)\,a_{n + 2} - \left(2 n + 3\right)\,a_{n + 1} + \left(n + 1\right)\,a_{n} = \nobreak 0$ & -- & 0.1 \\
\code{primes} & sequence & no-model & -- & -- & -- & 0.0 \\
\addlinespace[4pt]
\multicolumn{7}{@{}l}{\textit{moments} (9 cases)} \\*[1pt]
\code{moments n! (exponential)} & moments & verified & -- & $f' + f = \nobreak 0$ & $e^{- x}$ & 0.9 \\
\code{moments (n+2)!/2 (gamma 3)} & moments & verified & -- & $x\,f' + \left(x - 2\right)\,f = \nobreak 0$ & $\frac{1}{2}\,x^{2} e^{- x}$ & 0.7 \\
\addlinespace[2pt]
\code{moments Catalan (Marchenko-Pastur)} & moments & verified & -- & $x \left(x - 4\right)\,f' - 2\,f = \nobreak 0$ & $\displaystyle \frac{1}{2 \pi}\,\frac{\sqrt{4 - x}}{\sqrt{x}}$ & 2.5 \\
\addlinespace[2pt]
\code{moments binomial(2n, n) (arcsine on [0, 4])} & moments & verified & -- & $x \left(x - 4\right)\,f' + \left(x - 2\right)\,f = \nobreak 0$ & $\displaystyle \frac{1}{\pi \sqrt{4 x - x^{2}}}$ & 1.2 \\
\addlinespace[2pt]
\code{moments (n-1)!! (normal)} & moments & verified & -- & $f' + x\,f = \nobreak 0$ & $\displaystyle \frac{\sqrt{2} e^{- x^{2}/2}}{2 \sqrt{\pi}}$ & 0.4 \\
\addlinespace[2pt]
\code{moments of beta(2, 3)} & moments & verified & -- & $x \left(x - 1\right)\,f' - \left(3 x - 1\right)\,f = \nobreak 0$ & $12\,x \left(1 - x\right)^{2}$ & 0.1 \\
\addlinespace[2pt]
\code{moments of the semicircle law} & moments & verified & -- & $\left(x - 2\right) \left(x + 2\right)\,f' - x\,f = \nobreak 0$ & $\displaystyle \frac{1}{2 \pi}\,\sqrt{4 - x^{2}}$ & 1.9 \\
\addlinespace[2pt]
\code{moments 1/(n+1) (uniform)} & moments & verified & -- & $f' = \nobreak 0$ & $1$ & 0.1 \\
\code{Bell numbers (Poisson moments)} & moments & no-model & -- & -- & -- & 0.0 \\
\addlinespace[4pt]
\multicolumn{7}{@{}l}{\textit{later set 1} (25 cases)} \\*[3pt]
\code{Struve H0(x)} & callable & verified & -- & $x\,f''' + 2\,f'' + x\,f' + f = \nobreak 0$ & $\displaystyle \frac{2}{\pi}\,x \,\pFq{1}{2}\!\left(1;\,3/2, 3/2;\,- x^{2}/4\right)$ & 35.1 \\
\addlinespace[2pt]
\code{J\_\{1/3\}(x)} & callable & verified & -- & $9 x^{2}\,f'' + 9 x\,f' + \left(3 x - 1\right) \left(3 x + 1\right)\,f = \nobreak 0$ & $J_{1/3}\!\left(x\right)$ & 29.0 \\
\code{exp(-x) J0(x)} & callable & verified & -- & $x\,f'' + \left(2 x + 1\right)\,f' + \left(2 x + 1\right)\,f = \nobreak 0$ & $J_{0}\!\left(x\right) e^{- x}$ & 26.9 \\
\addlinespace[2pt]
\code{1F1(1/3; 2/3; x)} & callable & verified & -- & $3 x\,f'' - \left(3 x - 2\right)\,f' - f = \nobreak 0$ & $\displaystyle \frac{2^{2/3} \pi}{\Gamma\!\left(1/6\right)}\,x^{1/6} I_{- 1/6}\!\left(x/2\right) e^{x/2}$ & 24.6 \\
\addlinespace[2pt]
\code{2F1(1/3, 2/3; 1; x)} & callable & verified & -- & $9 x \left(x - 1\right)\,f'' + 9 \left(2 x - 1\right)\,f' + 2\,f = \nobreak 0$ & $\,\pFq{2}{1}\!\left(1/3, 2/3;\,1;\,x\right)$ & 46.7 \\
\code{sqrt(1+x) + sqrt(1-x)} & callable & verified & -- & $4 \left(x - 1\right) \left(x + 1\right)\,f'' + 4 x\,f' - f = \nobreak 0$ & $\sqrt{2 \sqrt{1 - x^{2}} + 2}$ & 21.5 \\
\addlinespace[2pt]
\code{(1 + x\textasciicircum{}2/3)\textasciicircum{}-2} & callable & verified & -- & $\left(x^{2} + 3\right)\,f' + 4 x\,f = \nobreak 0$ & $\displaystyle \frac{9}{\left(x^{2} + 3\right)^{2}}$ & 12.9 \\
\addlinespace[2pt]
\code{arctan(x)} & callable & verified & $y = \arctan x$ & $y\,g' - g = \nobreak 0$ & $\arctan\!\left(x\right)$ & 11.2 \\
\code{exp(1/x)} & callable & verified & $y = 1/x$ & $g' - g = \nobreak 0$ & $e^{1/x}$ & 19.7 \\
\code{log(x)\textasciicircum{}2} & callable & verified & $y = \log x$ & $y\,g' - 2\,g = \nobreak 0$ & $\log^{2}\!\left(x\right)$ & 24.7 \\
\code{cosh(sqrt(x))} & callable & verified & -- & $4 x\,f'' + 2\,f' - f = \nobreak 0$ & $\cosh\!\left(\sqrt{x}\right)$ & 20.8 \\
\code{Ei(x)} & callable & verified & -- & $x\,f'' - \left(x - 1\right)\,f' = \nobreak 0$ & $\Ei\!\left(x\right)$ & 24.2 \\
\code{Fresnel S(x)} & callable & verified & -- & $x\,f''' - f'' + \pi^{2} x^{3}\,f' = \nobreak 0$ & $S\!\left(x\right)$ & 29.1 \\
\code{Gumbel density exp(-x - exp(-x))} & callable & verified & $y = e^{x}$ & $y^{2}\,g' + \left(y - 1\right)\,g = \nobreak 0$ & $e^{- x - e^{- x}}$ & 34.6 \\
\addlinespace[2pt]
\code{logistic density} & callable & verified & $y = e^{x}$ ($k = 1$) & $\left(y + 1\right)\,g' + 2\,g = \nobreak 0$ & $\displaystyle \frac{e^{x}}{\left(e^{x} + 1\right)^{2}}$ & 34.2 \\
\addlinespace[2pt]
\code{2e6 draws, Rayleigh(1)} & draws & conjecture & $y = x^{2}$ ($k = 1$) & $2\,g' + g = \nobreak 0$ & $x e^{- x^{2}/2}$ & 25.9 \\
\code{2e6 draws, chi-square(4)} & draws & conjecture & -- & $2 x\,f' + \left(x - 2\right)\,f = \nobreak 0$ & $\frac{1}{4}\,x e^{- x/2}$ & 21.0 \\
\addlinespace[2pt]
\code{2e6 draws, Student t(5)} & draws & conjecture & -- & $\left(x^{2} + 5\right)\,f' + 6 x\,f = \nobreak 0$ & $\displaystyle \frac{200 \sqrt{5}}{3 \pi \left(x^{2} + 5\right)^{3}}$ & 30.4 \\
\addlinespace[2pt]
\code{2e6 draws, Cauchy} & draws & conjecture & $y = \arctan x$ ($k = 1$) & $g' = \nobreak 0$ & $\displaystyle \frac{1}{\pi \left(x^{2} + 1\right)}$ & 10.6 \\
\addlinespace[2pt]
\code{2e6 draws, arcsine} & draws & conjecture & -- & $2 x \left(x - 1\right)\,f' + \left(2 x - 1\right)\,f = \nobreak 0$ & $\displaystyle \frac{1}{\pi \sqrt{x - x^{2}}}$ & 33.5 \\
\addlinespace[2pt]
\code{Apery numbers} & sequence & verified & -- & $\left(n + 2\right)^{3}\,a_{n + 2} - \left(2 n + 3\right) \left(17 n^{2} + 51 n + 39\right)\,a_{n + 1} + \left(n + 1\right)^{3}\,a_{n} = \nobreak 0$ & -- & 0.2 \\
\code{derangements} & sequence & verified & -- & $a_{n + 2} - \left(n + 1\right)\,a_{n + 1} - \left(n + 1\right)\,a_{n} = \nobreak 0$ & -- & 0.0 \\
\code{large Schroeder numbers} & sequence & verified & -- & $\left(n + 3\right)\,a_{n + 2} - 3 \left(2 n + 3\right)\,a_{n + 1} + n\,a_{n} = \nobreak 0$ & -- & 0.2 \\
\code{central trinomial coefficients} & sequence & verified & -- & $\left(n + 2\right)\,a_{n + 2} - \left(2 n + 3\right)\,a_{n + 1} - 3 \left(n + 1\right)\,a_{n} = \nobreak 0$ & -- & 0.1 \\
\code{partition numbers} & sequence & no-model & -- & -- & -- & 0.0 \\
\addlinespace[4pt]
\multicolumn{7}{@{}l}{\textit{later set 2} (22 cases)} \\*[1pt]
\code{Y0(x)} & callable & verified & -- & $x\,f'' + f' + x\,f = \nobreak 0$ & $Y_{0}\!\left(x\right)$ & 27.0 \\
\code{erfc(x)} & callable & verified & -- & $f'' + 2 x\,f' = \nobreak 0$ & $1 - \erf\!\left(x\right)$ & 11.2 \\
\code{x K1(x)} & callable & verified & -- & $x\,f'' - f' - x\,f = \nobreak 0$ & $x K_{1}\!\left(x\right)$ & 80.9 \\
\code{sin(x)\textasciicircum{}2} & callable & verified & -- & $f''' + 4\,f' = \nobreak 0$ & $\sin^{2}\!\left(x\right)$ & 11.7 \\
\code{Si(x)} & callable & verified & -- & $x\,f''' + 2\,f'' + x\,f' = \nobreak 0$ & $\operatorname{Si}\!\left(x\right)$ & 31.5 \\
\code{(1+x)\textasciicircum{}(1/3) exp(-x/2)} & callable & verified & -- & $6 \left(x + 1\right)\,f' + \left(3 x + 1\right)\,f = \nobreak 0$ & $\left(x + 1\right)^{1/3} e^{- x/2}$ & 15.9 \\
\code{exp(-x\textasciicircum{}2/2) H3(x)} & callable & verified & -- & $x \left(2 x^{2} - 3\right)\,f' + \left(2 x^{4} - 9 x^{2} + 3\right)\,f = \nobreak 0$ & $\left(8 x^{3} - 12 x\right) e^{- x^{2}/2}$ & 7.4 \\
\addlinespace[2pt]
\code{parabolic cylinder D\_1/2(x)} & callable & verified & -- & $4\,f'' - \left(x - 2\right) \left(x + 2\right)\,f = \nobreak 0$ & $\displaystyle \frac{2^{1/4}}{4 \sqrt{\pi} \Gamma\!\left(1/4\right)}\,e^{- x^{2}/4} \allowbreak\bigl(x \Gamma\!\left(1/4\right)^{2} \,\pFq{1}{1}\!\left(1/4;\,3/2;\,x^{2}/2\right) + 4 \pi \,\pFq{1}{1}\!\left(- 1/4;\,1/2;\,x^{2}/2\right)\bigr)$ & 11.6 \\
\addlinespace[2pt]
\code{1/(x (1 + log(x)\textasciicircum{}2))} & callable & verified & $y = \log x$ & $\left(y^{2} + 1\right)\,g' + \left(y + 1\right)^{2}\,g = \nobreak 0$ & $\displaystyle \frac{1}{x \left(\log^{2}\!\left(x\right) + 1\right)}$ & 21.8 \\
\addlinespace[2pt]
\code{exp(-a tan(x)\textasciicircum{}2)/\allowbreak{}cos(x)\textasciicircum{}2} & callable & verified & $y = \tan x$ ($k = 1$) & $g' + a_{1} y\,g = \nobreak 0$ & $\left(\tan^{2}\!\left(x\right) + 1\right) e^{- a_{1} \tan^{2}\!\left(x\right)/2}$ & 21.1 \\
\code{x\textasciicircum{}(1/4) exp(-x\textasciicircum{}(5/4))} & callable & verified & $y = x^{5/4}$ ($k = 1$) & $g' + g = \nobreak 0$ & $x^{1/4} e^{- x^{5/4}}$ & 24.2 \\
\addlinespace[2pt]
\code{1/(1 + exp(-x))} & callable & verified & $y = e^{x}$ ($k = 1$) & $\left(y + 1\right)\,g' + g = \nobreak 0$ & $\displaystyle \frac{e^{x}}{e^{x} + 1}$ & 24.5 \\
\addlinespace[2pt]
\code{exp(sin(x))} & callable & no-model & -- & -- & -- & 54.2 \\
\code{Ai(x), float64 table} & table & verified & -- & $f'' - x\,f = \nobreak 0$ & $\Ai\!\left(x\right)$ & 2.7 \\
\code{J0 rounded to 12 digits, table} & table & verified & -- & $x\,f'' + f' + x\,f = \nobreak 0$ & $J_{0}\!\left(x\right)$ & 1.9 \\
\code{2e6 draws, beta(3, 2)} & draws & conjecture & -- & $x \left(x - 1\right)\,f' - \left(3 x - 2\right)\,f = \nobreak 0$ & $12\,x^{2} \left(1 - x\right)$ & 29.9 \\
\code{2e6 draws, Laplace} & draws & no-model & -- & -- & -- & 15.5 \\
\code{binomial(2n, n)\textasciicircum{}2} & sequence & verified & -- & $\left(n + 1\right)^{2}\,a_{n + 1} - 4 \left(2 n + 1\right)^{2}\,a_{n} = \nobreak 0$ & -- & 0.0 \\
\code{central Delannoy numbers} & sequence & verified & -- & $\left(n + 2\right)\,a_{n + 2} - 3 \left(2 n + 3\right)\,a_{n + 1} + \left(n + 1\right)\,a_{n} = \nobreak 0$ & -- & 0.1 \\
\code{Euler totient} & sequence & no-model & -- & -- & -- & 0.0 \\
\code{moments of the uniform density on [-1, 1]} & moments & verified & -- & $f' = \nobreak 0$ & $\frac{1}{2}$ & 0.0 \\
\addlinespace[2pt]
\code{moments binomial(2n, n)/\allowbreak{}4\textasciicircum{}n (arcsine on [0, 1])} & moments & verified & -- & $2 x \left(x - 1\right)\,f' + \left(2 x - 1\right)\,f = \nobreak 0$ & $\displaystyle \frac{1}{\pi \sqrt{x - x^{2}}}$ & 0.3 \\
\end{longtable}
}
\end{landscape}

\end{document}